\documentclass[journal]{rmaa}

\usepackage{amsmath,graphicx,natbib}

\usepackage{booktabs}
\usepackage{longtable}
\usepackage{multirow}
\usepackage{url}
\usepackage{graphicx}
\usepackage{color}
\usepackage{array}
\usepackage{soul} 

\usepackage{threeparttable}

\usepackage{hyperref}

\newcolumntype{L}[1]{>{\raggedright\let\newline\\\arraybackslash\hspace{0pt}}m{#1}}
\newcolumntype{C}[1]{>{\centering\let\newline\\\arraybackslash\hspace{0pt}}m{#1}}
\newcolumntype{R}[1]{>{\raggedleft\let\newline\\\arraybackslash\hspace{0pt}}m{#1}}

\newcommand{\Ha}{$\rm{H}\alpha$}
\newcommand{\Hb}{$\rm{H}\beta$}

\DeclareRobustCommand{\ION}[2]{%
\relax\ifmmode
\ifx\testbx\f@series
{\mathbf{#1\,\mathsc{#2}}}\else
{\mathrm{#1\,\mathsc{#2}}}\fi
\else\textup{#1\,{\mdseries\textsc{#2}}}%
\fi}
\newcommand{\lam}{$\lambda$}

\newcommand{\nii}{[\ION{N}{ii}]}

\newcommand{\oii}{[\ION{O}{ii}]}
\newcommand{\oiii}{[\ION{O}{iii}]}
\newcommand{\sii}{[\ION{S}{ii}]}

\def\msun{\mbox{$M_\odot$}}
\def\mbh{\mbox{$M_\bullet$}}
\def\ms{\mbox{$M_*$}}

\begin{document}

\title{SDSS IV MaNGA - Properties of AGN host galaxies} 

%\correspondingauthor{S.F. S\'anchez}
%\email{sfsanchez@astro.unam.mx}

\author{
S.F. S\'anchez\altaffilmark{1},
V. Avila-Reese\altaffilmark{1},
H. Hernandez-Toledo\altaffilmark{1},
E. Cortes-Su\'arez\altaffilmark{1},
A. Rodr\'\i guez-Puebla\altaffilmark{1},
H. Ibarra-Medel\altaffilmark{1},
M. Cano-D\'\i az\altaffilmark{2},
J.K. Barrera-Ballesteros\altaffilmark{3}, 
C. A. Negrete\altaffilmark{2},
A. R. Calette\altaffilmark{1},
A. de Lorenzo-C\'aceres\altaffilmark{1},
R. A. Ortega-Minakata\altaffilmark{1},
E. Aquino\altaffilmark{1},
O. Valenzuela\altaffilmark{1},
J. C. Clemente\altaffilmark{1},
T. Storchi-Bergmann\altaffilmark{4,5},
R. Riffel\altaffilmark{4,5},
J. Schimoia\altaffilmark{4,5},
R. A. Riffel\altaffilmark{6,5},
S. B. Rembold\altaffilmark{6,5}, 
J. R. Brownstein\altaffilmark{7},
K. Pan\altaffilmark{8},
R. Yates\altaffilmark{9},
N. Mallmann\altaffilmark{6,5},
T. Bitsakis\altaffilmark{10}
}
\altaffiltext{1}{Instituto de Astronom\'ia, Universidad Nacional Aut\'onoma de  M\'exico, A.~P. 70-264, C.P. 04510, M\'exico, D.F., Mexico}
\altaffiltext{2}{CONACYT Research Fellow - Instituto de Astronom\'{\i}a, Universidad Nacional Aut\'onoma de M\'exico, Apartado Postal 70-264, Mexico D.F., 04510 Mexico }
\altaffiltext{3}{Department of Physics \& Astronomy, Johns Hopkins University, Bloomberg Center, 3400 N. Charles St., Baltimore, MD 21218, USA}
\altaffiltext{4}{Departamento de Astronomia, IF, Universidade Federal do Rio Grande do Sul, CP 15051, 91501-970, Porto Alegre, RS, Brazil}
\altaffiltext{5}{Departamento de F\'isica, CCNE, Universidade Federal de Santa Maria, 97105-900, Santa Maria, RS, Brazil}
\altaffiltext{6}{ Laboratório Interinstitucional de e- Astronomia, Rua General José Cristino, 77 Vasco da Gama, Rio de Janeiro, Brasil, 20921-400}
\altaffiltext{7}{Department of Physics and Astronomy, University of Utah, 115 S. 1400 E., Salt Lake City, UT 84112, USA}
\altaffiltext{8}{Apache Point Observatory and New Mexico State
University, P.O. Box 59, Sunspot, NM, 88349-0059, USA}
\altaffiltext{9}{Max-Planck-Institut f\"{u}r Extraterrestrische Physik, Giessenbachstraße, 85748 Garching, Germany}
\altaffiltext{10}{Instituto de Radioastronom\'\i a y Astrof\'\i sica, UNAM, Campus Morelia, A.P. 3-72, C.P. 58089, Mexico}
%\author{A. Rodr\'\i guez-Puebla, A. R. Calette}
%\affiliation{Instituto de Astronom\'ia, Universidad Nacional Aut\'onoma de  M\'exico, A.~P. 70-264, C.P. 04510, M\'exico, D.F., Mexico}

%\\
%$^{1}$Instituto de Astronom\'ia, Universidad Nacional Aut\'onoma de  M\'exico, A.~P. 70-264, C.P. 04510, M\'exico, D.F., Mexico \\
%\\$^{2}$ CONACYT Research Fellow - Instituto de Astronom\'{\i}a, Universidad Nacional Aut\'onoma de M\'exico, Apartado Postal 70-264, Mexico D.F., 04510 Mexico \\
\shortauthor{S.F. S\'anchez et al.}
% Title for running header
\shorttitle{MaNGA, properties of AGN hosts}

% Full postal addresses (in alphabetical surname order!)
% plus email addresses in parentheses.
\fulladdresses{

\item Instituto de Astronom\'\i a,Universidad Nacional Auton\'oma de Mexico, A.P. 70-264, 04510, M\'exico,D.F. (sfsanchez@as\
tro.unam.mx)

}

%\begin{abstract}
\abstract{We present the characterization of the main properties of a sample of 98 AGN host galaxies, both type-II and type-I, in comparison with those of $\sim$2700 non-active galaxies observed by the MaNGA survey. We found that AGN hosts are morphologically early-type or early-spirals. For a given morphology, AGN hosts are, on average, more massive, more compact, more central peaked and rather pressure- than rotational-supported systems. We confirm previous results indicating that AGN hosts are located in the intermediate/transition region between star-forming and non-star-forming galaxies (i.e., the so-called green valley), both in the Color-Magnitude and the star formation main sequence diagrams. Taking into account their distribution in terms of the stellar metallicity and oxygen gas abundance and a rough estimation of their molecular gas content, we consider that these galaxies are in the process of halting/quenching the star formation, in an actual transition between both groups. The analysis of the radial distributions of the star-formation rate, specific star-formation rate, and molecular gas density show that the quenching happens from inside-out involving both a decrease of the efficiency of the star formation and a deficit of molecular gas. All the intermediate data-products used to derive the results of our analysis are distributed in a database including the spatial distribution and average properties of the stellar populations and ionized gas, published as a Sloan Digital Sky Survey Value Added Catalog being part of the 14th Data Release: \url{http://www.sdss.org/dr14/manga/manga-data/manga-pipe3d-value-added-catalog/}}

\resumen{Este art\'iculo presenta una caracterizaci\'on detallada de las propiedades de 98 galaxias que albergan AGN, tanto de Tipo-I como de Tipo-II, en comparaci\'on con $\sim$2700 galaxias no activas. Ambas muestras proceden del mapeado MaNGA. Desde el punto de vista morfol\'ogico, las galaxias con AGN son de tipo temprano y espirales tempranas. Para una morfolog\'ia dada las galaxias con AGN son, en promedio, m\'as masivas, compactas y concentradas en las partes centrales que las no activas, adem\'as de estar m\'as soportadas por presi\'on que por rotaci\'on. Nuestro an\'alisis confirma resultados previos que indicaban que las galaxias con AGN se encuentran en la zona de transici\'on entra las galaxias con formaci\'on estelar y las que no forman estrellas, es decir, en la regi\'on del valle verde. Esto ocurre tanto en el diagrama color-magnitud como en la representaci\'on formaci\'on estelar frente a masa.  Teniendo en cuenta las distribuciones relativas de metalicidad estelar y abundancia gaseosa de ox\'igeno, y una estimaci\'on del contenido en gas molecular, consideramos que estas galaxias est\'an en el proceso de inhibir/detener su formaci\'on estelar. El an\'alisis de las distribuciones radiales de la tasa de formaci\'on estelar, la tasa de formaci\'on estelar espec\'ifica y la densidad de gas molecular muestra que esta inhibici\'on ocurre desde dentro hacia fuera, lo que supone un descenso de la eficiencia de la formaci\'on estelar y un d\'eficit de gas molecular. Todos los datos y subproductos utilizados para derivar los resultados del presente an\'alisis, incluyendo las distribuciones espaciales y las propiedades promedio de las poblaciones estelares y el gas ionizado, est\'an disponibles p\'ublicamente mediante una base de datos {\emph{Sloan Digital Sky Survey Value Added Catalog}} como parte del {\emph{14th Data Release}}: \url{http://www.sdss.org/dr14/manga/manga-data/manga-pipe3d-value-added-catalog/}}
%\end{abstract}

\addkeyword{catalogs}
\addkeyword{galaxies: active}
\addkeyword{galaxies: evolution}
\addkeyword{galaxies: nuclei}
\addkeyword{galaxies: star formation}
\addkeyword{techniques: imaging spectroscopy}

\maketitle

\section{Introduction}

Active Galactic Nuclei (AGNs) are among the most energetic processes in the Universe.  Being powered by the accretion of matter into a super-massive black hole (SMBH; \mbh$>10^6$ \msun) that resides in the center of most galaxies, they can be as luminous as their host galaxies, or even more, outshining the light of all the stars together \citep[e.g][]{jahnke04}. In essence they are characterized by a luminous point-like source residing in the center of the host galaxy.

%The discovery of the AGNs, were based on the study of radio sources. It was in fact a fortuitous founding because around 90\% of AGNs are radio quiet. Since the first detection in the 60's \citep{arpetal67,burbidgeburbidge67},
%AGNs have been classified in many ways based on their observational features. As radio-loud sources they can exhibiting compact synchrotron radio emission or large
%structures with the shape of radio-jets or disrupted lobes with a huge range of scales (from a few parsecs to tens of Megaparsecs). 

In the optical range, the AGN spectra may exhibit a characteristic power-law continuum together with a set of strong nuclear emission lines, signatures of high ionization. The characteristics of the emission lines depend on the kind of AGN and allows their classification as follows: (i) Type-I AGNs: in the case of the permitted lines,
%also some semi forbidden lines are broad, however, as they are in the UV, they were not mentioned
they can present broad components with a width of several thousands of km/s ($\sim 1000 - 10000$ km/s), usually with a narrow component superposed to the broad one. (ii) Type-II AGNs: only narrow components with a width that does not exceed 1200 km/s. %due to the nature of this paper, I omitted the explanation of the intermediate type AGNs
(iii) Blazars: no lines except when a highly variable continuum is in a low phase (BL LAC objects and optically violently variable QSOs, OVVs).

%Based on their contrasts between the central source and its host galaxies, AGNs can be classified as (a) QSO (quasi-stellar objects), if the central source dominates the intensity in a way that the host galaxy is most invisible (based on ground-based images, i.e., $\sim$1$\arcsec$ resolution ones). In radio, they were labeled as quasars (quasi-stellar radio source), or (b) Seyferts, that has been classified as type-I and type-II too. Finally, when the central ionization is very weak the boundary between AGNs and other sources of  hard ionization (like post-AGB stars) becomes diffuse, in the regime of the so-called LINERs (Low-Ionization Nuclear Emission-line Regions), in particular now the LINER-like or LI(N)ER emission has been characterized along the full optical extension of many galaxies without any evidence of the presence of an active nuclei. 

{ In addition, many radio-loud AGNs do not present any evidence for the presence of the central source in the optical range, exhibiting a perfectly normal stellar-dominated spectrum.  The undoubtedly signature of the presence of an AGN is the hard X-ray radiation, which is signature of the thermal, synchrotron, and high energetic radiation processes that happen in the accretion disk surrounding the black hole. However, the shallow detection limit of many X-ray observations affects the detectability of that feature.}

The exotic emission shown in AGNs and the relatively small fraction of AGNs in the Local Universe ($\sim$1-3\% for type-I AGNs and $\sim$20\% for type-II ones, if we include LINERs) has constrained the scope of their study to the characterization of peculiar non-thermal sources in a limited number of objects. In other words, AGNs did not seem to play any significant role on the overall evolution of galaxies. However, three
observational results have changed that view in the last decades: i) the presence  of strong correlations between the mass of the central black hole and the properties of the host galaxy, such as bulge luminosity, mass and velocity dispersion \citep[see for recent reviews][]{Kormendy+2013,Graham2016};
%REFS: Magorrian et al. 1998; Ferrarese & Merritt 2000; Tremaine et al. 2002; Marconi & Hunt 2003; Ha ̈ring & Rix 2004; Gu ̈ltekin et al. 2009; Gebhardt et al. 2011; Kormendy & Ho 2013; McConnell & Ma 2013
(ii) the need of an energetic process able to remove or heat gas in massive galaxies in order to halt their growth by star formation (SF) and re-conciliate this way the high-mass end of the observed galaxy mass (luminosity) 
functions with those derived by means of semi-analytic models of galaxy evolution \citep[e.g.,][]{Kauffmann+2000,Bower+2006,Croton+2006,DeLucia+2007,Somerville+2008}
and cosmological simulations \citep[e.g.,][]{Sijacki+2015,Rosas-Guevara+2016,Dubois+2016}; 
and (iii) the need for a fast ($\lesssim 1$ Gyr) morphological transformation between
spiral-like star-forming galaxies and dead ellipticals in the last 8 Gyrs
based on the number counting and luminosity distributions of both families 
of galaxies by different surveys \citep[e.g.,][]{Bell+2004,Faber+2007,Schiminovich+2007}. 
All together these results strongly suggest that SMBHs co-evolve with galaxies or, at least, with their spheroidal components \citep[see e.g.][]{Kormendy+2013}, and therefore AGN feedback seems to be an important phase in galaxy evolution.
%(with its number fraction indicating the relatively shortness of that phase). That phase seems to have profounds connections with the galaxy evolution, and indeed
Indeed, AGN negative feedback has been proposed as a key process to heat/eject gas, halt SF, and transform galaxies between different families \citep{Silk+1998,Silk2005,Hopkins+2010}. 
Actually, it may explain the evolutionary sequence between central low-ionization emission-line regions (LIERs) and extended LIERs proposed by \citet{Belfiore17a}.

Different observational results seem to support the scenario mentioned above. \citet{kauffmann03} showed that type-II AGNs selected from the SDSS sample 
are located in the so-called "green valley" (GV) of the color-magnitude diagram (CMD), that is, in the expected location for transitory objects between the blue cloud
of star-forming galaxies (SFGs) and the red sequence of retired/passive ones (RGs). These results were confirmed with a more detailed analysis of the host galaxies at intermediate redshift by \citet[e.g.,][]{sanchez04}, showing that type-I AGNs seem to be at the same location too. These results have been updated by more recent studies \citep[e.g.][]{Schawinski+2010,torres-papaqui12,torres-papaqui13,ortega-minakata15}. Indeed, such results indicate that AGN hosts are located in the intermediate/transitory regions in other diagrams, like the SF vs. stellar mass \citep[for a recent study see e.g.,][]{mariana16}. However, the possibility that these galaxies are found in the reported location due to a contamination by the AGN itself cannot be ruled out, as this effect has not been studied in detail. Another caveat is that, in general, the simplistic picture that all AGN hosts present evidence of recent interactions is known not to be true for most Seyfert galaxies \citep[e.g.][]{hunt99}, nor even for the stronger type-I QSOs \citep[e.g.][]{sanchez04,bohm13}. { Finally, there is a fundamental problem when comparing the properties of active and non-active galaxies. If the AGN activity is a short-lived -- compared with Hubble time -- recurrent process in galaxies, as it is assumed today, then any galaxy without an AGN could have had one in the past. Thus, any comparison between both families is only restricted to the current effects of the AGN activity in the overall evolution, and it is not possible to determine which effect may have occurred  in the past. Therefore, the fact that AGN hosts are located in particular regimes of galaxy properties is even more puzzling considering its recurrent and transitory nature.}

%Finally it is important to highligh

%\ComV{Quite el ultimo parrafo porque si bien es cierto que el virial shock heating en halos mayores a 1e12 Msun inhibe la caida de gas frio y por ende la SF en las galaxias masivas, solo con esto no es posible reproducir la funcion de L a altas luminosidades (el item (ii) de arriba justamente), lo cual invoca la necesidad del feedback del AGN como se dice arriba. Entonces incluir este parrafo implica negar el item (ii) usado como argumento a favor del rol de los AGNs en la evolución.}
%Additionally, it is not clear whether the AGN phenomenom is the key driver to quench galaxies since it is expected that in dark matter halos above $M_{\rm vir}\sim10^{12}M_{\odot}$ the star formation becomes strongly inefficient in any case \citep[e.g.][]{WhiteFrenk1991,Keres+2005,DekelBirnboim2006}

In order to address these questions, we present here a study of the main properties of the galaxies with AGN detected in the MaNGA/SDSS-IV survey \citep[Mapping Nearby Galaxies at the Apache Point Observatory,][]{manga}. We study in detail their global and radial properties compared with those of the full sample of galaxies observed by this survey, focused on the comparison of their structural (e.g., morphology, concentration) and dynamical properties (rotational vs. pressure support), and in particular their state in terms of current and recent SF activity, and its relation with the molecular gas content in these galaxies.

{ Recently, Rembold et al. (2017) studied the AGNs on the MaNGA sample using a different approach. They selected a control sample of two galaxies for each active one. They matched the properties of the host galaxies, such as mass, distance, morphology and inclination, in order to investigate if there are any stellar population properties related to the AGN alone regardless of the galaxy type. They found a correlation of the galaxy stellar population properties -- such as the contribution from different age bins as well as the mean age -- with the luminosity of the AGN. This work can be considered complementary to ours, as in our paper, we aim to compare the host properties, including the stellar population, to those of all non-active galaxies of the MaNGA sample.}

This paper is also aimed to present a Value Added Catalog (VAC) that is part of the 14th Data Release of  SDSS \citep{SDSSDR14} for the MaNGA galaxies. The dataproducts presented in the VAC were produced by the {\sc Pipe3D} pipeline \citep{Pipe3D_II}. 

This article is structured in the following way: In section \ref{sample} we describe the sample and currently used dataset; Section \ref{ana} summarizes the main steps of the performed analysis. In Section
\ref{AGNs} we describe the AGN hosts selection and the different groups in which we have classified the sample of comparison galaxies. Section \ref{results} shows the main results, presented in the following subsections: (i) subsection \ref{sec:which} shows which kind of galaxies host  AGNs; (ii) subection \ref{sec:GV} demonstrates that they are located in the GV; (iii) subsection \ref{sec:gas} shows the deficit of molecular gas in these galaxies; (iv) subsection \ref{sec:rad} and \ref{sec:rad_morph} show the radial distribution of the SF rate (SFR) and molecular gas content, demonstrating that the quenching of SF happens from inside-out, and finally (v) subsection \ref{sec:GVGs} compares the AGN hosts with the non-active galaxies in the GV. The results are discussed in Section \ref{sec:disc}, and the main conclusions are presented in Section \ref{sec:con}. The content of the distributed dataproducts included in the SDSS-DR14 VAC are described in Appendix \ref{app:Pipe3D}, and the catalog of AGN candidates is included in Appendix \ref{app:agns}. 

Along this article we assume the standard $\Lambda$ Cold Dark Matter cosmology with the parameters: H$_0$=71 km/s/Mpc, $\Omega_M$=0.27, $\Omega_\Lambda$=0.73. Finally, Table \ref{T_acron} lists all the acronyms used in this paper, including
the ones of the surveys/catalogs mentioned here.

%TABLE: Lists all the acronyms used in this paper
\begin{table}
    \centering
    \caption{List of acronyms used in this paper} 
    \resizebox{8.0cm}{!} {
        \begin{threeparttable}
            \begin{tabular}{cc}
                \hline
                \hline
                AGN &  Active Galactic Nuclei \\
                BLR & Broad Line Region \\
                BPT & Baldwin, Phillips \& Terlevich diagram \\
                CMD & Color-Magnitude Diagram \\
                EW & Equivalent Width \\
                FoV & Field of View \\
                FWHM & Full Width at Half Maximum \\                
                GV & Green Valley \\
%                GVG & Green Valley Galaxy \\
                IFS & Integral Field Spectroscopy \\
                IFU & Integral Field Unit \\
                ISM & Interstellar Medium \\
                LINERs & Low-Ionization Nuclear Emission-line Regions\\
%                LW & Luminosity Weighted \\
                IMF & Initial Mass Function \\
%                MW & Mass Weighted \\
                MZR & Mass-Metallicity Relation \\
                NLR & Narrow Line Region \\
                PSF & Point Spread Function \\
                RG & Retired Galaxy \\
%                RGMS & Retired Galaxy Main Sequence \\
				S/N & Signal-to-noise ratio \\
                SFE & Star Formation Efficiency \\
                SFG & Star-Forming Galaxy \\
%                SFMS & Star-Forming Main Sequence \\
                SFMS & Star-forming Main Sequence \\
                SFR & Star Formation Rate \\
                sSFR & Specific Star Formation Rate \\
                SMBH & Super-Massive Black Hole \\
                SSP & Single Stellar Population \\
                \hline
                DR & Data Release \\
                CALIFA &  Calar Alto Legacy Integral Field spectroscopy Area survey\\ 
                MaNGA & Mapping Nearby Galaxies at APO \\
                NSA & NASA-Sloan Atlas \\
                SDSS & Sloan Digital Sky Survey \\
                VAC & Value Added Catalog \\
                \hline
            \end{tabular}
        \end{threeparttable}
    }
    \label{T_acron}
\end{table}

\section{Sample and data}
\label{sample}

We use the observed sample by the MaNGA \citep[][]{manga} survey until June 2016 (so called MPL-5 sample). MaNGA is part of the 4th version of the Sloan Digital Sky Survey \citep[SDSS-IV][]{blanton17}. The goal of the ongoing MaNGA survey is to observe approximately 10,000 galaxies; a detailed description of the selection parameters can be found in \citet{manga}, including the main properties of the sample, while a general description of the Survey Design is found in \citet{renbin16b}. The sample was extracted from the NASA-Sloan atlas (NSA, Blanton M. \url{http://www.nsatlas.org}). Therefore, all the parameters derived for those galaxies are available (effective radius, Sersic indices, multi-band photometry, etc.). The MaNGA survey is taking place at the 2.5 meter Apache Point Observatory \citep{2006AJ....131.2332G}. Observations are carried out using a set of 17 different fiber-bundles science integral-field units  \citep[IFU; ][]{2015AJ....149...77D}. These IFUs feed two dual channel spectrographs \citep{2013AJ....146...32S}. Details of the survey spectrophotometric calibrations can be found in \citet{2016AJ....151....8Y}. Observations were performed following the strategy described in \citet{law15}, and reduced by a dedicated pipeline described in \citet{2016AJ....152...83L}. These reduced datacubes are internally provided to the collaboration 
trough the data release MPL-5. This sample includes more than 2700 galaxies at redshift 0.03$<z<$0.17, covering a wide range of galaxy parameters (e.g, stellar mass, SFR and morphology), providing with a panoramic view of the properties of the population in the Local Universe. For details on the distribution of galaxies in terms of their redshifts, colors, absolute magnitude and scale-lengths, and a comparison with other on-going or recent IFU surveys, see \citet{sanchez2017a}.

\section{Analysis}
\label{ana}

We analyze the datacubes using the {\sc Pipe3D} pipeline \citep{Pipe3D_II}, which is designed to fit the continuum with stellar population models and to measure the nebular emission lines of IFS data. This pipeline is based on the {\sc FIT3D } fitting package \citep{Pipe3D_I}. The current implementation of {\sc Pipe3D} adopts the GSD156 library of simple stellar populations \citep[SSPs][]{cid-fernandes13}, that
comprises 156 templates covering 39 stellar ages (from 1Myr to 14.1Gyr), and 4 metallicities (Z/Z$\odot$=0.2, 0.4, 1, and 1.5). These templates have been extensively used within the CALIFA collaboration \citep[e.g.][]{perez13,rosa14}, and for other surveys. %\citep[e.g.][]{ibarra16}. Este no es otro survey, se usaron galaxias de MaNGA.
Details of the fitting procedure, dust attenuation curve, and uncertainties on the processing of the stellar populations are given in \citet{Pipe3D_I,Pipe3D_II}. 

In summary, a spatial binning is first performed in order to reach a S/N of 50 across the entire field of view (FoV) for each datacube. A
stellar population fit of the co-added spectra within each spatial bin is then computed. The fitting procedure involves two steps: first, the stellar velocity and velocity dispersion are derived, together with the average dust attenuation affecting the stellar populations (A$_{V,ssp}$). Second, a multi-SSP linear fitting is performed, using the library described before and adopting the kinematics and dust attenuation derived in the first step. This second step is repeated including perturbations of the original spectrum within its errors; this Monte-Carlo procedure provides with the best coefficients of the linear fitting and their errors, which are propagated for any further parameter derived for the stellar populations.

We estimate the stellar-population model for each spaxel by re-scaling the best fitted model within each spatial bin to the continuum flux intensity in the corresponding spaxel,
following \citet{cid-fernandes13} and \citet{Pipe3D_I}. This model is used to derive the average stellar properties at each position, including the actual stellar mass density, light- and mass-weighted average stellar age and metallicity, and the average dust attenuation. In addition, the same parameters are derived accross the look-back time, which comprises in essence the SF and chemical enrichment histories of the galaxy at different locations. In this analysis we followed \citet{Pipe3D_II}, but also \citet{cid-fernandes13}, \citet{rosa16a}, \citet{rosa17} and \citet{rgb17}. In a similar way as described in \citet{mariana16} it is possible to co-add, average or azimuthal average those parameters to estimate their actual (and/or time evolving) integrated, characteristics or radial distributions.  

%That estimation of the stellar mass has a typical error of 0.15 dex, as described in \citet{Pipe3D_II}.

The stellar-population model spectra are then subtracted to the original cube to create a gas-pure cube comprising only the ionised gas emission lines (and the noise). Individual emission line fluxes were then measured spaxel by spaxel using both a single Gaussian fitting for each emission line and spectrum, and a weighted momentum analysis, as described in \citet{Pipe3D_II}. For this particular dataset, we extracted the flux intensity and equivalent widths of the following emission lines: \Ha, \Hb, \oii\ \lam 3727,
\oiii\ \lam4959, \oiii\ \lam5007, [\ION{O}{i}] \lam6301, \nii\ \lam6548,
\nii\ \lam6583, \sii \lam6717 and \sii \lam6731 \citep[although a total of 52 emission lines are analyzed][]{Pipe3D_II}. All those intensities were corrected by dust attenuation. For doing so, the spaxel-to-spaxel \Ha/\Hb\, ratio is used. Then, it is assumed a canonical value of 2.86 for this ratio \citep{osterbrock89}, and adopting a \citet{cardelli89} extinction law and a R$_{\rm V}$=3.1 (i.e., a Milky-Way-like extinction law), the spatial dust attenuation in the V-band ($A_{V,gas}$) is derived. Finally, using the same extinction law and  derived attenuation, the correction for each emission line at each location within the FoV was applied.

All the parameters derived by {\sc Pipe3D} for the $\sim$2700 galaxies/cubes studied here, including the average, integrated and characteristic values and their spatial distributions, are publicly accessible through the SDSS-IV Value Added Catalog (VAC) web-site as described in  Appendix \ref{app:Pipe3D}. In addition to the parameters described before we have derived the following properties, also included in the distributed VAC.
%webpage\footnote{\url{http://testng.sdss.org/dr14/manga/manga-data/manga-pipe3d-value-added-catalog/}}.

\subsection{Star Formation Rate}  
\label{sec:SFR}

The SFR and SFR surface densities, $\Sigma_{\rm SFR}$, were derived using the \Ha\ intensities for all the spaxels with detected ionized gas. The intensities are transformed to luminosities (using the adopted cosmology) and corrected by dust attenuation as indicated below. Finally we apply the \citet{kennicutt98} calibration to obtain the spatially-resolved distribution of the SFR surface density. We use all the spaxels irrespectively of the origin of the ionization. By doing so, we take into account the PSF wings in the star-forming regions, that may present equivalent widths below the cut applied in \citet{sanchez2017a} and \citet{mariana16} (as we will explain the following sections). On the other hand, we are including in our SF measurement regions that are clearly not ionized by young stars. For SFGs that contribution is rather low, due to the strong difference in equivalent widths, as already noticed by \citet{catalan15}, and therefore the SFR is only marginally affected. However, for the RGs, the ionization comes from other sources, including AGN ionization, post-AGB stars, or rejuvenation in the outer regions \citep[e.g][]{sarzi10,papa13,sign13,Gomes16a,Gomes16b,Belfiore17a}. Therefore, the \Ha-based SFR for RGs should be considered as an upper limit. However, for the main goals of this study (comparing the properties of the AGN hosts with respect to the overall population) that value is good enough. In general, the reported SFRs (and densities) should be considered as just a linear transformation of the H$\alpha$ luminosity (or surface density luminosity).

\subsection{Oxygen abundances} 

The spatially-resolved oxygen abundances are derived only in those spaxels whose ionization is compatible with being produced by star-forming areas following \citet{sanchez13}. For doing so, we select those spaxel located below the \citet{kewley01} demarcation curve in the classical BPT diagnostic diagram \citep[\oiii/\Hb\, vs \nii/\Ha\, diagram,][]{baldwin81}, and 
with a EW(\Ha) larger than 6\,\AA. These criteria ensure that the ionization is compatible with being due to young stars \citep{sanchez14}. Then, we used different line ratios to derive the oxygen abundance using the so-called $t2$ calibration following \citet{sanchez2017a}. In essence, this calibrator averages the oxygen abundances derived with the R23 line ratio, O3N2 and N2 calibrators \citep{marino13}, and the ONS one \citep{pilyugin10}, and corrects them using a rough estimation of the effect of the temperature inhomogeneities in the ionized nebulae { following \citet{2006RMxAC..26R.163P}}. 
In addition, we have derived the oxygen abundance using a total of 7 calibrators, described in \citet{sanchez2017a}, for comparison purposes. However, along this article we will describe only the results based on the $t2$ calibrator. For the remaining ones, the results were quantitatively different but qualitatively similar.

\subsection{Molecular Gas estimation}
\label{gas}

The cold molecular gas is a very important parameter to understand the SF processes since it is the basic ingredient from which stars are formed \citep[see e.g.,][]{Kennicutt+2012,Krumholz+2012}. Indeed, the well known Schmidt-Kennicutt law that shows the correlation of the integrated gas mass (molecular+atomic) with the integrated star-formation rate \citep[e.g.][]{kennicutt98,Saintonge+2011} is maintained at kpc-scales only for the molecular gas \citep[e.g.,][and references therein]{Kennicutt07,leroy13}. Combining the information of the molecular gas content with that from IFS has proved to be a key tool in understanding the SF in galaxies and why
it halts \citep[e.g.][]{cappellari13}, and it is opening a new set of perspectives on how to explore these processes \citep[e.g.][]{utomo17,galbany17}. Despite of its importance there are few attempts to combine both datasets on a
large number of galaxies \citep{young11,bolatto17}. Unfortunately, there are molecular
gas data for just a handful of galaxies extracted from the MaNGA survey \citep{lin17}. However, it is still possible to make a rough estimation of the amount of molecular gas in galaxies based on the estimated dust attenuation and the dust-to-gas ratio \citep[e.g.][]{brin14}. The amount of visual extinction along the typical line of sight through the ISM is correlated with the total column density of molecular hydrogen \citep[e.g.][]{bohlin1978}, with a scaling factor that at a first order can be expressed in the following way:

\begin{equation}
 \Sigma_{gas} = 15 \left( \frac{A_V}{mag} \right) (M_\odot pc^{-2}),
\end{equation}
\label{eq1}
where  $\Sigma_{gas}$ is the molecular gas mass density and $A_V$ is the line-of-sight dust attenuation \citep{Heiderman2010}. It is known that the scaling factor between the two parameters may vary from galaxy to galaxy and
even within a galaxy, depending mostly on the gas metallicity and the optical depth that regulates
the amount of dust in a particular gas cloud \citep[e.g.][]{boquien13}. We
introduce a correction factor depending on the oxygen abundance with the form:

\begin{equation}
 \Delta\Sigma_{gas} = {\rm log}(O/H)-2.67\ (M_\odot pc^{-2}),
\end{equation}
\label{eq2}
where the factor {\it 2.67} was derived by comparing our estimation
of the molecular gas with the measurements based on CO presented by \citet{bolatto17} for the galaxies on that sample, making use of the IFU data provided by the CALIFA survey \citep[][Barrera-Ballesteros et al. in prep]{sanchez12b}. The estimated molecular gas densities based on the dust attenuation do not present a systematic difference in average, with a scatter of $\sim$0.3 dex when compared with measurements based on CO \citep[e.g.][]{galbany17}, and therefore they should
be considered as a first order approximation to the real values. %However, for the ultimate goal of this article, those estimations are good enough, since our main concern is to determine whether the halt of star formation is associated with a lack of gas or it is induced by other processes.

%\section{Anlysis}
%Summary of Pipe3D. Dataproducts in an appendix.

%\section{AGN selection}

\subsection{Morphological classification}
\label{sec:morph}

The morphological properties of the present (MPL-5) MaNGA sample have been directly estimated by a visual inspection regardless of any other morphological classification that may be available in different databases \citep[e.g., Galaxy Zoo][]{galzoo}. The $gri$-color composed images of all MaNGA galaxies were displayed through a link to the SDSS server. Different zoom and scale options were used to better judge both (i) the morphological details in the inner/outer parts of galaxies and (ii) the immediate apparent galaxy environment. The classification was carried out in various steps. In a first step, images were judged according to the standard Hubble morphological classification:

\begin{enumerate}
\item  Ellipticals as roundish/ellipsoidal featureless objects without obvious signs of external disk components. No estimate of
the apparent ellipticity was carried out.
\item Lenticulars as elongated ellipsoidals showing obvious signs of a external disk component that may or may not contain a bar-like 
structure. Edge-on galaxies without any sign of structure along the apparent disk were also considered as lenticular candidates.
\item E/S0 as galaxies showing characteristics as in 1) and 2), or they are not clearly distinguished between both of them.
\item S0a as galaxies showing characteristics as in 2) but with additional hints of tightly wounded arms
\item S for spirals considering transition types as a, ab, b, bc, c, cd, d, dm, m up to Irr for irregulars.
\item Clear bars (B) and apparent/oval bars (AB) 
\item For edge-on galaxies, a galaxy is classified as S only in case a dusty/knotty structure is recognized along the disk.
\item For nearly edge-on galaxies, a more detail classification other than S is provided only in cases were clear disk/bulge structures are recognized. 
\item The apparent compact-like nature of a galaxy is emphasized. Compact cases with hints of a disk are considered as S cases. 
Compact cases without hints of any disk component are considered as Unknown (U) cases.
\item Tidal features, apparent bridges and tails, the presence of nearby apparent companions and the location of a galaxy towards a group/cluster were 
all identified and highlighted with a comment.
\end{enumerate}

In a second step, an evaluation of the morphology is carried out after (i) applying some basic image processing to the $gri$-SDSS images and (ii) judging the geometric parameters (ellipticity, position angle, A4 parameter) after an isophotal analysis. A first goal in this second step is to isolate as much as possible lenticular galaxies masquerading as ellipticals. 

The results from this morphological classification are similar to other studies for the Local Universe. In general, $\sim$30\% of our galaxies
are early-types (E/S0), and $\sim$70\% are either spirals (Sa-Sdm) or irregulars (less than a 5\%), in agreement with previous results \citep[e.g.][]{wolf05,calvi12}. In $\sim$70\% of the spirals we do not find evidence of bars (A-type), while 2/3 of the remaining 30\% correspond to strong bars (B-type) and 1/3 correspond to weak bars, in agreement with the expectations \citep[AB-type; e.g.][]{jogee04}.
%{ REF in here}
%In general we found very similar fraction of galaxies 

\begin{figure*}
  \centering
    \includegraphics[width=7.5cm, angle=270]{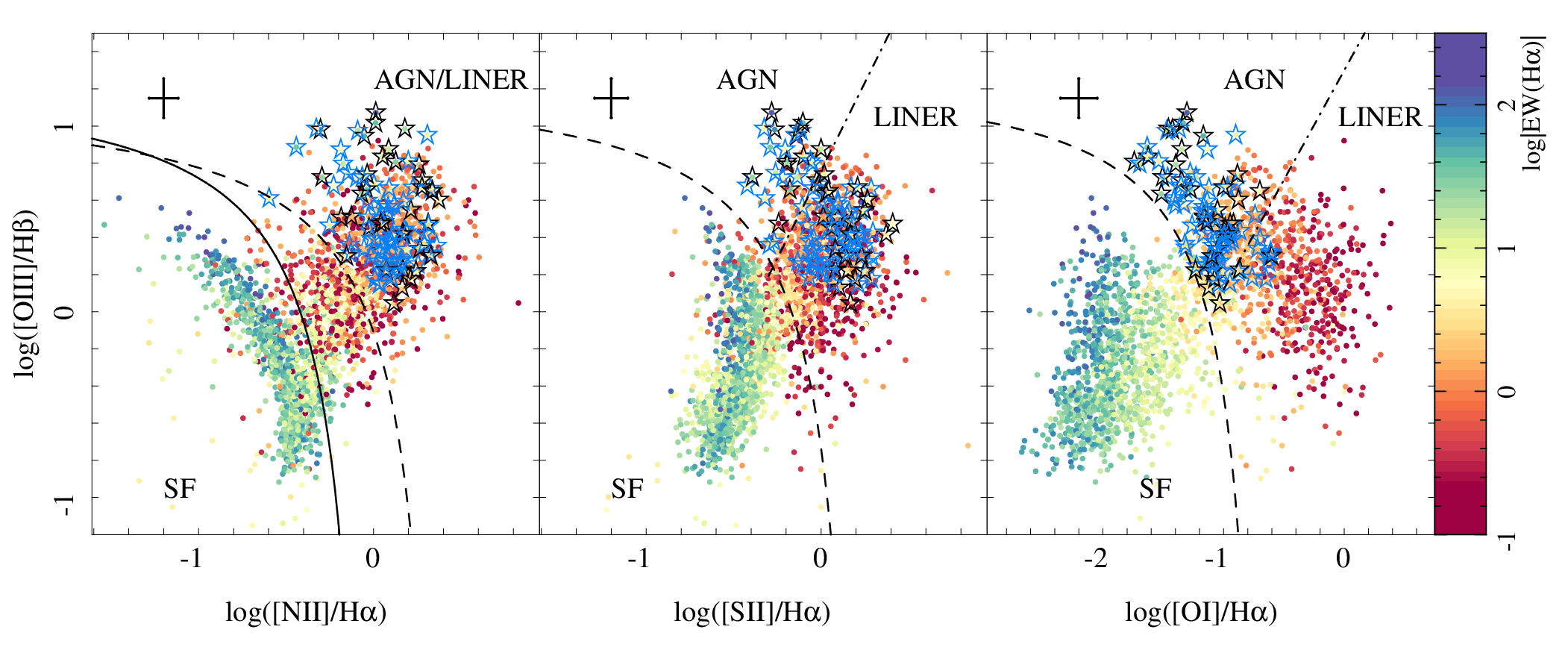}
  \caption{Diagnostic diagrams for the central ionized gas of the sample galaxies, including the distributions of the [\ION{O}{iii}/H$\beta$] vs. [\ION{N}{ii}]/H$\alpha$ line ratio (left panel), [\ION{O}{iii}/H$\beta$] vs. [\ION{S}{ii}]/H$\alpha$  (central panel), and [\ION{O}{iii}/H$\beta$] vs. [\ION{O}{i}]/H$\alpha$  (right panel). Each galaxy with ionized gas detected within an aperture of 3$\arcsec\times$3$\arcsec$ at the center of the galaxies is represented with a solid circle, color-coded by the logarithm of the equivalent with of H$\alpha$. Open stars indicate the location of our AGN candidates, colored in light-blue for the type-II and in black for the type-I candidates. The typical errors of the considered line ratios are indicated with an error-bar in each panel. Dashed lines indicate the demarcation lines adopted for our classification, as described by \citet{kewley01} and \citet{kewley06}. The solid line represents the location of the \citet{kauffmann03} demarcation line, and dashed-dotted lines represent the location of the Seyfert/LINER demarcation line. Both of them have been included only as a reference.}
  \label{fig:BPT_EW}
\end{figure*}

\subsection{AGN selection and the AGN sample}
\label{AGNs}

We select our sample of AGN candidates based on the spectroscopic properties 
of the ionized gas in the central region (3$\arcsec\times$3$\arcsec$) of the galaxies. The main goal of this selection is not to derive a sample of candidates
that include all possible galaxies with AGNs, but to select those ones that
we are more confident of being real ones. Thus, as we will see later, our selection
criteria is different than that of other studies using MaNGA data (Rembold et al., in prep), and they could be biased towards galaxies hosting strong AGNs.

Optical type-II AGNs are frequently selected based on the location of the line ratios between a set of strong forbidden lines sensible to the strength of the ionization (e.g., [\ION{O}{ii}],[\ION{O}{iii}],\ION{O}{i},[\ION{N}{ii}],[\ION{S}{ii}]) with the nearest (in wavelength) hydrogen emission line from the Balmer series (e.g., H$\alpha$, H$\beta$). That set of comparisons comprises the so-called diagnostic diagrams. The most widely used is the BPT diagram \citep{baldwin81}, that compares the [\ION{O}{iii}]/H$\beta$ versus the [\ION{N}{ii}]/H$\alpha$ line ratios. Other diagrams have been introduced later, like the ones that involve 
 [\ION{O}{iii}]/H$\beta$ versus [\ION{S}{ii}]/H$\alpha$ or [\ION{O}{i}]/H$\alpha$
\citep[e.g.][]{veil95}. \citet{kewley06} presented a summary of the most frequently used diagnostic diagrams. Figure \ref{fig:BPT_EW} shows the distribution of the line ratios for the central regions of the analyzed galaxies (2755), with a color code indicating the EW(H$\alpha$) on those regions. Only in 174 galaxies the considered emission lines were not detected, confirming previous results about the high fraction of galaxies with ionized gas detected by IFS surveys \citep[e.g.][]{Gomes16b}. Different demarcation lines have been proposed for
this diagram. The most popular ones are the \citet{kauffmann03} and
\citet{kewley01} curves. They are usually invoked to distinguish
between star-forming regions (below the Kauffmann et al. 2013 curve)
and AGNs (above the Kewley et al. 2001 curve) . The location between
both curves is normally assigned to a mixture of different sources of
ionization. Additional demarcation lines have been proposed for the
region above the \citet{kewley01} curve to segregate between Seyfert
and LINERs \citep[e.g.,][]{kewley06}. 

Although they are frequently used, the nature and meaning of the listed demarcation lines is largely ignored. The Kauffmann curve is a pure empirical tracing of the so-called classical location of star-forming/\ION{H}{ii} regions drawn to select the envelope of the galaxies that are supposed to form stars in the SDSS-DR1 catalog. Therefore, it is supposed to select the most secure higher envelope for star-forming regions: i.e., line ratios above that curve are unlikely to be produced by ionization due to young stars. However, below that curve one could still have many different sources of ionizations, contrary to the common understanding of this curve. The Kewley curve is a more physically-driven envelop, derived from the analysis of the expected line ratios extracted from photoionization models were the ionizing source is a set of young stars created along a continuous star-formation process of a maximum of 4 Myr (above that time there were found little differences). Thus, this demarcation line indicates that line ratios above it cannot be produced by ionization due to young stars (within the assumptions of the considered models). However, it says nothing regarding the nature of the ionization below it, again, contrary to the common understanding of this curve. Therefore, both lines could be used to segregate the nature of the ionization only at a first order, and in the following way: above the Kauffmann (Kewley) demarcation line the ionization is unlikely (impossible) to be produced by young stars.

In summary, to consider that all ionized regions below either the Kauffmann or Kewley demarcation lines are due to photoionization associated with OB stars is a frequent mistake. Indeed it is clearly appreciated that below both curves, and in particular the Kewley one, there is a large number of ionized regions with equivalent widths well below 6\AA\ (Fig. \ref{fig:BPT_EW}, left panel), a limit introduced by \citet{sanchez14} to impose the minimum contribution of young stars to explain the observed ionization. This limit has been recently confirmed using photoionization models by \citet{mori16}. However, it is true that
most of the ionized regions below this demarcation lines (and in particular the Kauffmann one) present larger EWs, being compatible with ionization associated with star-forming regions. On the other hand, most of the regions above the considered demarcation lines present equivalent widths of H$\alpha$ below 3\AA\ (and of the order of $\sim$1-2\AA), in particular above the Kewley curve. These values are the typical ones observed in ionization due to post-AGBs \citep[e.g.][]{binn94,sta08,sarzi10,papa13,Gomes16a,mori16}. It may be the case that there are still weak AGNs that present that low equivalent widths, but by construction they are indistinguishable from ionization due to that old stellar component, only based on the information provided by optical spectroscopy. Contrary to the common expectation, the ionization due to post-AGB stars is not only located above the described demarcation lines. It is frequently found in the bottom-right end of the classical location of \ION{H}{ii} regions, expanding to the area normally associated with the LINER-like emission \citep[e.g.][]{Gomes16b,mori16}. Finally, other sources of ionization, like shocks, are distributed well below and above the two demarcation lines. Therefore, they are in essence useless to distinguish the source of ionization in this regards unless they are combined with other information, like the morphology of the ionized area or its kinematics \citep[e.g.,][]{wild14,carlos16}. 

\subsubsection{The AGN selection procedure}
\label{sec:AGNselection}

In accordance to the discussion above, to select our AGN candidates we  apply a double criterion, imposing that (i) they have emission line ratios above the Kewley demarcation line indicated before (i.e., we exclude the star-forming regions) and (ii) the EW(H$\alpha$) is larger than 1.5\AA\ in the central regions, following \citet{cid-fernandes10}, but relaxing the criteria to include weaker AGNs.

Based on the three diagnostic diagrams shown in Fig. \ref{fig:BPT_EW} we find 683 galaxies with its central ionization above the Kewley curve in the first panel ([\ION{N}{ii}]/H$\alpha$). Out of them 142 have an equivalent width larger than 1.5 \AA. For those 683 galaxies, 629 are above the Kewley demarcation line for the central panel ([\ION{S}{ii}]/H$\alpha$), with 125 fulfilling the EW criterion. Finally, of those 629 only 302 are above the demarcation line for the right panel ([\ION{O}{i}]/H$\alpha$), with 97 fulfilling the EW criterion. Those ones represent the final sample of AGN candidates, being labeled as open stars in Fig. \ref{fig:BPT_EW}. It is worth noticing that our selected candidates are mostly above the Seyfert/LINER demarcation line for the [\ION{O}{I}]/H$\alpha$ diagram (with only 11 out of 97 objects below that curve). However, our selection still excludes one fourth of the objects above that demarcation line. This diagram presents a more clear bi-modality in the distribution of points, with a better segregation in terms of the EW(H$\alpha$) for galaxies above and below the Kewley demarcation line. This is a clear evidence that [\ION{O}{I}]/H$\alpha$ is a much better tracer of the ionization strength than the other two line ratios \citep[e.g.][]{Schawinski+2010}. On the other hand, our selection criteria disagree completely with the Seyfert/LINER demarcation line proposed for the  [\ION{N}{ii}]/H$\alpha$ and [\ION{S}{ii}]/H$\alpha$ diagnostic diagrams \citep[what can be appreciated in ][too]{Schawinski+2010} .

%%%%%%%%%%%%%%%%%%%%%%%%%%%%%%%%%%%%%%%%%%%%%%%%%%%%%%%%%%%%%%%%%%%%%
\begin{figure*}
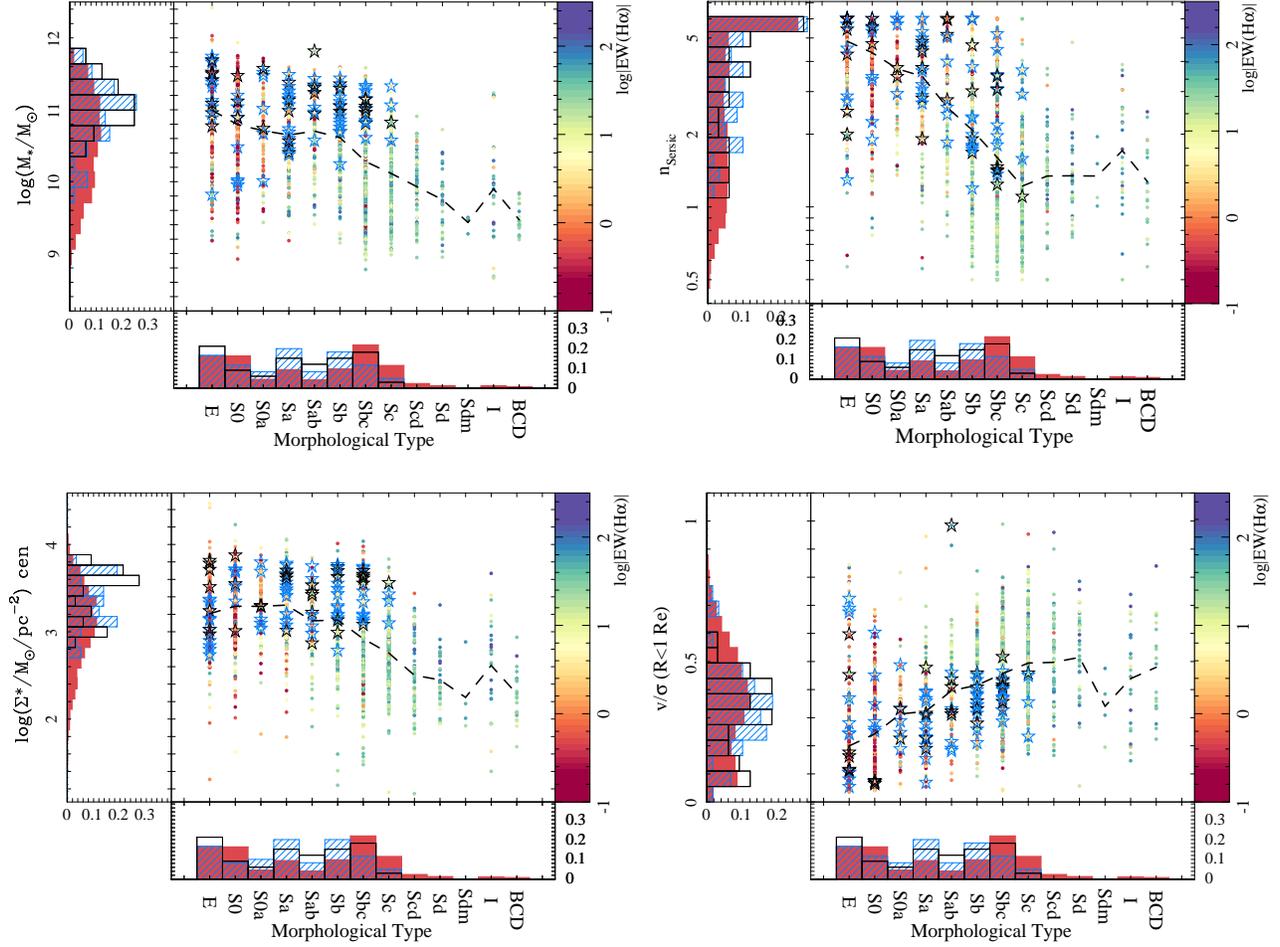

  \centering
\includegraphics[width=6.5cm, angle=270]{2bBPT_cen.ps}\includegraphics[width=6.5cm, angle=270]{2aBPT_cen.ps}
\includegraphics[width=6.5cm, angle=270]{2SMass-BPT_cen.ps}\includegraphics[width=6.5cm, angle=270]{2aaBPT_cen.ps}
\caption{Distribution of stellar masses (top-left panel), Sersic indices (top-right panel), central stellar mass density (bottom-left panel) and v/$\sigma$ ratio within one effective radius (bottom-right panel) versus the morphological type for the full sample of galaxies, using the same symbols as the ones used in Fig. \ref{fig:BPT_EW}. The dashed line in each panel represents the average stellar mass, Sersic index, effective radius and v/$\sigma$ ratio for each morphological bin respectively. The normalized histograms of each parameter for the full sample (solid red), type-II AGN hosts (hashed light blue), and type-I ones (open black), are also included. The slight differences in the histograms reflect the different number of galaxies for which the parameter shown in the y-axis has been accurately derived.}
  \label{fig:Morph}
\end{figure*}
%%%%%%%%%%%%%%%%%%%%%%%%%%%%%%%%%%%%%%%%%%%%%%%%%%%%%%%%%%%%%%%%%%%%%

%In Rembold et al. (2017) the authors have used the the same BPT and WHAM diagrams as we have used in the present paper, but for the nuclear spectrum only (from SDSS-III) to select the AGN sample from MPL-5, avoiding the "Transition Objects" (TO). Their sample contains 62 AGN, probably because they have eliminated the TOs while we did not, or due to the use of single aperture spectra.

\subsubsection{Type-I AGNs}
\label{sec:type_I}

The most broadly accepted classification in AGNs is the one that separates them between type-I and type-II depending on the presence of a broad (FWHM$\sim$1000-10000 km/s) component in the permitted emission lines \citep[e.g.][]{peterson04}. That broad component is explained within the classical Unification Scheme by the existence of ionized gas  near to the accretion disk of the SMBH, that moves fast on chaotic orbits due to the strong gravitational potential of the nucleus. This is the BLR region. The absence of broad forbidden lines is explained by the high density of the ionized gas. The classical explanation for the distinction between type-I (with an observed broad component) and type-II (without it), is the presence of a dense dust torus between this BLR and the region emitting the forbidden lines, less dense, far away and moving slowly from the nucleus, the so-called NLR, since the line-of-sight of the nucleus should be independent of it \citep{urry95}. 
%For that reason it is important to determine which AGNs in our sample have a signature of a broad component in the permitted lines.

The selection of type-I AGNs is done based on the presence of a broad component in H$\alpha$ (the strongest and easiest to analyze permitted line in our wavelength range). For doing so, we have fitted the stellar-subtracted spectrum in the central region of each galaxy (defined before) within the wavelength range covered by H$\alpha$ and the [\ION{N}{ii}] doublet by using four Gaussian functions: three narrow ones for each Nitrogen line and H$\alpha$ (FWHM$<$250 km/s), and an additional broad component for H$\alpha$ (1000$<$FWHM$<$10,000 km/s). { No     component is considered for the continuum emission by the AGN itself, since it is not relevant for this particular analysis. In a follow-up study (Cortes et al., in prep.) we are exploring in detail the properties by the type-I AGNs themselves. There we include different models for the AGN continuum emission.} The fitting was performed using {\sc FIT3D} \citep{Pipe3D_I}. Type-I AGN candidates were selected as objects for which the peak-intensity of the broad-component has a signal-to-noise larger than five. A total of 36 candidates were selected. 35 out of 36 were already selected as AGNs based on the diagnostic diagram criteria described in the previous section. The remaining one (manga-8132-6101) does not fulfill the EW cut for the narrow component, but it is above the three demarcation lines indicated before.

Our definition of type-I AGNs is broader than the more detailed classifications commonly used in the literature, in which there is a wide range of types between type II and type I, depending on the relative strength between the narrow and broad components. In here we  consider as type I any AGN with the presence of a detectable broad component, irrespectively of its relative strength.
%{ NOTE: Should we include the relative strength between the H$\alpha$ fluxes?}
% TBD in future articles

\subsubsection{The final sample of AGN galaxies}
\label{sec:final}

In summary, we have selected 98 AGNs out of 2755 galaxies ($\sim$4\%), a fraction very similar to the one reported by \citet{Schawinski+2010}. 
For 36 of them we have detected a possible broad component in the permitted lines, being classified as type-I AGNs (black-open stars in Fig. \ref{fig:BPT_EW}). The remaining (63 AGNs) are classified as type-II ones (light blue-open stars in Fig. \ref{fig:BPT_EW}). Table \ref{tab:agns} in the Appendix \ref{app:agns} presents the list of all the AGN candidates, including the main properties used to classify them. 

{ We should stress here that our selection is clearly biased towards gas rich, bright nuclear sources, such as any sample of optically selected  AGNs. Other active nuclei like (i) the radio-galaxies, which in many cases present weak or no emission lines \citep[e.g.][]{willott01}, (ii) the low frequent BL Lac or type-0 objects \citep[e.g.][]{urry95}, or (iii) the dusty/obscured AGNs \citep[e.g.][]{benn98} are excluded by the current selection. However, we consider that this selection does not impose any strong bias in our sample  for the final goals of this study. First, the time scales between radio emission and nuclear activity in radio-loud AGNs are different, in particular for those radio-galaxies without signatures of AGN activity and extremely large radio structures \citep[][]{Buttiglione10,Tadhunter12}. Therefore, they could be considered as the fossils of a past nuclear activity rather than a current on-going one. Second, the number of the exotic type-O objects is so low that excluding them would not compromise our results; and third, the fraction of obscured AGNs is known to be lower than what it was anticipated in the past, and there are few differences in the selection of optical and X-ray AGNs apart from in the range of weak AGNs, that in any case are excluded from our analysis \citep[e.g.][]{Georgantopoulos2010}. Thus, our selection is restricted to galaxies currently hosting an active and strong AGN with enough gas to present clear signatures of the activity in the optical emission lines.
%Finally, we would like to stress that our selection is 
}

%\Com{Aldo: Segun entiendo Alenka busco estos AGNs en el catalogo de Veron-Cetty, no seria buena idea reportar los que estan confirmados en ambos catalogos?}
% OK: Diselo a ALENKA!!!!

\section{Results}
\label{results}

%The main goal of this article is to characterize the main properties of the host galaxies of AGNs and to determine its location in the possible evolutionary sequence between star-forming and retired/passive galaxies, SFGs and RGs, respectively. For doing so, we will compare their main global and radial properties with those of the overall sample of galaxies.

\subsection{Which galaxies host an AGN?}
\label{sec:which}

Figure \ref{fig:Morph} shows the morphological distribution of the AGN hosts (type-II and type-I), compared with that of their non-active counterparts, with respect to different properties of those galaxies: (i) the integrated stellar mass, (ii) the Sersic index, (iii) the stellar mass density in the central region (defined before), and (iv) the rotation velocity-to-velocity dispersion ratio (v/$\sigma$) within one effective radius. The general trends found for the bulk of galaxies between their morphological type and the different analyzed properties follow in average the expected distributions. Late-type galaxies are in general less massive, less concentrated (lower Sersic indices), with smaller central stellar mass densities, and more frequently supported by rotation than by pressure (disordered motions). On the other hand, early-type galaxies are more massive, more concentrated, with larger stellar mass densities and they are more frequently supported by pressure. The trends are clearly defined for all the morphological types, in agreement with previous results on larger statistical samples \citep[e.g.][]{nair2010}. Only for the elliptical galaxies (E-type) we find a slightly larger distribution of the analyzed properties, in particular for the V/$\sigma$ ratio. 
%{ Comparison with previous results?}
Even more, there is a clear trend between the morphology of the galaxies and the H$\alpha$ EW in the central regions, with late-type galaxies prompt to present higher values than early-type ones, most likely as a consequence of the connection between morphology and ionization in galaxies.

Regarding stellar mass, AGN galaxies, specially type-I, present a distribution strongly biased to larger masses with respect to the distribution of non-active ones (see first panel in Fig. \ref{fig:Morph}). { For the} morphology, the fraction of elliptical galaxies is rather similar for non-active, type-I and type-II AGN galaxies, indicating no clear preference for AGNs to be located on this type of galaxies. For lenticular galaxies, there is a deficit of both AGN types in comparison with E-type or Sa galaxies.
Type-II AGNs are more frequently found in early-type spirals. In particular the fraction of these objects found in S0a, Sa and Sab, is almost twice than that of their non-active counterparts, at the expenses of a much lower number in Sbc, and none of them in more late type spirals. Type-I AGNs are found also in early-type spirals. However, they are far more frequent in Sb and Sbc galaxies. 
Taking into account that our morphological classification for type-I AGNs maybe affected by the presence of the point-like source itself we will take that distinction between both types with caution. Despite of this caveat, in general we can conclude that the morphological distribution of AGN hosts with respect to that one of non-active galaxies is biased to types S0a-Sbc ($\sim$70\%) and E/S0 ($\sim$30\%) and none in spirals later than Sc. { A similar result has been previously reported in the literature \citep[e.g.,][]{catalan17}.}

Regarding the presence of bars, we find that for the spiral AGN hosts $\sim$50\% of them do not present  evidence of bars. This fraction is clearly lower than the value found for all the galaxies \citep[$\sim$70\%, see e.g.][]{Menendez+2007,Sheth+2008,Cisternas15}. Indeed, AGN hosts show a larger fraction of strong bars ($\sim$40\%) and a similar fraction of weak bars. This result may indicate that AGNs are more frequently found in barred galaxies, a result that it is controversial since different authors have found different results in this regards \citep[e.g.][]{Cisternas15}. However, we have to take it with caution. The detectability of a bar is affected by many parameters, from the selected observed band to the resolution of the images. Considering the wide range of redshifts covered by the MaNGA sample we cannot be totally sure that our detectability is not affected by resolution effects. Even more, AGN hosts are biased towards more early-type spirals in our sample, and in this regime the fraction of barred galaxies increases. 
%Even more, the sample selection of MaNGA \citep[e.g.][]{manga} preclude us to make any statement on this regards without making a proper volume correction. 
A more detailed analysis of the bar fraction on a sub-set of well resolved galaxies will be presented elsewhere (Hernandez-Toledo et al., in prep.)

%AGN hosts are not equally distributed along the different properties we have compared with for the same morphological type. 
In Fig. \ref{fig:Morph} we represent with a dashed-line the mean value of the considered parameter for each morphological type. The location of AGN hosts (represented by open stars) is clearly asymmetrical with respect to this mean value. In general, they are more massive ($\sim$75\% above the mean value), more concentrated ($\sim$70\%), with larger stellar-mass densities in the central regions ($\sim$75\%), and less rotational-supported ($\sim$65\%). Moreover, like in the case of the morphological distribution, we find clear differences between type-I and type-II AGNs. The former ones are more massive in general ($\sim$84\%), with higher mass densities in the central regions ($\sim$79\%), and more dominated by pressure ($\sim$80\%).

%In summary, AGN hosts are mostly elliptical galaxies or early-type spirals, and on average more massive, compact, and less rotational-supported than the general population of those galaxies. In other words, AGNs are scarce in late-type spirals, and for early-type ones, AGNs are are much less frequent in the less massive, less compact, and more supported by rotation galaxies.

%%%%%%%%%%%%%%%%%%%%%%%%%%%%%%%%%%%%%%%%%%%%%%%%%%%%%%%%%%%%%%%%%%%%%%%%%%%
\begin{figure*}[th]
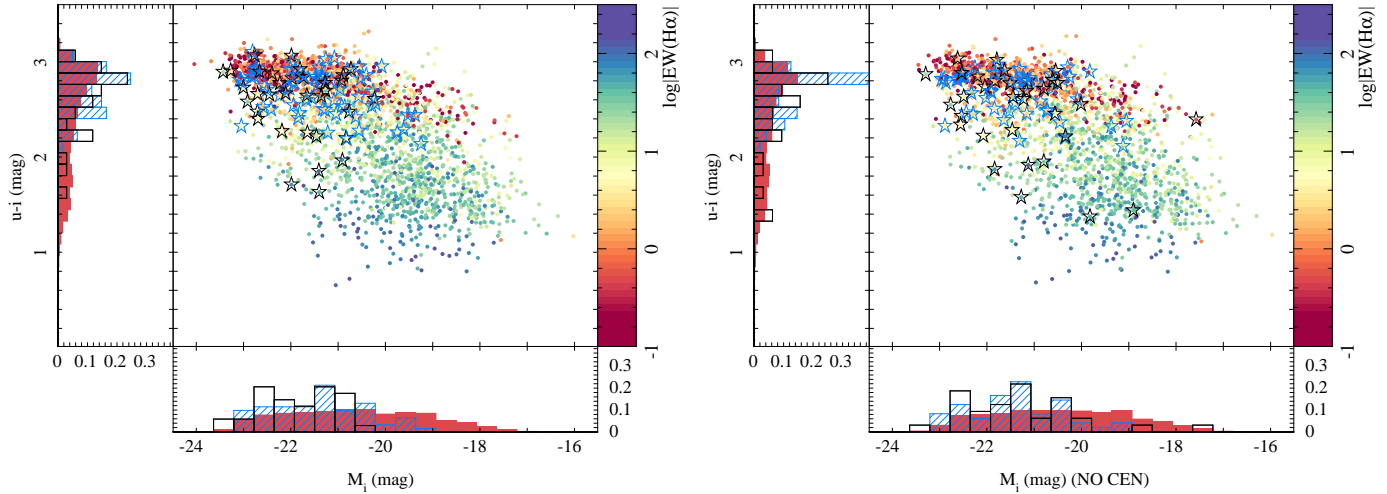

  \centering
    \includegraphics[width=7cm, angle=270]{3aBPT_cen.ps}\includegraphics[width=7cm, angle=270]{3bBPT_cen.ps}
  \caption{Distribution of the $u-i$ color versus the $i$-band absolute magnitude (CMD) for the full sample of galaxies analyzed in this study, using the same symbols than in Fig. \ref{fig:BPT_EW}. Left panel shows the color extracted from the original MaNGA datacubes, while right panel shows the same colors once subtracted the contribution of the central 3$\arcsec$$\times$3$\arcsec$ region. The normalized histograms for each respective parameter for the full sample (solid red), type-II AGN hosts (hashed light blue), and  type-I (open black) are included.}
  \label{fig:CMD}
\end{figure*}

%%%%%%%%%%%%%%%%%%%%%%%%%%%%%%%%%%%%%%%%%%%%%%%%%%%%%%%%%%%%%%%%%%%%%%%%%%%%

\subsection{Are AGN hosts in the green valley?}
\label{sec:GV}

During the last decade it has been clearly established that galaxies in the Local Universe and at least in the last $\sim$8-9 Gyrs ($z\sim$1) present a clear bi-modality in most of their properties \citep[][for a review, see \citealp{Blanton+2009}]{Strateva+2001,Baldry+2004,blanton03,Bell+2004,blanton05}. On one hand, early-type galaxies are mostly supported by  velocity dispersion, being more compact, more massive, and populated by older stars. They present lower gas fractions, and a lower degree or almost absence of SF. On the other hand, late-type galaxies are mostly supported by rotation, being less compact, less massive, and with younger stellar populations. They also present higher gas fractions, and a larger degree of SF. This separation in their main properties does not show as a continuous distribution, but it has a bimodal shape. This is evident when most of those properties are compared, and it was first highlighted in the CMDs. When integrated blue-to-red colors of galaxies are represented along their absolute magnitude, early- and late-type galaxies split clearly in two groups: (i) the red sequence \citep[already known for decades from the study of galaxy clusters, e.g.][]{butcher84,sanchez02,sanchez07b} and (ii) the blue cloud. In between them there is a region of low number density of galaxies, frequently known as the green valley, GV. This bimodal distribution and the scarcity of galaxies in the GV suggest that the transformation (if any) between both groups has to be fast compared with the Hubble time \citep[e.g.,][but see \citealp{Schawinski+2014} and \citealp{Smethurst+2015}]{Bell+2004,Faber+2007,Martin+2007,Goncalves+2012,Lian+2016}. { The fact that galaxies in low density groups, strongly affected by tidal interactions and with signatures of E+A spectra are more frequently found in the GV \citep[e.g.][]{Bitsakis+2016} supports the scenario that these are galaxies under transformation.}

The negative feedback produced by AGN has been proposed as a mechanism for halting SF (see references in the Introduction), and hence, for fostering the transition from the blue cloud to the red sequence { \citep[e.g.][]{catalan17}}. The fact that AGN hosts are mostly located in the transitory GV region supports this proposal \citep[e.g.][ but see \citealp{Xue+2010} and \citealp{Trump+2015}]{kauffmann03,sanchez04}.  Following, we will explore whether the AGN host galaxies in the MaNGA sample are in the GV or not.  

%%%%%%%%%%%%%%%%%%%%%%%%%%%%%%%%%%%%%%%%%%%%%%%%%%%
\begin{figure*}
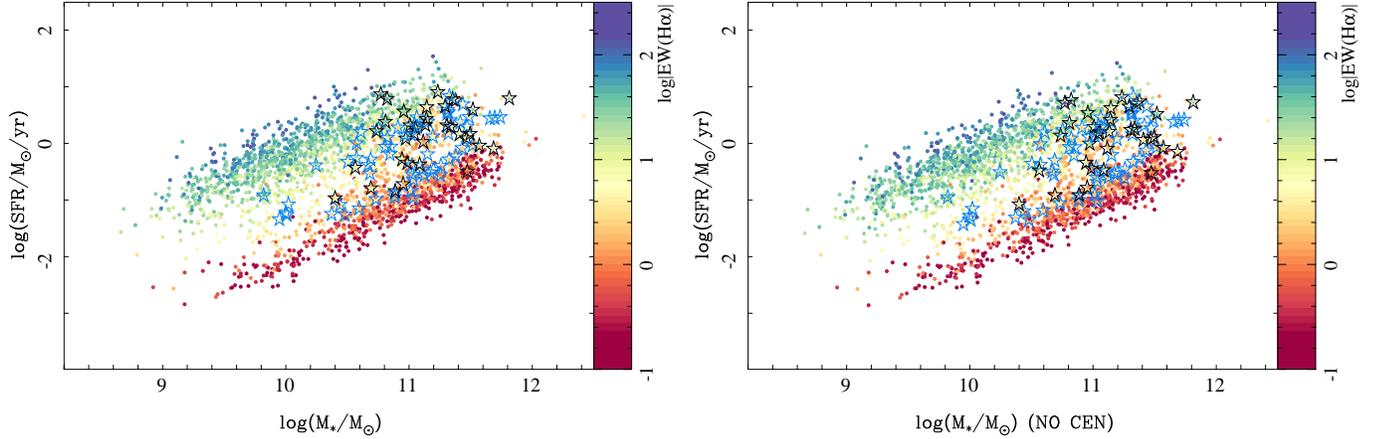

  \centering
    \includegraphics[width=6.1cm, angle=270]{1BPT_cen.ps}\includegraphics[width=6.1cm, angle=270]{1aBPT_cen.ps}
  \caption{Distribution of the SFR versus the stellar mass for the full sample of galaxies analyzed in this study, using the same symbols as in Fig. \ref{fig:BPT_EW}. We should note the SFR, as defined here, is just a linear transformation of the H$\alpha$ luminosity and for RGs it should be considered as an upper-limit to the real SFR (due to the contribution of other ionization sources). Left panel shows the SFR derived co-adding the H$\alpha$ luminosity across the entire FoV of the MaNGA datacubes, while right panel shows the same value once the contribution of the central 3$\arcsec$$\times$3$\arcsec$ region is subtracted.}
  \label{fig:SFMS}
\end{figure*}
%%%%%%%%%%%%%%%%%%%%%%%%%%%%%%%%%%%%%%%%%%%%%%%%%%%

\subsubsection{Color-magnitude diagram}
\label{sec:CMD}

Fig. \ref{fig:CMD}, left panel, shows the $u-i$ vs. M$_{abs,i}$ CMD for the full sample of galaxies, together with the AGN hosts. The magnitudes have been extracted from the MaNGA datacubes, co-adding the spectra within the FoV 
%of the datacubes 
and convolving them with the transmission of the considered SDSS filters, deriving the magnitudes in the AB photometric system. There is a clear bimodal distribution in the CMD, that it is better highlighted by the typical EW of H$\alpha$ in the central regions of the galaxies:  galaxies in the red sequence present a low EW of H$\alpha$ in most of the cases ($\sim$1-3 \AA), while galaxies in the blue cloud present much larger values ($\sim$10-500 \AA). AGN hosts are mostly located in the bluer end of the red-sequence towards the GV. This is also evident in the color-histograms of the same figure. In general type-II AGNs are more clearly packed just below the red sequence, covering a narrower range of colors. On the other hand, type-I ones are distributed covering a broader range of colors. This very same result was already noticed by \citet{sanchez04}. 

A basic criticism to the described location of AGN hosts in the CMD is that the nuclear source, intrinsically blue and potentially strong, may alter the over-all colors of the objects and shift them towards the GV. This could be particularly important in the case of type-I AGNs \citep[e.g.][]{sanchez03,jahnke04,jahnke04b,zhang16}. In order to explore that possibility we have repeated the derivation of the magnitudes and colors but subtracting the central spectra for each datacube. These central spectra corresponds roughly to an aperture of the size of the PSF, and in principle it should remove the strongest effects of the nuclear source. This procedure was performed for all the galaxies, irrespectively of the presence or not of an AGN. Fig. \ref{fig:CMD}, right panel, shows the CMD for the full sample of galaxies and the AGN hosts once this nuclear subtraction is performed. Despite of having subtracted the nuclear region, the new distribution looks very similar to the previous one. The colors of  type-II AGN hosts are more concentrated towards the region just below the red sequence, as it is clearly seen when comparing their histograms. Type-I hosts appear more dispersed, occupying the GV region. For some particular objects the contamination by the AGN is very strong, in particular for type-I hosts. In at least three cases they shift to very different locations than the initial ones, moving them from the red sequence to the blue cloud (two cases) or towards the lower luminosity end of the that sequence (one case). However, this does not affect the overall distribution.

This result indicates that most of AGN hosts are really located in the intermediate region between the blue cloud and the red sequence, and this preferential location does not seem to be induced by the photometric pollution of the AGN itself.

\subsubsection{Star Formation vs. stellar mass}
\label{sec:SFMS}

The location of AGN hosts in the GV of the CMD has induced other authors to explore whether they are located also in intermediate positions of other diagrams that exhibit the bimodal distribution described before. A major example is the SFR vs. integrated stellar mass \citep[][etc]{Brinchmann04,Salim07,Noeske+2007,Renzini15,Sparre15}. This diagram shows two clearly distinguished regions where galaxies concentrate \citep[e.g.,][]{mariana16}: (i) the star-forming Main Sequence (SFMS), which shows a linear correlation between the logarithm of the SFR and the logarithm of M$_{*}$, with a slope slightly lower than one ($\sim$0.8), and (ii) the Sequence of passive or retired galaxies, RGs, which shows another linear correlation but with a smaller normalization and a slope slightly larger than that of SFMS. Both correlations exhibit a tight distribution, with a dispersion of $\sim$0.2-0.3 dex, slightly larger in the case of the RGs. The slope of the SFMS seems to be rather constant over cosmological time. However, the zero-point presents a shift towards larger values in the past, following the cosmological evolution of the SFR in the Universe (see \citealp{Speagle14}, \citealp{Katsianis15}, and \citealp{Rodriguez-Puebla+2017} for a recent compilation of many works).

The nature of those correlations is intrinsically different, and it usually generates some confusion. The former correlation indicates that when galaxies form stars,  the integrated SFR follows a power of the look-back time (not an exponential profile as generally assumed), with the power being almost constant in at least the last 8 Gyrs. 
%
% SFS: Ok... asumamos esto por ahora.
% 
%\Com{Aldo: Esto claro si las galaxias evolucionan alrededor de la relacion promedio SFR-$M_*$ de las SFMS. VAR: de acuerdo con Aldo, cuidado con esta interpretacion, aunque si es cierto que las SFing galaxies no estan en una fase de decaimiento exponencial}. 
The later correlation does not reflect any kind of connection between the SFR and M$_{*}$, since actually the dominant ionizing source for galaxies in the RG sequence is not compatible with SF. As pointed out by \citet{mariana16}, their ionization is located in the so-called LINER-like (or LIER) area of the BPT diagram, being most probably dominated by some source of ionization produced by old-stars \citep[e.g., post-AGBs;][]{keel83,binn94,binn09,sarzi10,cid11,papa13,sign13,Gomes16a,Gomes16b,Belfiore17a}. The fact that its luminosity correlates with M$_{*}$ reinforces its stellar nature, indicating that they most probably present a characteristic EW(H$\alpha$) \citep[e.g.][]{mori16}. Actually, when the SFR is not derived from the H$\alpha$ ionized gas, like in our case, but it is extracted from the analysis of the stellar population using inversion methods, this second trend is less evident as pointed out by \citet{rosa17}. %Since both methods present limitations, we will continue using our derived SFR for RGs, taking into account that it is at the best an upper limit to the current SFR.

Fig. \ref{fig:SFMS} shows the SFR-$M_*$ diagram for the full sample of galaxies analyzed in here, together with the AGNs hosts.
%As expected, there is a clear bimodal distribution with the SFMS and the RG sequence clearly distinguished. For this later ones, we have already indicated that the H$\alpha$ luminosity cannot be directly associated with a SFR, and, in the best case, it will reflect the upper-limit to the current SFR in the considered galaxies. However, in order to compare with the literature and between star-forming and non star-forming galaxies we have just applied the linear transformation between both quantities. 
The { well known} bi-modal distribution is highlighted by the clear difference between the EW of H$\alpha$ in the central regions between the SFMS and the RG sequence, with a segregation more pronounced than in the case of the CMD. Actually, Cano-Diaz et al. (in prep.) have shown that by making a cut at the EW(H$\alpha$)=6 \AA\ it is possible to distinguish clearly not only between the SFMS and RGs but between star-forming and non star-forming regions within the galaxies. We will use this criterion to define star-forming/non star-forming regions and galaxies through this study. 

%%%%%%%%%%%%%%%%%%%%%%%%%%%%%%%%%%%%%%%%%%%%%%%%%%%
% Age-Mass
\begin{figure*}
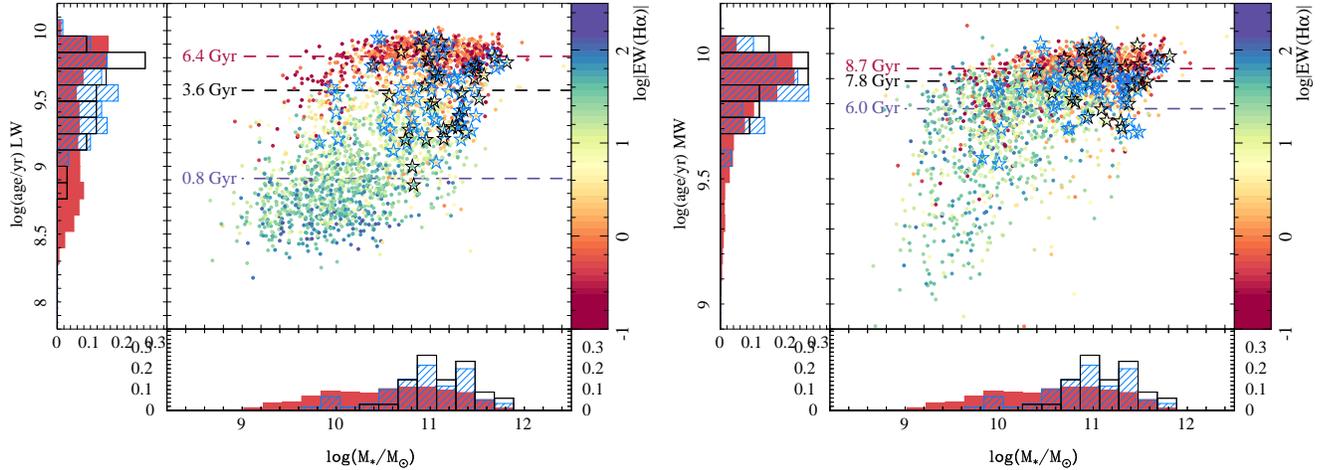

  \centering
    \includegraphics[width=6.5cm, angle=270]{13BPT_cen.ps}\includegraphics[width=6.5cm, angle=270]{14BPT_cen.ps}
  \caption{Distribution of the average age along the stellar mass for the full sample of galaxies analyzed in this study, using the same symbols as  in Fig. \ref{fig:BPT_EW}. Left panel includes the luminosity-weighted ages, while right-panel shows the mass-weighted ages. The dashed horizontal lines and the corresponding values indicate the average ages for the RGs (red), AGNs (black) and SFGs (purple).}
  \label{fig:Age_Mass}
\end{figure*}
%%%%%%%%%%%%%%%%%%%%%%%%%%%%%%%%%%%%%%%%%%%%%%%%%%%

%\subsubsection{AGN galaxies in the SFR--\ms\ diagram}

The location of the AGN hosts between the regions of the SFMS and the RGs sequence in the SFR--\ms\ diagram was first reported by \citet{mariana16}, already hinted by \citet{catalan15} { and confirmed by \citet{catalan17}}. Fig. \ref{fig:SFMS} confirms these results. Both type-I and type-II AGN hosts are clearly located in between the SFMS and RG regions. Like in the case of the CMD, the location of both AGN types is slightly different. Type-I hosts are more concentrated in the higher-mass range and more frequently found in the lower end of the SFMS. On the other hand, Type-II hosts are more broadly distributed in terms of their mass (what is also seen in Fig.\ref{fig:Morph}, upper-left panel), and they are found both in the lower-end of the SFMS and the upper-end of the RGs region. We cannot foresee any non-physical reason or selection bias that explains that separation. If true, it may indicate that both families of AGNs are intrinsically different, or at least that they host galaxies that evolve in a different way.

As in the case of the CMD, a possible reason why AGN hosts are located in the intermediate regions between star-forming and non star-forming galaxies could be the contamination of the nuclear source. In this particular case the strongest effect would be an increase of the H$\alpha$ luminosity, due to ionization by the AGN, that will shift galaxies in the RG sequence up towards the intermediate area. \citet{catalan15} already explored that possibility for type-II AGNs and found that the contamination is small and can be neglected in comparison with the overall integrated H$\alpha$ luminosity across the entire galaxy. This has not been tested yet for type-I AGNs. Despite the possible contamination by the central ionization through the entire optical extension of the host galaxy that has been observed in different AGNs \citep[e.g.][]{husemann10,bego05}, the strongest contribution is located in the central regions. Therefore, following the same procedure described for the CMD above, we estimate the decontaminated stellar-mass density and SFR by subtracting the contribution of the central region (PSF size) to both quantities. Fig. \ref{fig:SFMS}, right-panel, shows the result of this analysis. As in the case of the CMD, and in agreement with the results from \citet{catalan15} { and \citet{catalan17}}, the location within the  SFR--\ms\ diagram of AGN hosts is not significantly affected by the possible pollution by the nuclear source. This result indicates that indeed the AGNs are hosted by galaxies that are genuinely located in the intermediate/transition region between the SFMS and the RG regions in the SFR--\ms\ diagram.

{ It is still possible that the selected AGN hosts are located in the GV due to poor selection. As we stated in Sec. \ref{sec:final} our selection excludes weak AGNs that may reside in early-type, gas poor and mostly retired galaxies; in particular we have excluded all radio-galaxies. Those AGNs would reside most probably in the sequence of RGs. However, as stated before, the time-scales between the extended radio emission and the nuclear activity may be different, with the former being much longer, and here we are discussing the properties of the host galaxies of currently active AGNs. While most of radio-loud but optically inactive galaxies would reside in the RG region \citep[e.g., M87][]{Butcher1980}, we speculate that the optically active ones -- those that present strong optical emission lines \citep[e.g., 3C120][]{Sanchez2004,Begona2005} -- should be located in the GV as their radio-quiet counter parts. We intend to explore this possibility in a future  study. 

On the other hand, our optically selected AGN candidates may be out-shined by the intense circumnuclear SF in the case of the bright star-forming galaxies. \citet{elli17} has recently confirmed that galaxies with stronger integrated SFR are those that present stronger nuclear $\Sigma_{\rm SFR}$ compared to the average population. Based on this result it may be possible that our AGN detection is precluded for galaxies in the SFMS, and, in combination with our bias against gas poor/weak AGNs, we detect only those found in the GV due to poor selection. We explore that possibility by comparing the H$\alpha$ flux intensities and luminosities with the central aperture considered in this study between active and non-active galaxies. While we reproduce the results of \citet{elli17}, none of the SF galaxies has a stronger H$\alpha$ luminosity than the selected AGNs. Thus, being out-shined by a circumnuclear SF is highly unlikely. Even more considering that due to the strong differences in the line ratios an AGN would present clear signatures in the emission line ratios even if the H$\alpha$ luminosity is 10 times weaker than that produced by SF in the same aperture. We should stress out here that this holds for kpc-scale spatially resolved spectroscopic data. In the case of flux intensities integrated over much larger apertures, in particular for the full optical extension of the galaxies, the shading by SF should have a stronger effect.}

%%%%%% Age-Mass Discussion %%%%%%%
\subsubsection{The age--mass diagram}
\label{sec:age_mass}

Within most of the proposed scenarios, galaxies in the GV (like AGN hosts) are in transition between the SFMS and RG region. 
%It is required that that transition is shorter than the Hubble time, due to the low number of galaxies in this location (see above).
It is possible to estimate the amount of time required to complete the transition by comparing the average ages of the stellar populations for the SFGs, RGs, and AGN hosts. Fig. \ref{fig:Age_Mass} shows the characteristic luminosity- and mass-weighted ages for the stellar populations \citep[i.e., the value at the effective radius][]{rosa16a,Pipe3D_II} for all the analyzed galaxies, together with the AGN hosts. Like in the case of the CMD and the SFR--\ms\ diagram, AGN hosts are located in an intermediate region between the intermediate/young-age SFGs and the old RGs. %VAR: modifique el texto para tomar en cuenta edades en vez de colores
For the luminosity-weighted ages (normalized at 5500\AA\ \citealp{Pipe3D_I}), which are more sensitive to the young stellar populations, the AGN hosts are $\sim$3 Gyr older than the SFGs and $\sim$3 Gyr younger than the RGs. In the case of the mass-weighted ages, which are more sensitive to the bulk of stars and are on average formed at early cosmological times \citep[e.g.][]{per-gon08,perez13,ibarra16}, the respective differences are of $\sim$1 Gyr with respect to the RGs and $\sim$2 Gyr with respect to the SFGs. If we consider the offsets between the average ages for the different types of galaxies as a clock of the last massive SF event that contributed significantly to the light (and in less amount to the mass of the galaxy) we may consider that the quenching in local AGN hosts happened about 1-2 Gyrs ago.

Despite of these results, we should be cautious in making a causal connection between AGN activity and the transition between both groups. In particular, we should highlight the fact that only half to one-third of the galaxies in the so-called GV (either in the CMD or the SFR--\ms\ diagrams) host and AGN. For the remaining galaxies, either the AGN is too weak to be selected by restrictive EW cut or they do not host a nuclear source, and therefore, their transition either implies a different time-scale than the AGN activity or there is no mandatory need for an AGN to be active during the transition process. We should keep that in mind in order not to over-interpret the results. 

Finally, more recent studies have suggested that the preference of AGNs for the GV and bulge-dominated galaxies is the result of selection effects \citep[see e.g.,][ and references therein]{Xue+2010,Trump+2015}. This selection effect could be due to the fact that AGN signatures in the diagnostic diagrams can be hidden by \ION{H}{ii} regions in galaxies with significant levels of SF, particularly in the BPT diagram, and after accounting for this bias AGNs are most common in massive galaxies with high sSFRs \citep{Trump+2015}. Note, however, that this is not applicable to our analysis as we are using a more restrictive criterion for selecting AGNs, which results in a selection of strong AGNs. Indeed, strong AGNs are expected to affect more their host galaxies. Moreover, it is not clear whether the above mentioned studies could have aperture effects on their AGN detections, which is not our case.

%%%%%%%%%%%%%%%%%%%%%%%%%%%%%%%%%%%%%%%%%%%%%%%%%%%%%%%%%%%%%%%%%%%%%%%%%%
\begin{figure*}
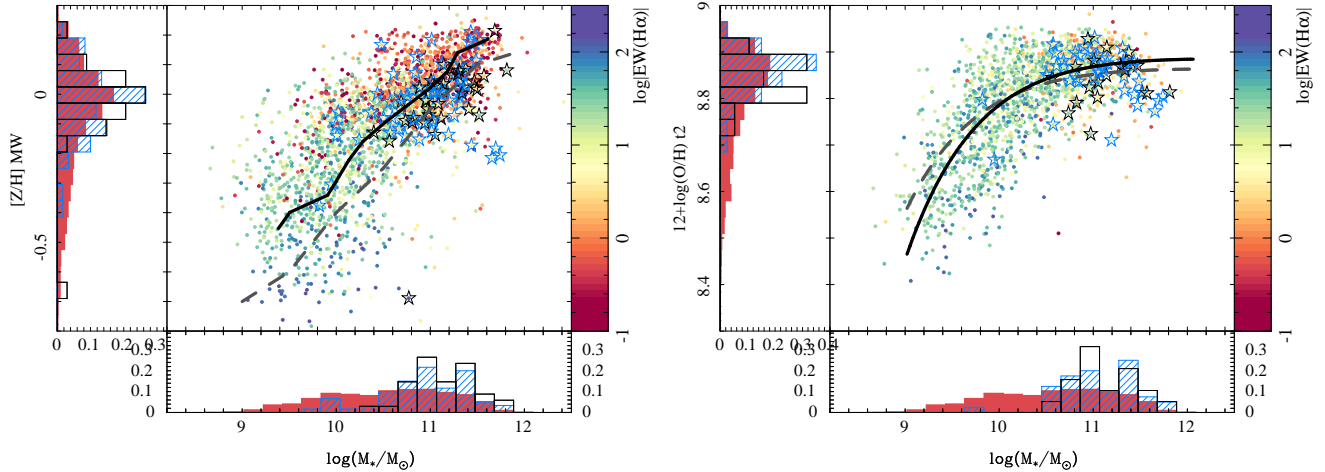

  \centering
    \includegraphics[width=6.5cm, angle=270]{16BPT_cen.ps}\includegraphics[width=6.5cm, angle=270]{4BPT_cen.ps}
  \caption{{\it Left panel:} Distribution of the mass-weighted stellar metallicity within one effective radius versus the stellar mass for the full sample of galaxies analyzed in this study, using the same symbols as in Fig. \ref{fig:BPT_EW}. The solid black line corresponds to the average value for different mass bins of 0.3 dex width, while the dashed grey line corresponds to the stellar mass-metallicity distribution reported by \citet{rosa14b}.{\it Right panel:} Distribution of the oxygen abundance within one effective radius using the so-called $t2$ calibrator, versus the stellar mass for those galaxies within the full sample that have enough coverage of gas emission lines compatible with being ionized by star-formation. The oxygen abundance has been computed following \citet{sanchez2017a}.  Each galaxy is represented with the same 
 symbols used in Fig. \ref{fig:BPT_EW}. The black solid line corresponds to the best fitted MZ relation for these using the formula described in \citet{sanchez13}, while the grey dashed line corresponds to the best fitted relation reported by \citet{bb17}. Normalized histograms of each respective parameter for the full sample (solid red), the type-II AGN hosts (hashed light blue), and the type-I ones (open black) are included as well.}
  \label{fig:MZR}
\end{figure*}
%%%%%%%%%%%%%%%%%%%%%%%%%%%%%%%%%%%%%%%%%%%%%%%%%%%%%%%%%%%%%%%%%%%%%%%%%%

\subsection{Metal content in AGN hosts}
\label{sec:metal}

%The most accepted scenario for the transformation between star-forming and non star-forming galaxies is that the first ones become the second by a sequence of processes that involves major mergers, fueling of gas towards the inner regions, and the switch-on of an AGN that finally halts the star formation \citep[e.g.][]{hopkins09}. However, based on the previous results we cannot determine 

{ We cannot determine if AGN hosts are in a transition from the SFMS towards the RG sequence (or from the blue cloud to the red sequence), or the other way around, based on the previous results}. Alternative scenarios involve a rejuvenation of already RGs by accretion of gas or capture/minor-merger with gas rich galaxies  \citep[e.g.,][]{Thomas+2010}. Actually, early-type galaxies with blue colors, recent SF activity, and even with faint spiral-like structures, have been { previously} detected \citep{Schawinski+2009, Kannappan+2009,Thomas+2010,McIntosh+2014,Schawinski+2014,Vulcani+2015,Lacerna+2016,Gomes16b}. The fraction  of these galaxies increases as the mass is smaller and the environment is less dense. A new gas fueling could activate the nuclear AGN too, increasing slightly the SFR and make the colors bluer and we will equally detect the host galaxy in the green valley.
%without involving a connection with the foreseen transition between late-type and early-type galaxies.
%
% SFS: Me parece una buena propuesta alternativa:
% Porque no redactas un texto y lo ponemos
% como has hecho antes ??
%
%\Com{Aldo: Puse un comentario.\footnote{Existe otra propuesta que es muy diferente a las anteriores. En esta propuesta se piensa que inestabilidades violentas en el disco debido a fusiones mayores o menores pueden detonar un ``starburts'' en la region central debido a grandes flujos de gas por procesos disipativos. Esta propuesta funciona mejor a altos z's cuando las fracciones de gas son mas altas. Mas aun, este modelo predice que la partes mas internas de las galaxias deberian incrementar su densidad superficial central a este fenomeno la gente lo ha denominado ``wet compaction''. Esto ocurre generalmente en galaxias massivas y es posible que a altos z's estas galaxias puedan ser alimentadas por los llamados ``cold streams'' lo cual puede ayuda a extender el duty cycle del AGN y a la formacion estelar de la galaxia. Quizas esto sea material para la Fig. 9}}

A possible way of distinguishing between these two scenarios is to explore the metal content in these galaxies. If the SF is quenched at a certain time, when the AGN is still observable (i.e., within the last 10$^8$ yr, the supposed timescale of an AGN), the oxygen abundance should be ``frozen'', since this is only increased by the production of short-lived massive stars that derive into supernovae type-II. A similar effect could be produced by a rejuvenation if the accreted gas is less metal rich (e.g., if the captured gas-rich galaxy is less massive than the host). If the rejuvenation is due to gas that has been recycled in the host, then no decrease of the oxygen abundance is expected. This scenario for the gas-phase oxygen abundance is different than the one expected for the stellar metallicity ([Z/H]). This parameter results from the combination of the two major groups of elements produced in stars: $\alpha$ (like O, Mg...) and non-$\alpha$ elements (like Ti, Fe...). The non-$\alpha$ elements are produced in stars of any mass, being its bulk production dominated by intermediate mass stars, and therefore it requires a longer period of time to be produced (as the stars last longer times at lower masses). If no new SF process happens, the stellar metallicity gets frozen too, since it measures the metals trapped in the stars. Therefore, in the case of a quenching of the SF both the oxygen abundance and the stellar metallicity should be lower than that of the average population of galaxies at the same mass range. However, for the rejuvenation, although the oxygen abundance may be lower (at least in some cases), the stellar metallicity should not be substantially modified. These events do not imply a SF process large enough to modify the average metallicity in a galaxy, dominated by the bulk of stars formed in early times \citep{perez13,ibarra16}, since they involve just a tiny fraction of the overall stellar mass in galaxies. Therefore, it is expected that they do not modify the stellar metallicity.

Fig. \ref{fig:MZR}, left panel, shows the central mass-weighted stellar metallicity versus the integrated stellar mass for all the galaxies explored in this analysis together with the distribution for AGN hosts. There is a clear correlation between both parameters, known as the stellar mass-metallicity relation (MZR), that in our case is well represented by the black solid line. For comparison purposes we have included the MZR presented by \citet{rosa14b} using IFS data from the CALIFA survey (dashed grey line). Both relations follow the same trend, with a clear offset towards lower metallicities ($\Delta[Z/H]\sim-$0.1 dex) in the case of the relation proposed by \citet{rosa14b}. This result is expected since the library of SSPs templates adopted in that study comprises stellar populations covering a much wider metallicity range, including very metal poor populations not considered in our adopted library. Despite of this offset the general trends are pretty similar.

The location of AGN hosts in this diagram covers the more massive range, as expected from the results seen in previous sections. More interestingly, AGN hosts are preferably located below the value of the mean stellar metallicity for each mass (stellar MRZ), with a 69\% in the lower-metallicity range compared to 31\% in the upper-metallicity one. This trend is sharper for type-I AGN hosts, with a 77\% of them located in the lower-metallicity range.

Fig. \ref{fig:MZR}, right-panel, shows the distribution of the characteristic oxygen abundance along the integrated stellar mass for the 1641 non-active galaxies within the sample with detected emission lines compatible with being ionized by SF and enough spatial coverage to derive the abundance at the effective radius \citep[following][]{sanchez14,laura16,sanchez2017a,bb17}. As indicated in Section \ref{ana}, we adopted the $t2$ calibrator for the oxygen abundance. However, no qualitative result would change if other calibrator is assumed. The average trend between the two parameters is described by the solid line, following the formalism described by \citet{sanchez2017a}. The dashed-line shows the relation described by \citet{bb17}, for a similar dataset. There are some differences, most probably due to  differences among the samples, since in \citet{bb17} the AGNs were not excluded for this particular analysis.

The location of the AGN hosts in Fig.\ref{fig:MZR} has been highlighted following the same symbols as in previous figures. As in the case of the stellar MZR, the galaxies hosting a nuclear source are preferably located in the low abundance regime for their considered stellar mass, although with a slightly lower fraction. A 61\%\ of  AGN hosts have an abundance lower than the average corresponding to their masses. As in the case of the stellar metallicity the trend is sharper for type-I AGNs, with a 70\%\ of them with an oxygen abundance lower than the average.

%These results seem to indicate that star formation fades/slows-down at a certain time in AGN hosts in comparison with the bulk population of galaxies at the same stellar mass, freezing both the stellar metallicity and the gas-phase oxygen abundance. A rejuvenation by either more pristine gas via a minor merger or by enhanced star formation due to in-situ gas recycling does not agree with the observed differences between AGN hosts and the remaining population of galaxies.
{ These results agree with a quenching scenario more than with the rejuvenation one, in agreement with the scenario presented by \citet{yates14}, based on the analysis of the gas and stellar metallicities of the SDSS-DR7 dataset.}
However, other processes may agree with the observed distributions. For example, a major merger that does not involve a strong increase in the SFR may increase the stellar mass without modifying significantly neither the stellar metallicities nor the gas-phase oxygen abundance. 

%%%%%%%%%%%%%%%%%%%%%%%%%%%%%%%%%%%%%%%%%%%%%%%%%%%%%%%
\begin{figure}
  \centering
    \includegraphics[width=6.1cm, angle=270]{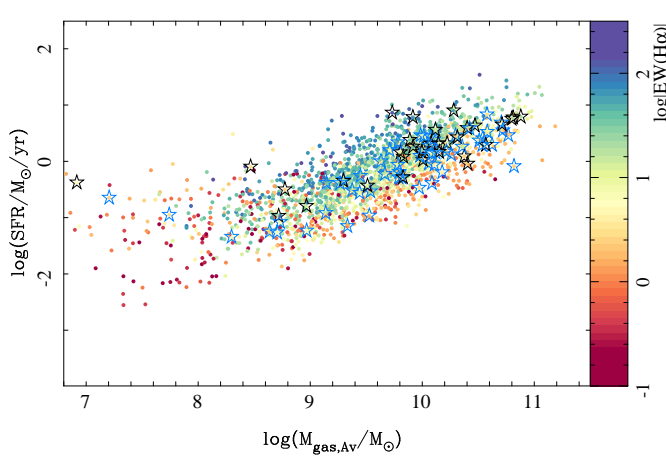}
  \caption{Integrated SFR along the estimated integrated molecular gas mass for the full sample of galaxies analyzed in this study, using the same symbols as in Fig. \ref{fig:BPT_EW}. }
  \label{fig:SKL}
\end{figure}
%%%%%%%%%%%%%%%%%%%%%%%%%%%%%%%%%%%%%%%%%%%%%%%%%%%%%%%

%In any case, either due to an internal process or an external one the star-formation seems to be halted at 

\subsection{Gas content: What halts star formation?}
\label{sec:gas}

Being established that most of the AGN hosts are located in the intermediate region between the blue/star-forming and the red/retired galaxies, and that most probably that transition is associated with a process that halts SF, we will explore now the possible reasons for that halting. In general there are two major possibilities: (i) a lack of molecular gas and (ii) the presence of gas in such conditions that SF is prevented. In order to explore those possibilities we have analyzed the dependence of our estimation of the molecular gas mass, described in Sec \ref{gas}, with both the SFR and the integrated stellar mass.

Fig \ref{fig:SKL} shows the distribution of the integrated SFR as a function of the estimated molecular gas mass for all the galaxies studied here, together with the AGN hosts. The correlation observed between both parameters was first proposed by \citet{schmidt59}, and it is generally known as the Schmidt-Kennicutt law \citep[e.g.][]{kennicutt98}. It is a direct consequence of the fact that stars are born in dense molecular gas regions. This relation was generalized for the atomic and molecular gas densities across entire galaxies by \citet{kennicutt98}, showing that the SFR density depends on a power of index $\sim$1.4-1.5 of the neutral gas mass density. The slope of this relation can be explained based on a simple self-gravitational picture in which the large-scale SFR is presumed to scale
with the growth rate of perturbations in the gas disk \citep[e.g.][]{kennicutt98}. Despite that possible explanation, different studies have derived slopes covering a wide range of values, between 1 and 2 \citep[e.g.,][]{gao04,nara12}, and with a wide range of dispersions too, ranging between $\sim$0.05 dex and $\sim$0.09 dex \citep[e.g.][]{komu12}. These differences are related to (i) the assumed IMFs for the derivation of the SFR, (ii) the conversion factor between the observed molecular transitions and the H$_2$ molecular mass \citep[e.g.][]{bolatto13,bolatto17}, and (iii) the selected tracer of that molecular gas \citep[e.g., CO, HCN ][]{gao04}. However, when the same IMF, tracer and conversion factor are applied similar trends are derived.

We have found a strong correlation with a coefficient of $r=0.76$ between the two parameters for the full sample of 
galaxies shown in Fig. \ref{fig:SKL}, with a slope lower than 
one ($\alpha=$0.62$\pm$0.02), and a dispersion of $\sigma=$0.43 dex. If we restrict the analysis to the subsample of SFGs, with EW(H$\alpha$)$> 6\AA$, the correlation is stronger ($r=0.81$), the slope shifts towards a value near to one ($\alpha=0.83\pm0.02$), and the dispersion decreases ($\sigma=$0.32 dex). This corresponds to an average depletion time of $\sim$4 Gyr, slightly larger than the most recently reported values \citep[e.g. $\sim$2.2 Gyr,][]{leroy13,utomo17,colombo17}. 

We must recall that our estimation of the molecular gas mass is based on an indirect calibration being derived from the dust attenuation, and that for the RGs the SFR is an upper-limit at best (being just a linear transformation from an H$\alpha$ luminosity whose ionization source most probably is not young stars). In general, the distribution of SFR vs. molecular gas mass in Fig. \ref{fig:SKL} is different for SFGs, defined as galaxies with EW(H$\alpha$)$>$6 \AA, and non-SFGs (or RGs), defined as galaxies with EW(H$\alpha$)$<$6 \AA (we use this definition for SFGs and RGs throughout this study). The former present larger SFRs at the same molecular gas mass ($\sim$0.5 dex higher), and a larger amount of that molecular gas mass ($\sim$0.3 dex larger). This indicates that the SFGs present a global SF efficiency (SFE$=$SFR/M$_{gas}$) larger than the RGs. However, if we compare this distribution with those of Fig. \ref{fig:BPT_EW}, \ref{fig:CMD} and \ref{fig:SFMS}, we do not see a clear bi-modality in this case, while SFGs and RGs are well separated in previous plots. This indicates that the H$\alpha$ flux is more directly proportional to the amount of gas than to any other physical process, like the source of the ionization. 

Regarding the AGN hosts, they are not distributed in any preferential region in this space of parameters, being indistinguishable for the overall population, and not being located in any transition region between SFGs and RGs (a region that in any case it is not clearly seen in this figure). In summary, we can conclude that if AGN hosts are forming stars they do follow a similar scaling relation than the rest of the galaxies with respect to the available amount of molecular gas. { This is consistent with the results presented by \citet{husemann17}, where they analyze the molecular gas content in a sample of QSOs. They found that when QSOs are hosted by disk (SFG) galaxies there is no significant difference in the gas fraction. On the other hand, for early-type galaxies they present lower gas fractions, but shorter depletion times.}

%%%%%%%%%%%%%%%%%%%%%%%%%%%%%%%%%%%%%%%%%%%%%%%%%%%%%%%
\begin{figure}
  \centering
    \includegraphics[width=6.1cm, angle=270]{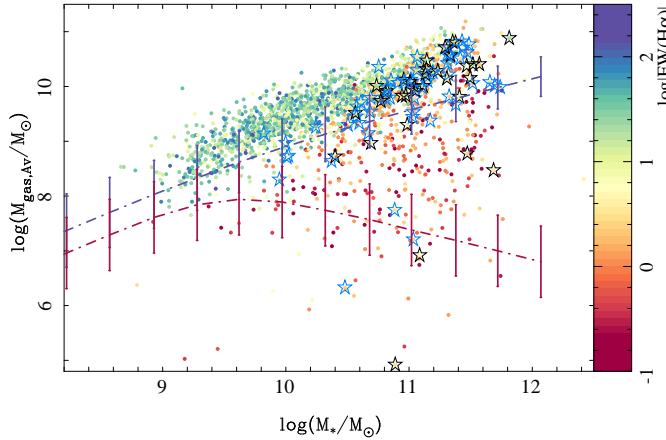}
  \caption{Integrated molecular gas mass vs. the integrated stellar mass for the same galaxies shown in Fig. \ref{fig:SKL}, and using the same symbols. The dash-and-dotted violet and red lines corresponds to the correlations found in \citet{Calette+2017} for an extensive compilation and homogenization from the literature for late- and early-type galaxies, respectively. The error-bars corresponds to the $\pm$1$\sigma$ intrinsic scatter of the distributions.}
  \label{fig:MM_gas}
\end{figure}
%%%%%%%%%%%%%%%%%%%%%%%%%%%%%%%%%%%%%%%%%%%%%%%%%%%%%%%

Figure \ref{fig:MM_gas} shows the distribution of the estimated molecular gas mass as a function of the integrated stellar mass. Over-plotted are the results from the compilation and homogenization of data from the literature recently presented by \citet{Calette+2017}, who were able to separate the data sets into late- and early-type galaxies and to take into account reported upper limits in the case of CO non-detections (CO is used as the main tracer of molecular hydrogen). Contrary to the SFR vs. M$_{gas}$ distribution shown in Fig. \ref{fig:SKL}, here a clear different pattern for SFGs and RGs is seen. A similar segregations is seen for the \citet{Calette+2017} results, if we assume that most SFGs are late-type galaxies and most RGs are early-type ones. The main difference found is that the molecular gas masses for late-type galaxies in \citet{Calette+2017} are lower than the ones reported here. This could be an effect of the molecular gas estimator adopted in our study or the result of the different selection criteria: although most of their late-type galaxies are surely SFGs, it is known that that there is no one-to-one correlation between both galaxy properties.

The SFGs in our sample present a strong ($r=$0.84) correlation between the two parameters, in the form:
\begin{equation}
\log{M_{gas}} = 0.86_{\pm 0.01}\log{M_*} + 0.75_{\pm 0.15},
\label{eq_gas}
\end{equation}
with a dispersion of $\sigma$= 0.30 dex. This means that for SFGs, the amount of molecular gas correlates tightly with the stellar mass, as reported also in \citet{Calette+2017} for late-type galaxies. If we consider that the stellar mass is a good tracer of the gravitational potential within the optical extension of these galaxies, we can interpret that result as the consequence of the ability of a potential to retain a certain amount of gas if it was not previously consumed. 
Under this scenario SFGs form stars as fast as they can with the available amount of molecular gas (following a SK-law), and the amount of gas is somehow regulated by the potential, following a scheme similar to the one proposed in the bathtub model of \citet{2013ApJ...772..119L}.

%%%%%%%%%%%%%%%%%%%%%%%%%%%%%%%%%%%%%%%%%%%%%%%%%%%%%%%%%%%%%%%%%%%%%%%%
\begin{figure*}
  \centering
\includegraphics[width=17.5cm,clip, trim=40 230 150 110]{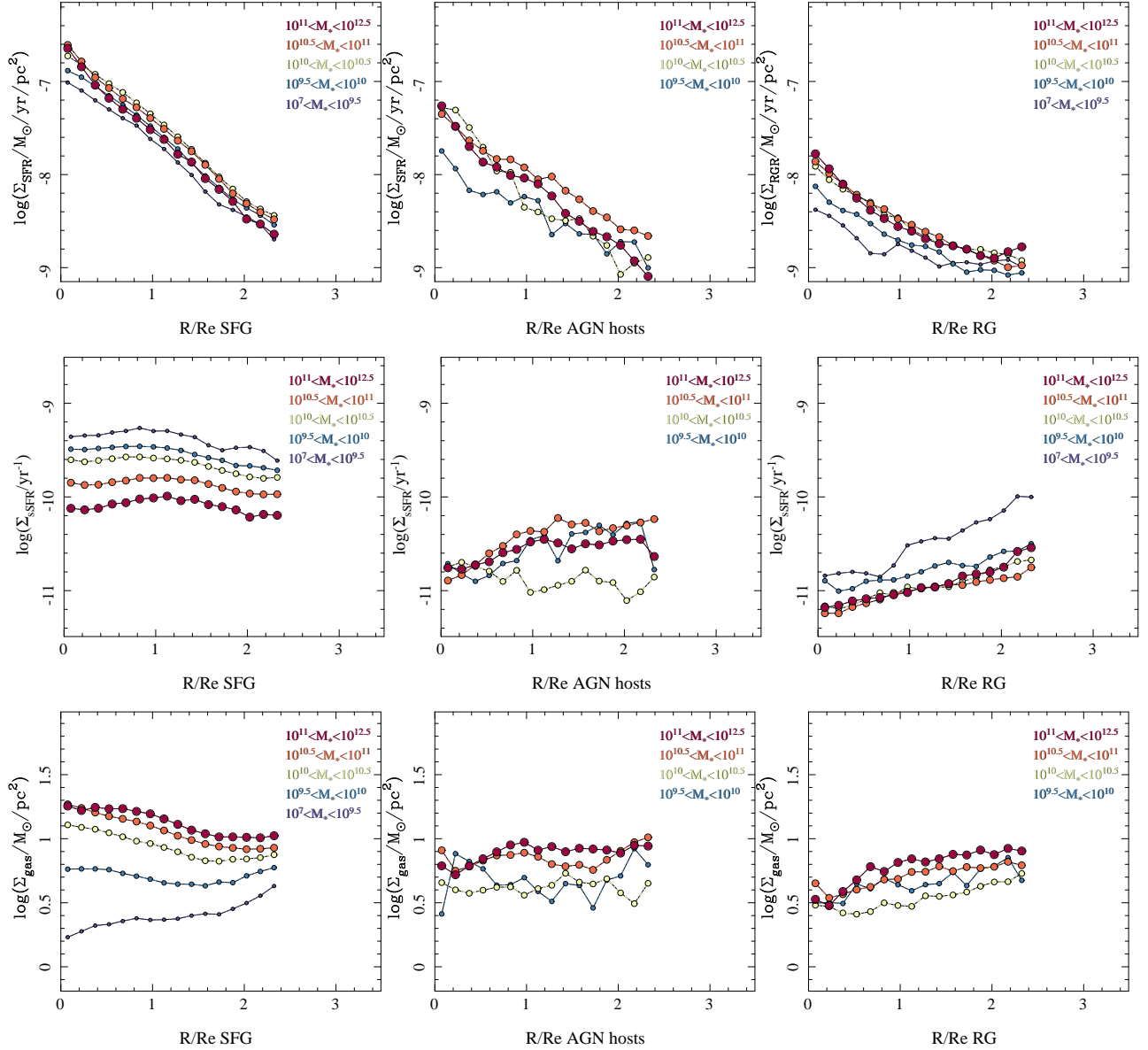}
\caption{From top to bottom each panel shows the radial distributions of (i) the SFR, (ii) sSFR and  (iii) molecular gas surface densities for three categories of galaxies: SFGs (left panels), AGN hosts (central panels) RGs (right panels). In each panel it is shown the average distribution of the considered parameter for the galaxies within four different mass bins: 10$^7$-10$^{9.5}$ (dark blue),  10$^{9.5}$-10$^{10}$ (light blue),  10$^{10}$-10$^{10.5}$ (yellow), 10$^{10.5}$-10$^{11}$ (orange) and 10$^{11}$-10$^{12.5}$ (red) in solar masses, with symbol size increasing with mass.}
  \label{fig:rad_Mass}
\end{figure*}
%%%%%%%%%%%%%%%%%%%%%%%%%%%%%%%%%%%%%%%%%%%%%%%%%%%%%%%%%%%%%%%%%%%%%%%%

Non star-forming (retired) galaxies present a totally different distribution.
%, qualitatively similar to the results for early-type galaxies by \citet{Calette+2017}, though with average values larger than those found by these authors. This is expected since those authors took into account the non detections in their derivation of the average distribution. 
For a given integrated stellar mass, non star-forming galaxies show a wide range of molecular gas masses that spread from an upper envelope defined by the loci of the SFGs (eq. \ref{eq_gas}) towards lower values that can be as low as 10$^4$M$_\odot$ (for the galaxies with detected ionized gas, that is the majority of the galaxies in our study, Sec. \ref{ana}). This indicates that these galaxies do not form stars at the same speed as the SFGs for their corresponding stellar mass due to a general lack molecular gas. However, it is not only the lack of gas what prevents the SF since, as we have seen when analyzing the SFR vs. M$_{gas}$ distribution, those galaxies form stars at a lower efficiency than the SFGs, although that difference is less sharp than the difference found in the amount of molecular gas.

AGN hosts are found in a transition region between SFGs and RGs in Fig \ref{fig:MM_gas}, being located preferably at the higher-mass end (what we have seen already in previous sections), mostly at the lower end of the sequence defined by the SFGs, and spread towards lower values of M$_{gas}$ for a given stellar mass. The main difference between type-I and type-II AGNs seems to be the range of stellar masses, without a clear difference in the distribution in this diagram. Like in previous cases we refrain ourselves to make a causal connection between the AGN activity and the process that has quenched the star formation. 
%However, these results may indicate that both processes are connected, igniting the AGN, removing molecular gas from the galaxy and decreasing the SFR efficiency. 

We should note that although RGs and AGN hosts present a lower amount of molecular gas and a lower SFE at the current epoch time, this does not preclude for having a stronger SFE and more molecular gas in the past. Indeed, a strong star-burst process, like the one predicted by the scenario outlined by \citet{Hopkins+2010}, could consume a substantial amount of gas in the past. This does not explain the lower SFE, but it could fit into the observations. We will explore the SF
histories of these galaxies in future studies in order to clarify that possibility (Ibarra-Medel et al., in prep.).

\subsection{Radial distributions: Inside-out quenching?}
\label{sec:rad}

%In previous sections we have established that  AGN hosts are found in the transition region between the blue/SFGs and red/RGs, and that this transition is associated with a process that quenches star formation in a fast way (compared to the Hubble time), decreasing the star-formation efficiency and more strongly removing the molecular gas from the galaxies. 

In the current Section we explore whether { the transition hinted in previous sections }happens in an homogeneous way in galaxies or it happens from the outer to the inner parts, or the other way around. For this analysis we will consider all AGN types together in order to increase the statistical numbers in the different analyzed bins.

Figure \ref{fig:rad_Mass}, top panels, show the azimuthally average radial profiles (in units of the effective radii) of the SF surface density ($\Sigma_{SFR}$) for the SFGs (left panel), AGN hosts (middle panel), and RGs (right panel), averaged by galaxy type in four different ranges of stellar masses. As expected, the SFGs have larger values of $\Sigma_{SFR}$ at any radius, with a clear inverse gradient following almost a pure exponential profile, with a slope of $\sim\-$1 dex/Re, similar for all stellar mass bins. On average, the  $\Sigma_{SFR}$ for the less massive galaxies ($M_*\sim 10^{9} M_\odot$) is $\sim$0.4 dex weaker than for the more massive ones ($M_*\sim 10^{10}-10^{11} M_\odot$), as a consequence of the local and global SFMS \citep[e.g.][]{mariana16}. However, the most massive galaxies seem to present a slightly lower $\Sigma_{SFR}$, which may indicate that for these galaxies their global SFR has started to deviate from the MS towards the GV as already pointed out by different authors \citep[e.g.][and references therein]{catalan15,rosa16a}. The RGs present a SF surface density one dex weaker than SFGs at any stellar mass range and at any galactocentric distance. Their profiles are less steep than that of the SFGs, with a shape that resembles more a de Vaucouleurs \citep{deVauc59} or Sersic \citep{sersic68} profile than a pure exponential one \citep{free70}. The difference between the less massive and more massive galaxies is larger than in the case of the SFGs, being of the order of $\sim$0.6 dex, reflecting this the fact that the spatial resolved RG sequence has a steeper slope than the SFMS (Cano-Diaz et al., in prep.). Like in previous cases we must recall that the $\Sigma_{SFR}$ for the RGs should be interpreted as purely H$\alpha$ luminosity densities, whose ionization nature should not be directly associated with young stars, and as an upper limit of the real $\Sigma_{SFR}$ in any case.

Finally, AGN hosts are somehow in between the SFMS and RG sequence, as expected from the results in Sec. \ref{sec:GV}. Their $\Sigma_{SFR}$ values are slightly lower than those of the SFGs for any stellar mass bin and at any radial distance, reflecting the fact that they are galaxies for which the SFR is somehow halted at a certain degree. However, from these distributions we cannot establish if the SF is stopped homogeneously in these galaxies, or if it describes a pattern of outside-in or inside-out quenching/halting. We should note that the number of AGN hosts in each mass bin is rather different, reflecting the fact that this kind of objects are more frequently found in massive galaxies (Sec. \ref{sec:which}). There is no AGN host in the lowest mass bin ($M_*\sim 10^{9} M_\odot$), just three in the second bin ($M_*\sim 10^{9.75} M_\odot$), seven in the next one ($M_*\sim 10^{10.25} M_\odot$), and a total of 54 in the most massive range ($M_*\sim 10^{11.5} M_\odot$). This affects the statistical significant in the comparison between different bins.

Figure \ref{fig:rad_Mass}, second row panels, show the azimuthal average radial distribution of the specific SFR (sSFR) for the same three different galaxy types (SFGs, AGN hosts and RGs), averaged for the same stellar-mass bins that in the previous panels. In the case of the SFGs the sSFR presents a rather constant distribution for each different mass bin, with larger values for less massive galaxies than for the more massive ones ($\sim$0.6 dex). This result may indicate that the spatial resolved SFMS presents a similar slope for the different stellar mass bins, but with a slightly different zero-point (Cano-D\'\i az et al., in prep.). A similar result was found by \citet[][see their Fig. 10]{rosa16a}, when segregating their sample of galaxies in different morphological types. For RGs the picture is totally different. While the sSFR has lower values for the most massive galaxies than for the less massive ones, the distribution presents a clear positive gradient in all mass bins: the central regions form stars at a much lower rate than the outer parts when compared to the already formed stellar mass density. This result can be interpreted as a differential decrease in the current SFR compared with the historical SFR in the galaxy from the inner to the outer parts. In other words, it is a clear indication that the quenching happens from the inner to the outer regions. This agrees with the scenario in which quenching is related to internal processes in galaxies \citep[e.g.][]{bundy06}.

Like in previous cases, AGN hosts are located in the intermediate regime between both groups of galaxies. Despite of the fact of their lower number, there are some clear differences in the radial distribution of the sSFR. Contrary to both SFGs and RGs, in AGN hosts the trend with mass is not preserved. More massive galaxies do not have lower sSFR than less massive ones. We lack of information for the lower-mass bin, in which we do not find AGNs, however, for the other two the trend is not present. Actually, the most massive and less massive bins with AGNs exhibit very similar sSFR radial distributions, with the bin in the middle showing a lower sSFR than the other two for most of the galactocentric distances. The limited number of galaxies in the two lowest mass bins may affect that result. In general the radial profiles are either flat or present a drop towards the inner regions (R$<1$R$_e$) with a flat distribution towards the outer ones (R$>1$R$_e$), showing a mixed trend between the SFGs and RGs.

Figure \ref{fig:rad_Mass}, third row panels, show the distribution of the azimuthally averaged molecular mass density ($\Sigma_{gas,Av}$) for different galaxy types (SFGs, AGN hosts, and RGs) and for the four different mass bins. For SFGs we have a rather flat distribution in the molecular gas density for the less massive galaxies, with a value of $\sim$2$M_\odot/pc^2$, and a negative gradient for the remaining stellar mass bins. These latter gradients follow an almost exponential profile, with a slope clearly smaller than the one found for the $\Sigma_{SFR}$ (upper panels). The most massive galaxies present a drop in the $\Sigma_{gas,Av}$ towards the central regions. This drop is much more evident for the RGs: for the four stellar mass bins with detected dust attenuation the inner regions present a clear deficit of molecular gas. The radial distribution present a positive non-linear gradient from undetected or nearly undetected molecular gas, several order of magnitudes lower than the values found for the SFGs at the same galactocentric distances, towards values of the same order to those of the SFGs in the outer regions ($R\sim 1.5-2 R_e$). This result agrees with those presented in Sec. \ref{sec:gas}, illustrating the fact that there is not only a drop in the efficiency of the SFR in these galaxies, but a lack of molecular gas to fuel that SF. Even more, it  clearly establishes that this deficit of molecular gas is stronger in the inner regions.

Regarding AGN hosts, the lack of a similarly large statistical number of galaxies and detected molecular gas at every galactocentric distance limits the interpretation of the results. However, despite of those limitations it is clear that these galaxies exhibit a similar deficit of molecular gas, and that deficit is stronger in the inner regions. {\it Per se} this result does not support the idea of these objects being in the transition phase between SFGs and RGs. However, considering that in terms of the radial distribution of the $\Sigma_{SFR}$ and $\Sigma_{sSFR}$ these galaxies seem to be {\it under transition}, we may speculate that they first lost the molecular gas from the inner to the outer regions and that causes the observed decrease in the absolute (SFR) and relative (sSFR) way. Since the radial distributions of $\Sigma_{gas,Av}$ for AGN hosts are more similar to those of RGs while the other two radial distributions are clearly in an intermediate step, we speculate that the decrease of molecular gas in the central regions happens before that of the SFR and sSFR.

%Fig \ref{fig:rad_Mass}, bottom panels, show the distribution of the azimuthally averaged luminosity weighted (luminosity-weighted) ages of the stellar populations in logarithm scale, averaged for the different galaxy types (SF, AGN hosts and RG), and for the four different mass bins. We find a negative gradient the luminosity-weighted ages for all the ranges of masses and all galaxy types, that is more clear in the non-starforming (RG) galaxies. In general the SFGs present younger ages in their stellar populations at any galactocentric distance than either the AGNs or the RG. They present younger stellar populations in the lower stellar mass bins, increasing with the mass, with a sharper gradient in the inner regions (R$<$R$_e$) than in the outer ones. Indeed the gradient is flat or even inverse (for the low mass range) beyond R$>$1.5-2 R$_e$. For the RG galaxies we find a negative gradient too, but in this case it extends along all the explored galactocentric distances. Like in the case of the SFGs the average age depends on the mass. Indeed, for the same stellar mass the stellar populations of the RG are always older at any galactocentric distance. In general they present a difference of $\sim$2 Gyrs in the ages of their stellar populations, what may indicate that this is the time-scale of the quenching processes described before.

%%%%%%%%%%%%%%%%%%%%%%%%%%%%%%%%%%%%%%%%%%%%%%%%%%%%%%%%%%%%%%%%%%%%%%%%%%%%
\begin{figure*}
  \centering
\includegraphics[width=17.5cm,clip, trim=40 230 150 110]{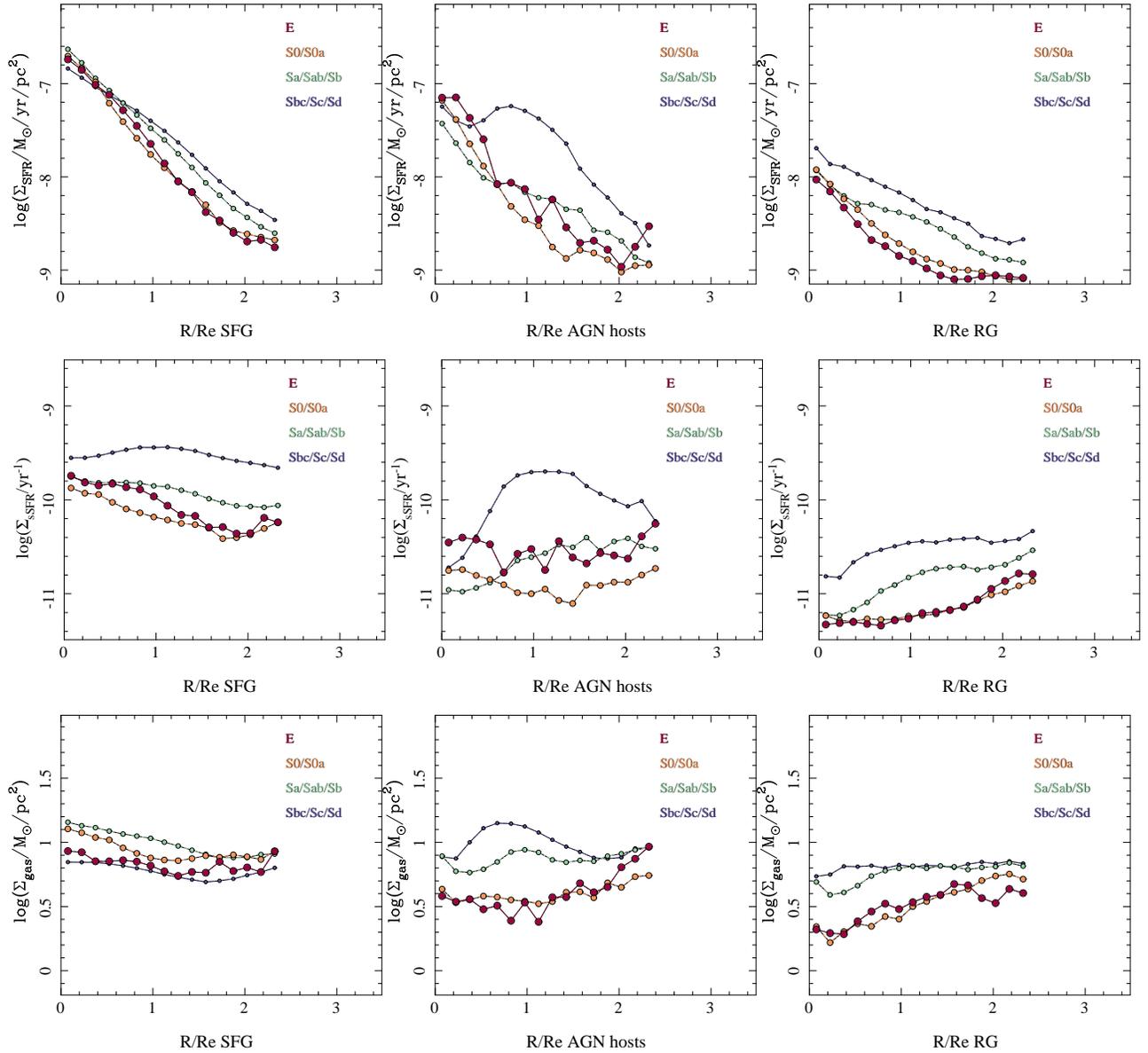}
\caption{From top to bottom each panel shows the radial distributions of (i) the SFR, (ii) sSFR, and (iii) molecular gas surface densities for three categories of galaxies: star-forming galaxies (left panels), AGN hosts (central panels) and retired galaxies (right panels). In each panel it is shown the average distribution of the considered parameter for the galaxies within four different morphological bins: Sbc/Sc/Sd (dark blue),  Sa/Sab/Sb (light blue),  S0/S0a (orange) and Elliptical galaxies (red), with symbol size increasing for more early-type galaxies.}
  \label{fig:rad_Morph}
\end{figure*}

All together, AGN hosts seem to be in a transition phase between SFGs and RGs regarding the analyzed radial gradients in different mass bins.

%However, the main goal of this study is to characterize the evolution of AGN hosts in comparison of that of the bulk galaxy population, not to study in detail what influence the evolution of galaxies in general. 

%In order to do so we repeated the same analysis presented in the current section for the different morphological types described before. The results of that analysis are presented in Appendix \ref{app:rad}. In summary we find very similar results of the ones described in this section: AGN hosts seems to be at the transition stage between SF and RG galaxies.

%\section{AGN hosts evolution by morphology}
%\label{app:morph}

%In Sec. \ref{sec:rad} and \ref{sec:SFH} we explored the radial distributions and star-formation histories of AGN hosts in comparison with the bulk population of SFG and RG dividing the galaxies in stellar mass bins. Mass and stellar-mass density are the primary drivers (or consequence) of the evolution of galaxies or regions in galaxies, as stated clearly by different authors \citep[e.g.][]{rosa15a,zibetti17}. However, morphology is a direct proxy of the dynamical state of a galaxy, and it influence its evolution in a clear way. It is out of the scope of this article to explore in detail the dependence of the analyzed parameters with the morphology. Regarding AGN hosts we have already shown that they cannot be found in galaxies later than Sc (Sec. \ref{sec:which}). In this section we compare the radial properties and SFHs of these galaxies compared to the bulk population for the different morphological types.

\subsection{Radial distributions by morphology}
\label{sec:rad_morph}

So far we have described the behavior of the radial profiles for different integrated stellar mass bins. Several previous studies have demonstrated that the radial properties of galaxies dependent more on the morphology than on mass
%the morphological type has a stronger impact in galaxy evolution than  stellar mass 
\citep[e.g.][and more references therein]{ibarra16,rosa14,rosa14b,rosa15a,rosa16a}. \citet{Schawinski+2010} found clear differences in the effects of AGNs in the evolution for different morphological types of their hosts. They propose a different transition to quiescence for early-type galaxies, that is assumed to be faster than for late-types. We explore that possibility by comparing the radial distribution of the properties described in the previous section for different morphological-type galaxies.

Figure \ref{fig:rad_Morph} shows the radial distributions of the same properties shown in Fig. \ref{fig:rad_Mass} ($\Sigma_{SFR}$, $\Sigma_{sSFR}$ and $\Sigma_{gas,Av}$, in each row), for the different groups of galaxies explored in this study (SFG, AGN hosts and RG, in each column). Radial gradients on these quantities segregated by morphology have been previously studied by \citet{rosa15a} and \citet{rosa16a} using IFU data with better physical spatial resolution than here. However, they did not split their sample into SFGs and RGs, and therefore it is not possible a simple/straight-forward comparison. Regarding $\Sigma_{SFR}$, the trends found by \citet{rosa16a} (their Fig. 6) are very similar to the ones seen in the first row of panels of Fig. \ref{fig:rad_Morph}, if we consider that most of the E/S0 galaxies are found in the RG sequence and most of the Sa--Sd galaxies are found in the SFMS. In general, the $\Sigma_{SFR}$ of the disk-dominated galaxies (later than Sa) presents an inverse gradient, similar to the one found for SFGs at any mass bin (Fig. \ref{fig:rad_Mass}, upper panels). Retired galaxies present a lower $\Sigma_{SFR}$ at any galactocentric distance and for any morphological type, apart from the later-type ones. 

As in the case of the segregation by stellar mass, AGN hosts seem to be in a { intermediate location} between SFGs and RGs, with a $\Sigma_{SFR}$ slightly lower than that of the SFGs and slightly larger than that of the RGs for any morphological type. Similar results are found for $\Sigma_{sSFR}$. For any morphological type, the SFGs present a larger sSFR than those in the RGs, with the AGN hosts being in a transition between the two groups. When comparing with \citet{rosa16a} (Fig. 7), we find a slightly different shape for the radial distribution. We reproduce their results only if we assume that most of the early-type galaxies are located in the RG sequence and most of the late-types in the SFMS. As in their case, late-type galaxies present a flatter distribution and larger values of sSFR than early-type ones for any group. However, the drop in the inner regions observed by them is only appreciated in the RGs. We find a clear difference in the shape of the sSFR for SFGs and RGs. In particular for S0/S0a and E-type SFGs, the sSFR presents a negative gradient while the gradient tends to be positive  when they are RGs. For any morphological type, the transition between the SFGs and the RGs involves a drop in the sSFR from the inner to the outer regions (apart from the S0/S0a group of AGN hosts). This result agrees with the one found for the stellar masses.

The $\Sigma_{gas,Av}$ presents a similar distribution/behavior for different morphological types and different stellar masses. For SFGs, all morphological types show a shallow negative gradient, with a similar gas content in all cases. The distribution is rather different in the case of RGs, with a clear drop/lack of molecular gas in the inner regions, that it is stronger/sharper for earlier-type galaxies than for later ones. Indeed, very few molecular gas is found in ellipticals RGs. As in previous cases, the AGN hosts seem to be in a transition regime between both groups.

%A similar pattern is found for the age distributions. At any morphological type and at any galactocentric distance AGN hosts present younger stellar populations than RGs, with later-type galaxies being younger, and more early-type ones older. This trend is broken for star-forming elliptical galaxies, that present a positive gradient in the age. This is the only case where we found this kind of gradients described by \citet{godd15}. This gradient is in agreement with the slightly rise of the sSFR towards the inner regions in these galaxies compared with S0/S0a ones (Fig. \ref{fig:rad_Morph}). If true it may indicate that early-type galaxies in the SFG group present some-kind of central rejuvenation. Finally, the stellar ages of AGN hosts galaxies are clearly older than that of the SFGs. However, when comparing with RGs we found similar or slightly younger stellar populations. 

A detailed analysis of the observed gradients indicates that the transition between SFG and RGs seems to  happen in a different way for early-type and late-type galaxies. In the first case, the transition seems to be more abrupt (stronger change in the SFR, sSFR and gas content). In the second case, the transition is clearly smoother. This different behavior is driven by the morphology, and is not observed in the mass trends discussed in the previous Section.

%%%%%%%%%%%%%%%%%%%%%%%%%%%%%%%%%%%%%%%%%%%%%%%%%%%%%%%%%%%%%%%%%%%%%%%%
\begin{figure}
  \centering
\includegraphics[clip, trim=40 230 435 170, width=7cm]{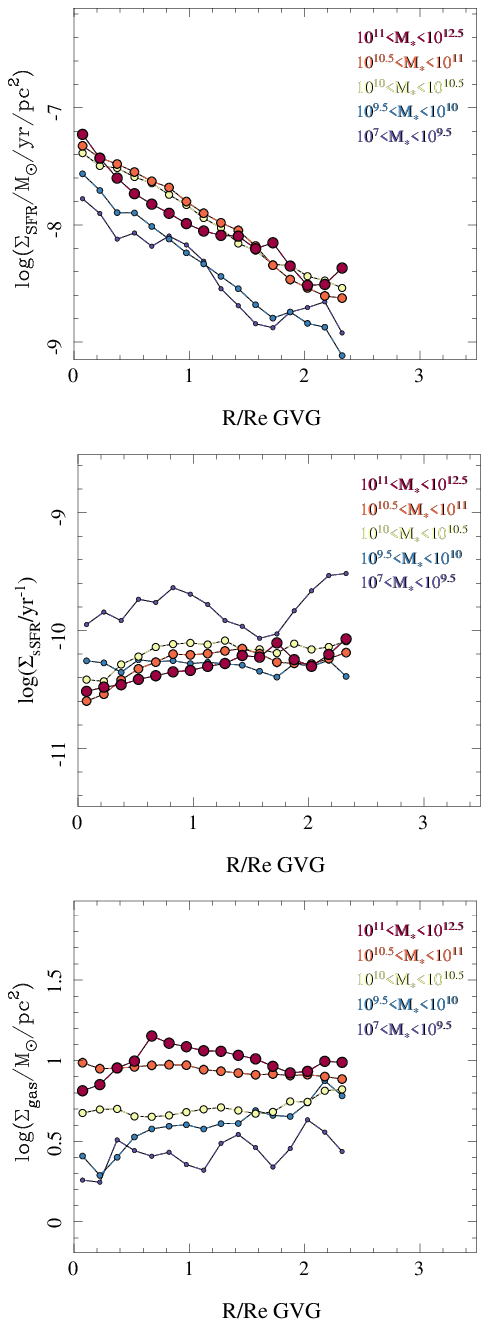}
\caption{From top to bottom each panel shows the radial distributions of (i) the SFR, (ii) sSFR, and  (iii) molecular gas surface densities for GV galaxies. In each panel it is shown the average distribution of the considered parameter for the galaxies within four different mass bins: 10$^7$-10$^{9.5}$ (dark blue),  10$^{9.5}$-10$^{10}$ (light blue),  10$^{10}$-10$^{10.5}$ (yellow), 10$^{10.5}$-10$^{11}$ (orange) and 10$^{11}$-10$^{12.5}$ (red) in stellar masses, with symbol size increasing with mass.}
  \label{fig:rad_GV_Mass}
\end{figure}
%%%%%%%%%%%%%%%%%%%%%%%%%%%%%%%%%%%%%%%%%%%%%%%%%%%%%%%%%%%%%%%%%%%%%%%%

%%%%%%%%%%%%%%%%%%%%%%%%%%%%%%%%%%%%%%%%%%%%%%%%%%%%%%%%%%%%%%%%%%%%%%%%
\begin{figure}
  \centering
  \includegraphics[clip, trim=40 230 435 170, width=7cm]{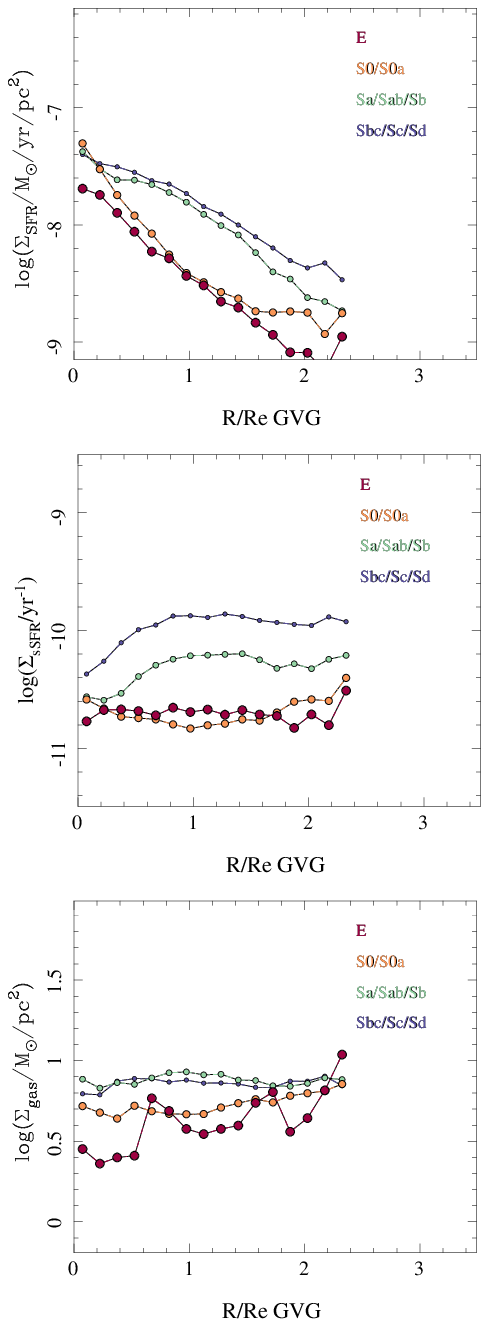}
\caption{From top to bottom, each panel shows the radial distributions of (i) the SFR, (ii) sSFR, and  (iii) molecular gas surface densities for GV galaxies. In each panel it is shown the average distribution of the considered parameter for the galaxies within four different morphological bins: Sbc/Sc/Sd (dark blue),  Sa/Sab/Sb (light blue),  S0/S0a (orange) and elliptical galaxies (red), with symbol size increasing for more early-type galaxies.}
  \label{fig:rad_GV_Morph}
\end{figure}
%%%%%%%%%%%%%%%%%%%%%%%%%%%%%%%%%%%%%%%%%%%%%%%%%%%%%%%%%%%%%%%%%%%%%%%%

\subsection{Non-active galaxies in the green valley}
\label{sec:GVGs}

In previous sections we established that AGN hosts are located in the GV in different distributions that highlights the bimodal distribution of the overall population of galaxies. 
%In particular, they are clearly located in the intermediate regions between SFGs and RGs in the SFMS, and they show evidence of being in a transit between both families of galaxies. Even more, they present evidence of being quenched from inside-out. 
However, it is also clear that not all galaxies in the GV host an AGN (to our detection limit and selection criteria). Thus, it is important to know if the described properties are privative of AGN hosts or they are common to the remaining GV galaxies.

To make this comparison we need to define GV galaxies. Being an intermediate type of galaxies, located in the area of low density in either the CMD, the SFR--$\ms$ or the age--mass diagrams, it is difficult to define clear boundaries to select them. In general, those limits would be somehow arbitrary and subject to fine tuning, sample completeness, refinement or argumentations regarding their selection. Following our own classification scheme, based on the distribution along the SFR--$\ms$ diagram and the EW(H$\alpha$) in the central regions, we classified intermediate/GV galaxies as those galaxies whose EW(H$\alpha$) ranges between 3\AA\ and 10\AA. The lower limit, 3\AA\ corresponds to the threshold established by \citet{cid-fernandes10} to distinguish between RGs and SFGs, and it corresponds to the location of the 1$\sigma$ upper limit of the RG sequence \citep{mariana16}. In a similar way, the upper-limit, 10\AA\ corresponds roughly to the 1$\sigma$ lower-limit of the SFG sequence. Thus, this regime corresponds to galaxies that are at least 1$\sigma$ off and in between the SFG and RG sequences. { These limits are similar to the ones defined by \citet{lacerda17} to distinguish between DIG and SF dominated regimes.} This range corresponds to the location in the SFMS diagram where a significant fraction of the AGN hosts are located. We find 220 galaxies in this GV area among the non-active galaxies (galaxies not hosting an AGN). This means that $\sim$1/3-th of all galaxies in the GV host and AGN, a proportion clearly larger than the general fraction in the total population ($\sim$3-4\%).

Figure \ref{fig:rad_GV_Mass} shows the radial distributions of the same quantities shown in Fig. \ref{fig:rad_Mass} %for the SFGs, AGN hosts and RGs, 
but for the GV galaxies: the average SFR, sSFR and molecular gas mass surface density profiles. Qualitatively these galaxies present similar radial gradients than the AGN hosts, being clearly in an intermediate/transit regime when compared to SFGs and RGs: (i) they present lower $\Sigma_{SFR}$, with a negative gradient from inside out; (ii) lower $\Sigma_{sSFR}$, with a positive gradient, indicating that the SFR is smaller in the center than in the outer parts in relation with their stellar mass densities; and (iii) there is a deficit of molecular gas in the inner regions compared with SFGs, but with larger amounts than the ones found in RGs. However, despite of these qualitatively similarities, there are quantitative differences. In general, all the characteristics that define them as intermediate galaxies listed before are smoother in the non-active GV galaxies than in AGN hosts: (a) Their $\Sigma_{SFR}$ and $\Sigma_{sSFR}$ are slightly larger than those of AGN hosts for all mass bins and galactocentric distances, with shallower gradients in the sSFR, and (b) the deficit of molecular gas in the inner regions is less pronounced, and they present similar gas content in the outer regions than that of SFGs. However, their stellar populations are older in average, with a luminosity-weighted (mass-weighted) age of 5.8 Gyr (8.5 Gyr) compared to the one reported in Sec. \ref{sec:age_mass} for AGN hosts: 3.6 Gyr (7.8 Gyr). In general they look like if they were in a more early phase of the transition between the two families of galaxies regarding the SFR, sSFR, and molecular gas content. On the ther hand, for the stellar ages, they seem to be in a more advances stage.

We should refrain from deriving a strong conclusion on this regards, since, as we indicated before, (i) the selection of the regime on which we classified galaxies as GV galaxies may require refinements that could affect the qualitative results, and (ii) by construction, our GV galaxies include galaxies from both our SFGs and RGs samples, and maybe we are contaminating the results somehow. Despite of all these caveats, it is clear that AGN hosts are more similar to non-active GV galaxies than to any other family of galaxies regarding the analyzed radial distributions, and both kind of objects show evidence of a transition between star-forming and non-star-forming galaxies that happens from inside-out, being associated with both a deficit of the molecular gas and decrease of the sSFR.

%\subsection{Radial distributions for Green Valley galaxies}
%\label{sec:rad_GV_morph}

Figure \ref{fig:rad_GV_Morph} shows the radial distributions of the same properties shown in Fig. \ref{fig:rad_GV_Mass} ($\Sigma_{SFR}$, $\Sigma_{sSFR}$ and  $\Sigma_{gas,Av}$, in each row) segregated by morphology, like in Fig. \ref{fig:rad_Morph}, but for GV galaxies. In this particular case we see a clear segregation between early- and late-type galaxies, already observed in the transition between SFGs and RGs and the AGN hosts. For the Sa to Sd galaxies, the shape of the profiles are very similar to those found for the same kind of galaxies in the RG subsample, although in general they seem to have slightly larger SFRs and sSFR, a lower decrease in the molecular gas content in the central regions, and younger stellar populations. The S0/S0a deviates from that trend in the sSFR, which show a rising towards the inner regions, not appreciated in the RGs, being more similar in  shape to  AGN hosts (Fig. \ref{fig:rad_Morph}), although with slightly larger sSFR values. They also deviate in the molecular gas content, with a clearly larger amount of gas, in particular in the central regions (even comparing with AGN hosts).

%Regarding the age, they present a negative gradient, like RGs and AGN hosts, bue their stellar populations are clearly younger (by $\sim$3 Gyrs). 

For elliptical GV galaxies, their trends are totally different than the ones presented in RGs or AGN hosts with the same morphology. They present a sharper $\Sigma_{SFR}$ distribution than the RGs (as expected due to the selection criteria), but with lower SFR values than the S0 in the GV and the AGN hosts in general. Their $\Sigma_{sSFR}$ have a shallow gradient, similar to the one shown by the AGN hosts, but with slightly lower values. Finally, their molecular gas does not show the sharp drop in the central regions found in AGN hosts of any morphology. Its surface density is larger than the one of the rest of the GV galaxies in the central regions 
and similar in the outer ones. When compared with retired ellipticals is seems that there is an increase of molecular gas in the inner regions.
%, and they have an inverse gradient in the stellar ages, with a drop in the central regions (R$<$R$_e$) and a flat distribution in the outer parts. 
%All together this result indicate that inactive elliptical galaxies enter in the green valley due to rejuvenation in the central regions. 
This may indicate that these elliptical galaxies enter in the GV due to rejuvenation, in particular in the central regions. Their nature seem to be different to that the AGN hosts in general. However, the low number of these galaxies (13, compared with a range of 34-91 for the three remaining morphological groups) prevents us for making a stronger conclusion on this regards.

In summary, GV galaxies seem to be in an intermediate stage between SFG and RGs for the different morphological types. At least for non-elliptical galaxies, that transition involves a quench of the SFR and/or a deficit of molecular gas in the inner regions. However, the evident sequence appreciated when these galaxies are segregated by mass (Fig. \ref{fig:rad_GV_Mass}, Sec. \ref{sec:GVGs}) is not present when segregating by morphology. The differences introduced by morphology are sharper, less sequential. In this regards the apparent smooth negative gradient in the $\Sigma_{SFR}$ and shallow positive gradient in $\Sigma_{sSFR}$ seen in Fig. \ref{fig:rad_GV_Mass} may be the result of averaging less smooth gradients through different morphologies. We will require an study with an even larger number of galaxies in order to have enough statistics to segregate non-active GV galaxies in morphology and mass bins simultaneously.

%We have also shown that AGN hosts clearly show properties of intermediate galaxies regarding the radial distributions of the SFR, sSFR, gas content and average age of the stellar populations. Based on these results it is tempting to consider that AGN activation is the physical driver for the quenching process. The fact that halting of SF happens from inside-out may support that idea. 

%To clarify a bit more this issues

%\begin{figure}
%  \centering
%    \includegraphics[width=6.5cm, angle=270]{5BPT_cen.ps}
%  \caption{TBW}
%  \label{fig:alpha_M}
%\end{figure}

%\subsection{Abundance Gradient}

%\Com{SFS:} Do they have a different abundance gradient?

%\subsection{Av vs. Mass}

%Av is a hint of the Gas mass: Are they are a different location? Gas starvation or not?
%\begin{figure}
%  \centering
%    \includegraphics[width=6.5cm, angle=270]{8BPT_cen.ps}
%  \caption{TBW}
%  \label{fig:FJ}
%\end{figure}

%\subsection{TF, FJ, RMS vs Mass}

%\Com{Erik: Re-do the plots for your article showing the location of AGNs}

%\begin{figure}
%  \centering
%    \includegraphics[width=6.5cm, angle=270]{11BPT_cen.ps}
%  \caption{TBW}
%  \label{fig:vel_sigma}
%\end{figure}

%\begin{figure}
%  \centering
%    \includegraphics[width=6.5cm, angle=270]{12BPT_cen.ps}
%  \caption{TBW}
%  \label{fig:vel_sigma}
%\end{figure}

%\subsection{Lambda ellipticity}
%\Com{SFS? Chucho, a hint?}

%\subsection{Presence of Bars}
%\Com{Adriana:} Do they have Bars more frequently

%\subsection{Environment}
%\Com{Aldo:} Do they live in central or satellite galaxies?

\section{Discussion}
\label{sec:disc}

Along this article we have explored the main properties of the AGN host galaxies 
extracted from the MaNGA survey in comparison with those of the non-nactive galaxies. 
%%VAR: En aras de acortar, creo que este resumen puede evitarse 
%We selected active galaxies based on the optical spectroscopic properties of the nuclear regions, following an spectral analysis performed using the {\sc Pipe3D} pipeline \citep{Pipe3D_II}. A total of 98 AGNs (36 type-I) were selected out of $\sim$2700 analyzed galaxies. The result of the analysis performed for the full sample of galaxies comprises a set of data-products that are distributed as a VAC of the SDSS DR14.
We have found that optically selected AGNs are hosted by mostly early-type galaxies or early-type spirals, and that for a given morphology, their hosts are in the regime of the more massive, more compact, denser in the central regions, and more pressure-supported than the average population of galaxies. These results { seem to} agree with the findings that the presence of a bulge is a mandatory condition for the presence of an active nucleus \citep[e.g.][]{magorrian98,gebhardt2000,haring04,Kormendy+2013}. Our results { seems to indicate} that it is not only mandatory the presence of a bulge, but the presence of a more compact, massive and %turbulent
dynamically hot mass concentration than the average at the same morphology. { Despite the many caveats that can be applied to our results}, in particular in the case of type-I AGNs (that disturb considerably the light-profile and therefore, the mass derivation), this result { may open} a new perspective in the exploration of the activation of a nuclear active region, that should be explored in the future. The activation of a SMBH into an AGN { may involve/require} not only the presence of a central compact object, but also the presence of material -mainly gas- supply large enough to feed it. Different scenarios have been proposed for the AGN feeding, involving secular processes in gas rich galaxies \citep{Hopkins_Hernquist2009a}, galaxy interactions and mergers \citep{Sanders+1988,Hopkins2006ApJS..163....1H}, and different kind of instabilities \citep{Dekel+2009}. However, most of them can explain only how gas is transported to the central $\sim$100 pc $-$ $\sim$1 kpc regime. 

Our results { seem to indicate} that whatever is the mechanism that transports  gas towards the central regions, AGN activation only happens in a particular set of galaxies, those that show a larger concentration of stellar mass in the central regions, with larger disordered motions. The projected spatial resolution of the MaNGA IFU data, $\sim$2.5$\arcsec$/FWHM, combined with the large redshift range coverage, implies that more than half of the galaxies are observed with a physical resolution between 2-4.5 kpc/FWHM. This prevents us to explore physical conditions in the very central regions, what will require detailed observations with much higher spatial resolutions. { This limits some how of our results.}

Even more, the segregation { seems to be} stronger for type-I AGNs. They { seem } to be hosted more frequently by the more massive, more central-concentrated, and more pressure-supported galaxies at any morphological type. If confirmed, this result { may} indicate that both families of AGNs cannot be unified only by a simple inclination/line-of-sight scheme, as suggested by \citet{urry95}. Other processes like possible evolutionary steps between the two AGN types could play a role in explaining the observed distributions \citep[][]{krongold02,villarroel14}
%\Com{SFS to Alenka: Help here, please}. 

\subsection{On the evolutionary stage of AGN hosts} 

We further discuss possible evolutionary implications from our study of the AGN hosts as compared with the non-active galaxies, both at the level of global properties and radial distributions. { We should stress here that these possible scenarios, based on our results, should have been confirmed by larger samples and more unbiased sample selections, including both radio-loud and X-ray selected AGNs.}

{ The loci of AGN hosts in different observational diagrams.-} We confirm that AGN hosts are mostly located in the so-called GV regime between the blue/SFGs and red/RGs groups, in all the different analyzed distributions: (i) the color-magnitude (or mass) diagram; (ii) the SFR--\ms\ diagram; and (iii) the age-mass diagram. Indeed, one-third of the galaxies in the GV host an AGN (based on our selection criteria), while this fraction is much smaller in the SFG and RG regimes. This result is independent of the contamination of the AGN in the considered parameter. Thus, the location in the GV does not seem to be induced by a contamination by the nuclear source. That result was already suggested by several previous studies \citep[e.g.][]{Kauffmann+2003,sanchez04,Schawinski+2014}, although  only a few of them took into account the contamination by the central source and type-I AGNs \citep[e.g.][]{sanchez04}. 

As we have indicated before, galaxies present a bimodal distribution with two well separated groups of SFGs (blue, young, gas-rich, star-forming, and rotational-supported in general) and RGs (red, old, gas-poor, non-star-forming and pressure-supported in general) for at least the last 8 Gyr \citep[e.g.][]{Bell+2004}. While stars are formed mostly in the first group \citep[e.g.][]{wolf05}, the stellar mass is accumulated in the second one \citep[e.g.][]{bundy09,bundy10}. This implies {\it per se} that should be a transformation from the first group (SFGs) towards the second one (RGs). That transformation requires of (1) some mass/environment-dependent evolutionary processes or the input of a large amount of energy that quenches the SF, and (2) a mechanism that transforms the morphology. The activation of an AGN \citep[e.g.][]{lipari94,sanders96,Hopkins+2009} and major mergers have been proposed to explain the previous first and second transforming steps, respectively.  The location of AGN hosts in the transition regime between both families of galaxies seems to { reinforce} that scenario, and the idea of having a causal connection between quenching and the ignition of the AGN activity is tempting. However, the time scales of both processes may be totally different. The best estimates of the duty cycle of an AGN indicate that most probably its activity lasts of the order of $\sim$0.1 Gyrs \citep[e.g.,][]{parma07,shu15}. However, the quenching process does not seem to have a similar time scale. The age differences indicate that it can last for a few Gyrs, at least when averaged across the galaxy. Therefore, more than a causal connection we can determine that both the AGN activity and quenching happen under similar conditions in galaxies, and the former, if any, may enhance the quenching process. Another possibility is that AGNs are activated in particular periods of the quenching/transformation process. This is supported by the fact that two-thirds of the GV galaxies do not host an AGN, and by the differences in the time scales of the different processes.

There is still the possibility that the GV is populated both by galaxies in a transition between SFG towards RGs or by galaxies that suffer a rejuvenation, presenting a slightly younger stellar population (mostly luminosity-weighted) due to a recent ignition of the SF by the capture of either pristine gas or a gas-rich satellite. However, the analysis of the location of AGN hosts in the Mass--Stellar Metallicity and Mass--gas-phase Metallicity relations (see subsection \ref{sec:metal}) suggests that this is not the main mechanism for populating the GV. Most AGN hosts present a stellar metallicity and gas oxygen abundance below the average value for their corresponding stellar mass. While the injection of more pristine gas and the ignition of SF may somehow reproduce the gas-phase abundance distribution, it can hardly modify the mass-weighted stellar metallicity of the galaxy, dominated by the old stellar population, whose bulk in mass was formed long time ago in almost all galaxies \citep[e.g.][]{perez13,ibarra16,rosa17}. The only way to decrease the stellar metallicity is that SF continues for a period at a lower rate, and therefore there is a lack of a new population of stars more enriched than the previous one. A similar scenario as this one was already proposed by \citet{yates14}. At least in the case of the AGN hosts this seems to be the scenario compatible with the observations.

The decrease of the SFR seems to have been primarily driven by a reduction of the available amount of molecular gas. Although we have derived just a rough estimation of this quantity, it is clear that RGs present a general deficit of gas at any mass range (subsection \ref{sec:gas}. While SFGs exhibit a tight correlation between the integrated stellar mass and the molecular gas, RGs present a wide range of molecular masses, spanning from that correlation (as an upper envelop) towards several orders of magnitudes lower levels (see also Calette et al. 2017). Although this is clearly the main reason for the low SFRs in these galaxies, we have found that RGs have a SFE at least a 50\%\ lower than SFGs for the same amount of molecular gas and stellar mass. Therefore, it is not only a lack of molecular gas what prevents/reduces the SFR but a process that precludes this SF for the same amount of molecular gas available. The lack of molecular gas could be  connected somehow to the AGN activity if this is strong enough to produce outflows that expel a substantial amount of gas outside the gravitational potential towards the galactic halo \citep[e.g.][]{rich10,kraft12,mingo12}. However, in this case we should consider that those GV galaxies without an AGN should be the remnants or fossils of previous AGN hosts. It is also possible that the decline in the SFE is related to the nuclear activity if we consider that an AGN can inject energy that heat the gas, increasing the dispersion and preventing the formation of stars. Again, under this scenario, non-active GV galaxies should be in an evolutionary sequence towards RGs after the AGN phase(s). However, it is not clear that our results support that scenario, as we will see later.

%\subsection{On the evolutionary status of AGN hosts from their radial distributions}

{ Radial distributions.} The radial distributions of the SFR, sSFR, and  molecular gas content analyzed in Secs. \ref{sec:rad} and \ref{sec:GVGs} agree with an scenario in which AGN hosts (and non-active GV galaxies) are in a transition between SFGs and RGs, with a decrease/quench of the SF, at any range of stellar masses. In detail, our results indicate that the decline of the SFR happens the inside out, being stronger in the nuclear regions. This decrease is already appreciated in the most massive SFGs, that present a soft decline of the sSFR and molecular gas content in the very central regions. That decline seems to evolve towards a generalized drop of the SFR for AGN hosts, and, for the central regions, a strong decay in the sSFR and molecular gas content. Those trends are just sharper in RGs, following a clear trend. 
%Finally, the stellar ages of AGN hosts are clearly older for any galactocentric distance and stellar mass bin than that of the SFGs, being more similar to the ones found in the RGs. 
However, contrary to what we have discussed in the previous paragraph, GV galaxies do not seem to be in a evolutionary step after AGNs in this sequence. Indeed, they present a pattern in the analyzed radial distributions more similar to that of the SFGs than to the RGs, with (i) $\Sigma_{SFR}$ and $\Sigma_{sSFR}$ values slightly larger than those of AGN hosts, and (ii) sSFR and molecular gas distributions with softer drops towards the inner regions. On the other hand, their ages are more similar to RGs than to AGN hosts, contrary to the previous results. 
%and (iii) younger stellar populations than that of the AGN hosts and slightly older than that of the SFGs. 

In summary, GV galaxies do not seem to fit in an scheme in which they are evolving from AGN hosts towards RGs, but rather being before active galaxies in this proposed sequence. If this is the case, then the causal connection between the ignition of the AGN activity and the quenching does not seem to fit with the observations. On the contrary, if all galaxies are ordered in the considered sequence, it seems that the process that stops the SF, producing a decline of molecular gas content and a decrease of the SFE in the central regions, happens before the ignition of the AGN activity. Under this scenario, the nuclear activity may speed-up the quenching process, but it does not seem to be its origin. Even more, there is still a third possibility, in which the quenching mechanisms for active and non-active GV galaxies are totally different and there is no evolutionary path between these two families of objects.

%\subsection{Morphological dependence}

{ Morphological dependence.} When analyzing the possible evolutionary paths segregating the galaxies by morphology it is found a very  similar general trend. Both AGN hosts and non-active GV galaxies seem to be in an intermediate step between SFGs and RGs with a general decrease of the SFR and sSFR, and a stronger decline in the sSFR and the molecular gas content in the central regions, coincident with older stellar populations. In the same way, non-active GV galaxies seem to have sSFRs and  molecular gas radial distributions more similar to the SFGs than to the RGs compared with those ones of AGN hosts. Thus, the evolutionary sequence, if exists, indicates that SFE and gas decline happen before the ignition of the AGN, and not after (or the two events are not causal or evolutionary connected). The main difference with respect to the sequence described for different stellar masses is that early- and late-type galaxies seem to present clearly distinguished evolutionary paths. Early-type galaxies show a sharper transition, in particular for non-active GV galaxies, with a larger drop in the SFR and sSFR, even with a lower or similar decline of the molecular gas content. Late-type galaxies present a lower decline in the SFR and sSFR, and mostly concentrated in the central regions, e.g., Sa-Sd GV galaxies, Fig. \ref{fig:rad_GV_Morph} and Sbc-Sd AGN hosts, Fig. \ref{fig:rad_Morph}. This result may indicate that there exist a single mechanism to explain the quenching and that bulge growth plays a role in the declining of
the SFR. For bulge-dominated systems (E/S0), the decline affects the whole galaxy, while for galaxies with smaller bulges (Sa--Sd), it is clearly associated to the central regions.

The most accepted scenario for the transformation between late-type SFGs and early-type RGs involves a retrograde major merger between gas rich galaxies that leads to instabilities in the gas. This gas falls towards the central regions igniting a violent star-burst and producing the formation of a pressure-supported spheroid. The gas infalling to the center ignites an AGN that expels and heats the remaining gas, thus quenching the SF from the inside-out \citep[][]{lipari94,sanders96,hopkins09,Hopkins+2010}. However, this scenario predicts that AGN hosts evolve into non-active GV galaxies as the timescale of the nuclear activity is considerably shorter than the transformation timescale. As we showed before, our results indicate the contrary: non-active GV galaxies seem to be in an earlier evolutionary stage than AGN hosts. \citet{martig09} proposed an alternative scenario in which the SF is halted without requiring a deficit of gas content. They proposed that the growth of a stellar spheroid can stabilize the gas disk, and quench SF by preventing the fragmentation of bound gas clumps.
This mechanism may fit somehow in the observed transition for late-type galaxies, and in the fact that we see a decrease of the SFE that seems to be stronger in the central regions. { This scenario agrees with the recent results by \citet{colombo17} in these regards.} However, we should stress that in all the cases we observe a decrease of the gas content in those regions too, and that seems to be the primary reason of why there is no SF. Therefore, gas is expelled or consumed somehow, although the efficiency may be affected by the extra stabilization introduced by the stellar spheroid.

It remains still open the question of what does produce the drop of molecular gas content in the inner regions and why the efficiency in the SF decreases. The lack of molecular gas could be induced by the energy injection from the AGN itself. Outflows associated with nuclear activity are frequently found in strong AGNs \citep[e.g.][]{rich10}. However, we largely ignore the effective energy input of this feedback on galaxies, despite the advances on the topic \citep[e.g.][]{fabian12}. On the other hand, central SF events may also drive outflows \citep[e.g.][]{carlos16}, without the requirement of an AGN or a violent process. In many cases those outflows are not detected due to projection effects, and their frequency are still unclear. How much gas is expelled during this events is still under debate, although it is considered that it is proportional to the SFR \citep[e.g.][]{Lilly+2013}, and therefore their effect should be larger in the more massive SFGs. Whether these events may be strong enough to induce an inside-out quenching that propagates through all the galaxy is unclear and dubious. In the case of the decrease of the SFE a plausible explanation could be the scenario proposed by \citet{martig09}.
Maybe the outflows themselves trigger the creation of an spheroidal seed and the stabilization of the disk halts the SF in a more gentle way for late-type galaxies, matching with the observations by \citet{Schawinski+2010}. This could be an scenario to be tested with simulations.

\section{Conclusions}
\label{sec:con}
For the sample of galaxies currently observed by the MaNGA survey (DR14), we have selected AGN host galaxies based on the optical spectroscopic properties of the nuclear regions, following an spectral analysis performed using the {\sc Pipe3D} pipeline \citep{Pipe3D_II}. A total of 98 AGN (36 type-I) were selected out of $\sim$2700 analyzed galaxies. We have explored the main global and radial properties of these AGN galaxies with those of the non-active galaxies having in mind whether the galaxies are star-forming, retired or from the green valley region. Our main conclusions can be summarized in the following way:

\begin{itemize}

\item AGN { seem to be }located mostly in early-type galaxies or early-type spirals. For a given morphology, AGN hosts { seem to be } more frequently found in the regime of the more massive, more compact, and more pressure-supported galaxies.

\item AGN hosts are { preferentially} located in the so-called GV and intermediate regime between blue-cloud/star-forming galaxies and red-sequence/retired galaxies. Their locations do not seem to be affected by the contamination from the AGN on the global properties, { or by a clear bias in the selection procedure of the AGN candidates.}

\item The population of active and non-active galaxies in the GV appears to be in a transition between SFGs towards RGs due to a decrease of the amount of molecular gas and a lower SFE. Rejuvenation, although it is still possible for particular galaxies, does not seem to be the main scenario to explain the properties of these galaxies. 
%{ This result should be explored with a more detailed analysis of the star-formation histories and different AGN selections.}

\item The decline/quenching of the SFR in the evolution from the SFGs towards the RGs happens inside-out, both in AGN hosts and non-active GV galaxies. This decline { seems to be} primarily induced by a drop in the amount of molecular gas in the inner regions. { Since our estimation of the molecular gas content is based on a dust-to-gas ratio that is valid only in an statistical manner, these result should be confirmed by more direct estimations of the molecular gas, like the one provided by CO measurements \citep[e.g.][]{bolatto17}. Indeed, recent estimations of the molecular gas based on CO observations of MaNGA GV galaxies seem to confirm our conclusions \citep{lin17}}

\item Non-active GV galaxies do not seem to be in an evolutionary stage between AGN hosts and RGs. If there is an evolutionary sequence between these two families in the transition between SFG and RGs, non-active ones seem to be in an earlier evolutionary stage, regarding their SFR, sSFR, and gas content. That would imply that AGN activity does not seem to be the only driver for the inside-out quenching. Even more, it may imply that there are different processes that produce the decline or quenching in the SFR. The results, regarding their stellar ages point towards the opposite scenario.

\item There is evidence of different evolutionary paths between SFGs and RGs for different morphological types. In particular, for non-active GV galaxies, the transition seems to be smoother for late-type galaxies and sharper for earlier-type ones.

\end{itemize}

In addition, we have publicly distributed all the dataproducts produced by the {\sc Pipe3D} pipeline used along this work as a Value Added Catalog corresponding to the DR14 of the SDSS survey (Appendix \ref{APP1}). To our knowledge, this is the largest distribution of this kind of IFU dataproducts produced so far.

\section{Acknowledgements}

We thank the CONACyT programs CB-180125 and DGAPA-UNAM IA100815 and IA101217 grants for their support to this project. CAN thanks CONACyT programs CB-221398 and DGAPA-UNAM grant IN107313. 
The data products presented in this paper  benefited from support and resources from the HPC cluster Atocatl at IA-UNAM. TB would like to acknowledge support from the CONACyT Research Fellowships  program.

Funding for the Sloan Digital Sky Survey IV has been provided by
the Alfred P. Sloan Foundation, the U.S. Department of Energy Office of
Science, and the Participating Institutions. SDSS-IV acknowledges
support and resources from the Center for High-Performance Computing at
the University of Utah. The SDSS web site is www.sdss.org.

SDSS-IV is managed by the Astrophysical Research Consortium for the 
Participating Institutions of the SDSS Collaboration including the 
Brazilian Participation Group, the Carnegie Institution for Science, 
Carnegie Mellon University, the Chilean Participation Group, the French Participation Group, Harvard-Smithsonian Center for Astrophysics, 
Instituto de Astrof\'isica de Canarias, The Johns Hopkins University, 
Kavli Institute for the Physics and Mathematics of the Universe (IPMU) / 
University of Tokyo, Lawrence Berkeley National Laboratory, 
Leibniz Institut f\"ur Astrophysik Potsdam (AIP),  
Max-Planck-Institut f\"ur Astronomie (MPIA Heidelberg), 
Max-Planck-Institut f\"ur Astrophysik (MPA Garching), 
Max-Planck-Institut f\"ur Extraterrestrische Physik (MPE), 
National Astronomical Observatories of China, New Mexico State University, 
New York University, University of Notre Dame, 
Observat\'ario Nacional / MCTI, The Ohio State University, 
Pennsylvania State University, Shanghai Astronomical Observatory, 
United Kingdom Participation Group,
Universidad Nacional Aut\'onoma de M\'exico, University of Arizona, 
University of Colorado Boulder, University of Oxford, University of Portsmouth, 
University of Utah, University of Virginia, University of Washington, University of Wisconsin, 
Vanderbilt University, and Yale University.

\appendix
\label{APP1}

\section{Pipe3D Data products}
\label{app:Pipe3D}

As stated in Sec.\ref{ana}, all the analysis performed along this article is based on the dataproducts produced by the {\sc Pipe3D} pipeline \citep{Pipe3D_II}. These dataproducts have been distributed as part of the 14th Data Release of the SDSS-IV survey (Masters et al., submitted), as part of a Value Added Catalogue (VAC) webpage \footnote{\url{http://www.sdss.org/dr14/manga/manga-data/manga-pipe3d-value-added-catalog/}}. This VAC comprises a single FITs file per galaxy/datacube  within the current MaNGA data release (v2\_1\_2), named  
manga-[plate]-[ifudsgn].Pipe3D.cube.fits.gz, where [plate] is the plate number, [ifudesign] is the design IFU size and number. Each FITs file comprises five extensions, that comprises the following information:

\begin{itemize}
\item HDU0 (ORG\_HDR): Header of the original MaNGA datacube.
\item HDU1 (SSP): Main parameters derived from the analysis of the stellar populations, including the luminosity-weighted and mass-weighted ages, metallicities, dust attenuation and stellar kinematics properties.
\item HDU2 (SFH): Weights of the decomposition of the stellar population for the adopted SSP library. It can be used to derive the spatial resolved star-formation and chemical enrichment histories of the galaxies and the luminosity-weighted and mass-weighted properties included in the NAME.SSP.cube.fits.gz dataproducts.
\item HDU3 (FLUX\_ELINES): Main parameters of 52 strong and weak emission lines derived using a weighted momentum analysis based on the kinematics of Halpha. It includes the flux intensity, equivalent width, velocity and velocity dispersion, and the corresponding errors for the different analyzed emission lines \footnote{ The EWs are a factor two larger than real ones due to a bug in version 2.1.2 of the code.}.
\item HDU4 (INDICES): Set of stellar absorption indices derived for each spaxel once subtracted the emission line contribution.
\end{itemize}

Apart from HDU0 (ORG\_HDR), the remaining extensions have the format of a datacube with the X- and Y-axis corresponding to the position in the sky, with the same sampling, format and WCS of the original MaNGA datacubes (stored in HDU0). In each channel/slice in the Z-axis is stored a different data-product (physical or observational parameter) derived by {\sc Pipe3D}, following the nomenclature fully described in \citet{Pipe3D_II}, with the information required to recover the corresponding parameter and its units stored in the header. The full headers are described in the SDSS-DR14 {\sc Pipe3D} VAC webpage.
%\footnote{\url{https://internal.sdss.org/dr14/datamodel/files/MANGA_PIPE3D/MANGADRP_VER/PIPE3D_VER/PLATE/manga.Pipe3D.cube.html}}. 

These dataproducts have been used to derive the integrated, characteristic and central properties presented in Sec. \ref{ana} and \ref{results}, and in particular in Fig. \ref{fig:BPT_EW}, \ref{fig:SFMS}, \ref{fig:MZR} and \ref{fig:SKL}. The full set of dataproducts is distributed through the SDSS-VAC webpages. 
%\url{https://data.sdss.org/sas/dr14/manga/spectro/pipe3d/v2_1_2/2.1.2/}.

%%%%%%%%%%%%%%%%%%%%%%%%%%%%%%%%%%%%%%%%%%%%%%%%%%%%%%%
\begin{figure*}
  \centering
    \includegraphics[width=6.5cm, angle=270]{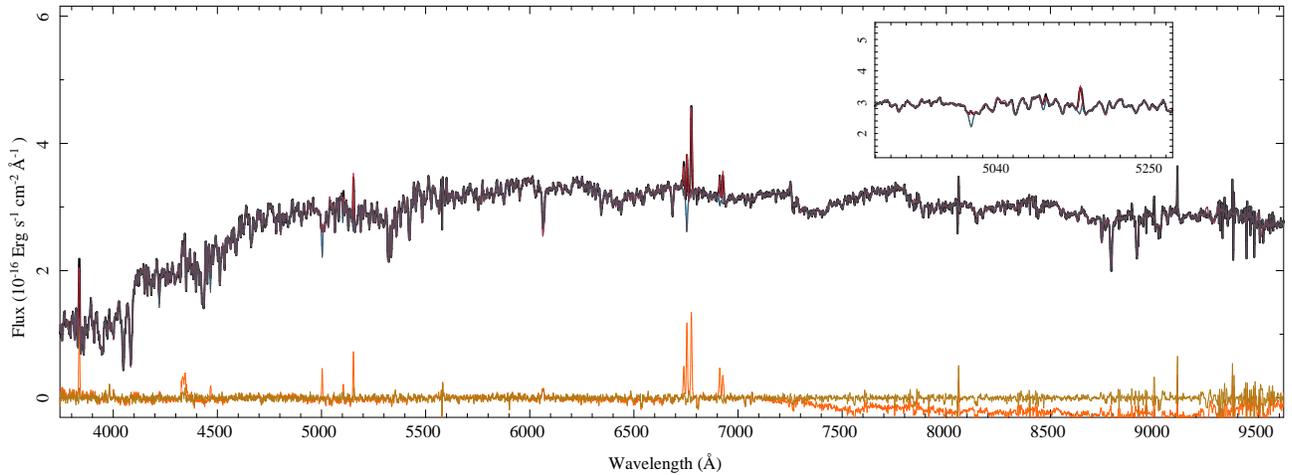}
  \caption{An example of the stellar-population and emission line analysis performed using {\sc Pipe3D} for the central spectrum (3\arcsec/diameter) of the galaxy manga-7495-12704 (black solid line). The best fitted stellar-population model is shown as a grey solid-line, and the best fitted stellar and ionized-gas model is shown as a red one. The original data (black) are plotted with three times the width of the other two to high-light the differences. In light-blue is shown the stellar component once subtracted the best fitted emission line models. The residual once subtracted the stellar population, without considering possible deviations in the spectrophotometric calibration of the stellar-library and the data is shown as an orange solid line \citep[see ][for more details]{Pipe3D_I}. Finally, the residuals from the full modeling (stellar population, ionized gas and spectrophotometric miss-match) is shown as a olive-green solid line. The overall residuals is smaller than the $\sim$5\% for most of the spectra \citep[][]{ibarra16,Pipe3D_II}. Similar plots for all the galaxies are available in the following webpage:\url{https://sas.sdss.org/resources/dr14/manga/spectro/pipe3d/v2_1_2/2.1.2/list/}}
  \label{fig:cen_spec}
\end{figure*}
%%%%%%%%%%%%%%%%%%%%%%%%%%%%%%%%%%%%%%%%%%%%%%%%%%%%%%%

In addition it is delivered a single FITs table with an entry per cube comprising the integrated properties of those galaxies (e.g., stellar mass, star-formation rate...), the characteristic values (e.g., oxygen abundance at the effective radius), some relevant information about the galaxy (e.g., ionization conditions in the center of the galaxy), together with some information extracted from the MaNGA-cube header to allow the identification of the target either in the sky or in the survey. The full list of parameters derived for each single galaxy are listed in Table \ref{tab:model}, and the FITs table can be downloaded from the SDSS-DR14 webpages\footnote{\url{https://data.sdss.org/sas/dr14/manga/spectro/pipe3d/v2_1_2/2.1.2/manga.Pipe3D-v2_1_2.fits}}.

%%%%%%%%%%%%%%%%%%%%%%%%%%%%%%%%%%%%%%%%%%%%%%%%%%%%%%%
\begin{figure*}
  \centering
    \includegraphics[width=16cm]{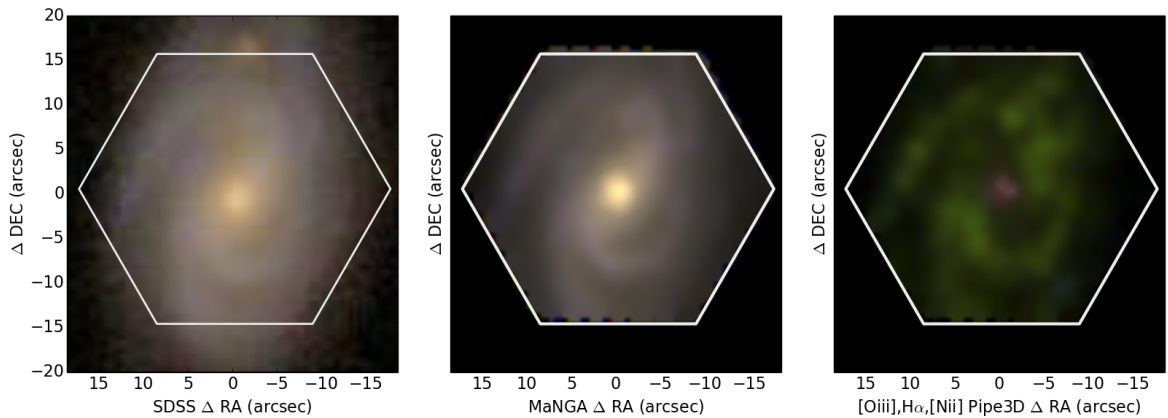}
  \caption{{\it Left panel:} Color image created from the SDSS $g$-,$r$- and $i$-band post-stamp images of 40$\arcsec$$\times$40$\arcsec$ FoV centred in the galaxy manga-7495-12704 (which central spectrum is shown in Fig. \ref{fig:cen_spec}). {\it Central panel:} Similar color image created using the same band images extracted from the original MaNGA cubes, by convolving each spectra at each spaxel with the corresponding filter response curve. {\it Right-Panel:} Color image created combining the [\ION{O}{iii}] (blue), H$\alpha$ (green) and [\ION{N}{ii}] emission line intensity maps extracted from the {\sc FLUX\_ELINES} extension of the {\sc Pipe3D} datacube of the same galaxy. Similar plots for all the galaxies are available in the following webpage: \url{https://sas.sdss.org/resources/dr14/manga/spectro/pipe3d/v2_1_2/2.1.2/list/}}
  \label{fig:maps}
\end{figure*}
%%%%%%%%%%%%%%%%%%%%%%%%%%%%%%%%%%%%%%%%%%%%%%%%%%%%%%%

\subsection{Quality Control}
\label{app:QC}

A visual inspection was applied to (1) the central spectrum (2.5 arcsec/diameter) and best fitted model (Figure \ref{fig:cen_spec}), and (2) the continuum and emission line maps (Figure \ref{fig:maps}) in order to identify critical/evident problems that may affect the quality of the data for the $\sim$2800 analyzed datacubes. When there is a clear issue with the data or the fitting the galaxy/cube is marked and it is removed from the list. Prior to any removal a new attempt to analyze the data modifying the input parameter of the pipeline is performed (in essence the initial guess for the redshift or the location of the centroid of the galaxy). Only for less than 20 cubes it is needed to adjust the input parameters in order to improve the quality of the analysis. In total less than 100 galaxies were removed due to critical issues with the input data. 

In addition to the visual inspection we test the accuracy of the derived quantities by performing a few simple comparisons between them and the ones included in the NSA catalog. For the Quality Control analysis we compared two  basic parameters: (i) the integrated stellar mass and (ii) the redshift. In general the number of galaxies with clear offsets between the NSA and the {\sc Pipe3D} results is very reduced ($\sim$ 3\%), and so far it is not clear the nature of these discrepancies. 

The QC-table includes the results of this basic Quality Control (QC) analysis. It includes for each of the analyzed cubes/galaxies a QC-flag. We label as OK those galaxies/cubes for which we do not find any problem (flag=2). Initially, we labeled as WRONG\_REDSHIFT (flag=1), those galaxies for which the fitting provided an erroneous automatic derivation of the redshift (mostly galaxies not centered in the FoV). There were about $\sim$100 of those cases. However, after different iteration modifying the input parameters there was no final object under this category. We label as BAD (flag=2) those cubes/galaxies for which {\sc Pipe3D} is unable to derive a data-product (4 cubes, with 19 more directly removed from the catalog being impossible to derive the average properties of the galaxies due to their low S/N). Finally, we consider WARNINGs (flag=3) those ones that either have a potential problem i<n the fitting and had to be refitted manually (~107 cubes/galaxies), and those that have differences either in the redshift (22 galaxies) or the stellar mass (55 galaxies) derivation when comparing to the NSA results beyond 2$\sigma$. In summary, of the 2812 original cubes, only for 2810 it was possible to perform the analysis. Of them in 19 the fitting process did not converge (mostly empty fields or very low-S/N targets). In addition there are 4 BAD cubes and 162 possible warnings. The final QC table was distributed together with the dataproducts in the SDSS-VAC webpage indicated before.

%%%% Table of delivered parameters
\begin{table*}
\caption{Parameters distributed in the Pipe3D VAC table}
\label{tab:model}
\begin{tabular}{llllL{9cm}}
\hline
COLUMN  & NAME & TYPE & UNITS & Description \\ 
\hline
1 &  MaNGAID& string&  & MaNGA name of the cube \\ 
2 &  objra& float& degree  & RA of the object  \\ 
3 &  objdec& float& degree  & DEC of the object  \\ 
4 &  redshift& float&   & redshift derived by Pipe3D form the center of the galaxy \\ 
5 &  re\_arc& float& arcsec& adopted effective radius in arcsec \\ 
6 &  PA& float& degrees& adopted position angle in degrees \\ 
7 &  ellip& float& & adopted elipticity \\ 
8 &  DL& float& & adopted luminosity distance in Mpc \\ 
9 &  re\_kpc& float& kpc& derived effective radius in kpc \\ 
10 &  log\_Mass& float& log(Msun)& Integrated stellar mass in units of the solar mass in logarithm scale \\ 
11 &  e\_log\_Mass& float& log(Msun) & Error of the integrated stellar mass in units of the solar mass in logarithm scale \\ 
12 &  log\_SFR\_Ha& float& log(Msun/yr)& Integrated star-formation rate derived from the integrated Halpha flux in logarithm scale \\ 
13 &  e\_log\_SFR\_Ha& float& log(Msun/yr)& Error of the Integrated star-formation rate derived from the integrated Halpha flux in logarithm scale \\ 
14 &  log\_SFR\_ssp& float& log(Msun/yr)& Integrated star-formation rate derived from the amount of stellar mass formed in the last 32Myr in logarithm scale \\ 
15 &  e\_log\_SFR\_ssp& float& log(Msun/yr)& Error of the Integrated star-formation rate derived from the amount of stellar mass formed in the last 32Myr in logarithm scale \\ 
16 &  log\_Mass\_gas& float& log(Msun)& Integrated gas mass in units of the solar mass in logarithm scale estimated from the dust attenuation \\ 
17 &  e\_log\_Mass\_gas& float& log(Msun)& Error in the integrated gas mass in units of the solar mass in logarithm scale estimated from the dust attenuation \\ 
18 &  Age\_LW\_Re\_fit& float& log(yr)& Luminosity weighted age of the stellar population in logarithm of years at the effective radius of the galaxy \\ 
19 &  e\_Age\_LW\_Re\_fit& float& log(yr)& Error in the luminosity weighted age of the stellar population in logarithm of years at the effective radius of the galaxy \\ 
20 &  alpha\_Age\_LW\_Re\_fit& float& & slope of the gradient of the LW log-age of the stellar population within a galactocentric distance of 0.5-2.0 r\_eff \\ 
21 &  e\_alpha\_Age\_LW\_Re\_fit& float& & Error of the slope of the gradient of the LW log-age of the stellar population within a galactocentric distance of 0.5-2.0 r\_eff \\ 
22 &  age\_MW\_Re\_fit& float& log(yr)& Mass weighted age of the stellar population in logarithm of years at the effective radius of the galaxy \\ 
23 &  e\_age\_MW\_Re\_fit& float& log(yr)& Error in the Mass weighted age of the stellar population in logarithm of years at the effective radius of the galaxy \\ 
\hline
\end{tabular}
\end{table*}
\clearpage
\addtocounter{table}{-1}
\begin{table*}
\caption{Parameters distributed in the Pipe3D VAC table ({\it continue})}
\begin{tabular}{llllL{9cm}}
\hline
COLUMN  & NAME & TYPE & UNITS & Description \\ 
\hline
24 &  alpha\_Age\_MW\_Re\_fit& float& & slope of the gradient of the MW log-age of the stellar population within a galactocentric distance of 0.5-2.0 r\_eff \\ 
25 &  e\_alpha\_Age\_MW\_Re\_fit& float& & Error of the slope of the gradient of the MW log-age of the stellar population within a galactocentric distance of 0.5-2.0 r\_eff \\ 
26 &  ZH\_LW\_Re\_fit& float& log(yr)& Luminosity weighted metallicity of the stellar population in logarithm normalized to the solar one at the effective radius of the galaxy \\ 
27 &  e\_ZH\_LW\_Re\_fit& float& log(yr)& Error in the luminosity weighted metallicity of the stellar population in logarithm normalized to the solar one at the effective radius of the galaxy \\ 
28 &  alpha\_ZH\_LW\_Re\_fit& float& & slope of the gradient of the LW log-metallicity of the stellar population within a galactocentric distance of 0.5-2.0 r\_eff \\ 
29 &  e\_alpha\_ZH\_LW\_Re\_fit& float& & Error of the slope of the gradient of the LW log-metallicity of the stellar population within a galactocentric distance of 0.5-2.0 r\_eff \\ 
30 &  ZH\_MW\_Re\_fit& float& log(yr)& Mass weighted metallicity of the stellar population in logarithm normalized to the solar one at the effective radius of the galaxy \\ 
31 &  e\_ZH\_MW\_Re\_fit& float& log(yr)& Error in the Mass weighted metallicity of the stellar population in logarithm normalized to the solar one at the effective radius of the galaxy \\ 
32 &  alpha\_ZH\_MW\_Re\_fit& float& & slope of the gradient of the MW log-metallicity of the stellar population within a galactocentric distance of 0.5-2.0 r\_eff \\ 
33 &  e\_alpha\_ZH\_MW\_Re\_fit& float& & Error of the slope of the gradient of the MW log-metallicity of the stellar population within a galactocentric distance of 0.5-2.0 r\_eff \\ 
34 &  Av\_ssp\_Re& float& mag& Dust attenuation in the V-band derived from the analysis of the stellar populations at the effective radius \\ 
35 &  e\_Av\_ssp\_Re& float& mag& Error of the dust attenuation in the V-band derived from the analysis of the stellar populations at the effective radius \\ 
36 &  Av\_gas\_Re& float& mag& Dust attenuation in the V-band derived from the Ha/Hb line ratios at the effective radius \\ 
37 &  e\_Av\_gas\_Re& float& mag& Error of the dust attenuation in the V-band derived from the Ha/Hb line ratios at the effective radius \\ 
38 &  OH\_Re\_fit\_O3N2& float& & 12+log(O/H) oxygen abundance at the effective radius derived using the Marino et al. 2013 O3N2 calibrator \\ 
39 &  e\_OH\_Re\_fit\_O3N2& float& & Error of 12+log(O/H) oxygen abundance at the effective radius derived using the Marino et al. 2013 O3N2 calibrator \\ 
40 &  alpha\_OH\_Re\_fit\_O3N2& float& & Slope of the oxygen abundance derived using the Marino et al. 2013 O3N2 calibrator within a galactocentric distance of 0.5-2.0 r\_eff \\ 
41 &  e\_alpha\_OH\_Re\_fit\_O3N2& float& & Error of the slope of the oxygen abundance derived using the Marino et al. 2013 O3N2 calibrator within a galactocentric distance of 0.5-2.0 r\_eff \\ 
42 &  OH\_Re\_fit\_N2& float& & 12+log(O/H) oxygen abundance at the effective radius derived using the Marino et al. 2013 N2 calibrator \\ 
43 &  e\_OH\_Re\_fit\_N2& float& & Error of 12+log(O/H) oxygen abundance at the effective radius derived using the Marino et al. 2013 N2 calibrator \\ 
44 &  alpha\_OH\_Re\_fit\_N2& float& & Slope of the oxygen abundance derived using the Marino et al. 2013 N2 calibrator within a galactocentric distance of 0.5-2.0 r\_eff \\ 
45 &  e\_alpha\_OH\_Re\_fit\_N2& float& & Error of the slope of the oxygen abundance derived using the Marino et al. 2013 N2 calibrator within a galactocentric distance of 0.5-2.0 r\_eff \\ 
46 &  OH\_Re\_fit\_ONS& float& & 12+log(O/H) oxygen abundance at the effective radius derived using the Pilyugin et al. 2010 ONS calibrator \\ 
47 &  e\_OH\_Re\_fit\_ONS& float& & Error of 12+log(O/H) oxygen abundance at the effective radius derived using the Pilyugin et al. 2010 ONS calibrator \\ 
\hline
\end{tabular}
\end{table*}
\clearpage
\addtocounter{table}{-1}
\begin{table*}
\caption{Parameters distributed in the Pipe3D VAC table ({\it continue})}
\begin{tabular}{llllL{9cm}}
\hline
COLUMN  & NAME & TYPE & UNITS & Description \\ 
\hline
48 &  alpha\_OH\_Re\_fit\_ONS& float& & Slope of the oxygen abundance derived using the Pilyugin et al. 2010 ONS calibrator within a galactocentric distance of 0.5-2.0 r\_eff \\ 
49 &  e\_alpha\_OH\_Re\_fit\_ONS& float& & Error of the slope of the oxygen abundance derived using the Pilyugin et al. 2010 ONS calibrator within a galactocentric distance of 0.5-2.0 r\_eff \\ 
50 &  OH\_Re\_fit\_pyqz& float& & 12+log(O/H) oxygen abundance at the effective radius derived using the pyqz calibrator \\ 
51 &  e\_OH\_Re\_fit\_pyqz& float& & Error of 12+log(O/H) oxygen abundance at the effective radius derived using the pyqz calibrator \\ 
52 &  alpha\_OH\_Re\_fit\_pyqz& float& & Slope of the oxygen abundance derived using the pyqz calibrator within a galactocentric distance of 0.5-2.0 r\_eff \\ 
53 &  e\_alpha\_OH\_Re\_fit\_pyqz& float& & Error of the slope of the oxygen abundance derived using the pyqz calibrator within a galactocentric distance of 0.5-2.0 r\_eff \\ 
54 &  OH\_Re\_fit\_t2& float& & 12+log(O/H) oxygen abundance at the effective radius derived using the t2 calibrator \\ 
55 &  e\_OH\_Re\_fit\_t2& float& & Error of 12+log(O/H) oxygen abundance at the effective radius derived using the t2 calibrator \\ 
56 &  alpha\_OH\_Re\_fit\_t2& float& & Slope of the oxygen abundance derived using the t2 calibrator within a galactocentric distance of 0.5-2.0 r\_eff \\ 
57 &  e\_alpha\_OH\_Re\_fit\_t2& float& & Error of the slope of the oxygen abundance derived using the t2 calibrator within a galactocentric distance of 0.5-2.0 r\_eff \\ 
58 &  OH\_Re\_fit\_M08& float& & 12+log(O/H) oxygen abundance at the effective radius derived using the Maiolino et al. 2008 calibrator \\ 
59 &  e\_OH\_Re\_fit\_M08& float& & Error of 12+log(O/H) oxygen abundance at the effective radius derived using the Maiolino et al. 2008 calibrator \\ 
60 &  alpha\_OH\_Re\_fit\_M08& float& & Slope of the oxygen abundance derived using the Maiolino et al. 2008 calibrator within a galactocentric distance of 0.5-2.0 r\_eff \\ 
61 &  e\_alpha\_OH\_Re\_fit\_M08& float& & Error of the slope of the oxygen abundance derived using the Maiolino et al. 2008 calibrator within a galactocentric distance of 0.5-2.0 r\_eff \\ 
62 &  OH\_Re\_fit\_T04& float& & 12+log(O/H) oxygen abundance at the effective radius derived using the Tremonti et al. 2004 calibrator \\ 
63 &  e\_OH\_Re\_fit\_T04& float& & Error of 12+log(O/H) oxygen abundance at the effective radius derived using the Tremonti et al. 2004 calibrator \\ 
64 &  alpha\_OH\_Re\_fit\_T04& float& & Slope of the oxygen abundance derived using the Tremonti et al. 2004 calibrator within a galactocentric distance of 0.5-2.0 r\_eff \\ 
65 &  e\_alpha\_OH\_Re\_fit\_T04& float& & Error of the slope of the oxygen abundance derived using the Tremonti et al. 2004 calibrator within a galactocentric distance of 0.5-2.0 r\_eff \\ 
66 &  NO\_Re\_fit\_EPM09& float& & log(N/O) nitrogen-to-oxygen abundance at the effective radius derived using the Perez-Montero et al. 2009 calibrator \\ 
67 &  e\_NO\_Re\_fit\_EPM09& float& & Error of log(N/O) nitrogen-to-oxygen abundance at the effective radius derived using the Perez-Montero et al. 2009 calibrator \\ 
68 &  alpha\_NO\_Re\_fit\_EPM09& float& & Slope of the nitrogen-to-oxygen abundance derived using the Perez-Montero et al. 2009 calibrator \\ 
69 &  e\_alpha\_NO\_Re\_fit\_EPM09& float& & Error of the slope of the nitrogen-to-oxygen abundance derived using the Perez-Montero et al. 2009 calibrator \\ 
70 &  NO\_Re\_fit\_N2S2& float& & log(N/O) nitrogen-to-oxygen abundance at the effective radius derived using the Perez-Montero et al. 2009 calibrator \\ 
71 &  e\_NO\_Re\_fit\_N2S2& float& & Error of log(N/O) nitrogen-to-oxygen abundance at the effective radius derived using the Perez-Montero et al. 2009 calibrator \\ 
\hline
\end{tabular}
\end{table*}
\clearpage
\addtocounter{table}{-1}
\begin{table*}
\caption{Parameters distributed in the Pipe3D VAC table ({\it continue})}
\begin{tabular}{llllL{9cm}}
\hline
COLUMN  & NAME & TYPE & UNITS & Description \\ 
\hline
72 &  alpha\_NO\_Re\_fit\_N2S2& float& & Slope of the nitrogen-to-oxygen abundance derived using the Dopita et al. N2/S2 calibrator \\ 
73 &  e\_alpha\_NO\_Re\_fit\_N2S2& float& & Error of the slope of the nitrogen-to-oxygen abundance derived using the Dopita et al. N2/S2 calibrator \\ 
74 &  log\_NII\_Ha\_cen& float& & logarithm of the [NII]6583/Halpha line ratio in the central 2.5arcsec/aperture \\ 
75 &  e\_log\_NII\_Ha\_cen& float& & error in the logarithm of the [NII]6583/Halpha line ratio in the central 2.5arcsec/aperture \\ 
76 &  log\_OIII\_Hb\_cen& float& & logarithm of the [OIII]5007/Hbeta line ratio in the central 2.5arcsec/aperture \\ 
77 &  e\_log\_OIII\_Hb\_cen& float& & error in the logarithm of the [OIII]5007/Hbeta line ratio in the central 2.5arcsec/aperture \\ 
78 &  log\_SII\_Ha\_cen& float& & logarithm of the [SII]6717+6731/Halpha line ratio in the central 2.5arcsec/aperture \\ 
79 &  e\_log\_SII\_Ha\_cen& float& & error in the logarithm of the [SII]6717/Halpha line ratio in the central 2.5arcsec/aperture \\ 
80 &  log\_OII\_Hb\_cen& float& & logarithm of the [OII]3727/Hbeta line ratio in the central 2.5arcsec/aperture \\ 
81 &  e\_log\_OII\_Hb\_cen& float& & error in the logarithm of the [OII]3727/Hbeta line ratio in the central 2.5arcsec/aperture \\ 
82 &  EW\_Ha\_cen& float& & EW of Halpha in the central 2.5arcsec/aperture \\ 
83 &  e\_EW\_Ha\_cen& float& & error of the EW of Halpha in the central 2.5arcsec/aperture \\ 
84 &  ion\_class\_cen& float& & Classification of the central ionization \\ 
85 &  sigma\_cen& float& km/s& Velocity dispersion (i.e. sigma) in the central 2.5 arcsec/aperture derived for the stellar populations \\ 
86 &  e\_sigma\_cen& float& km/s& error in the velocity dispersion in the central 2.5 arcsec/aperture derived for the stellar populations \\ 
87 &  sigma\_cen\_Ha& float& km/s& Velocity dispersion (i.e. sigma) in the central 2.5 arcsec/aperture derived for the Halpha emission line \\ 
88 &  e\_sigma\_cen\_Ha& float& km/s& error in the velocity dispersion in the central 2.5 arcsec/aperture derived for the  Halpha emission line \\ 
89 &  vel\_sigma\_Re& float&  & Velocity/dispersion ratio for the stellar populations within 1.5 effective radius \\ 
90 &  e\_vel\_sigma\_Re& float&  & error in the velocity/dispersion ratio for the stellar populations within 1.5 effective radius \\ 
91 &  Lambda\_Re& float&  & Specific angular momentum (lambda parameter) for the stellar populations within 1.5 effective radius \\ 
92 &  e\_Lambda\_Re& float&  & error in the specific angular momentum (lambda parameter) for the stellar populations within 1.5 effective radius \\ 
93 &  plateifu& string&  &  \\ 
94 &  plate& int&  & plate ID of the MaNGA cube \\ 
95 &  ifudsgn& int&  & IFU bundle ID of the MaNGA cube \\ 
\hline
\end{tabular}
\end{table*}

%%%%%%%%%% TABLE of AGN candidates
\begin{table*}
\caption{List of AGN candidates}
\label{tab:agns}
\begin{tabular}{lllrrrrrrc}
\hline
MaNGAID & RA & DEC & [\ION{N}{ii}]/H$\alpha$ & [\ION{O}{iii}]/H$\beta$ & [\ION{S}{ii}]/H$\alpha$ & [\ION{O}{i}]/H$\alpha$ & EW(H$\alpha$) & S/N & AGN\\ 
        & (deg) & (deg)  & log10  & log10 & log10 & log10 & \AA & H${\alpha} B$ & type \\ 
\hline
7443-6104 & 232.158069 & 42.442017 &  0.09$\pm$ 0.05 &  0.52$\pm$ 0.23 &  0.15$\pm$ 0.06 & -0.74$\pm$ 0.20 & -3.2$\pm$ 0.4 &  3.9 & 2 \\ 
7495-1902 & 205.044769 & 26.841041 & -0.60$\pm$ 0.03 &  0.61$\pm$ 0.08 & -0.32$\pm$ 0.03 & -1.41$\pm$ 0.25 & -9.9$\pm$ 0.4 &  0.0 & 2 \\ 
7815-6104 & 319.193099 & 11.043741 & -0.30$\pm$ 0.03 &  0.98$\pm$ 0.01 & -0.27$\pm$ 0.02 & -1.42$\pm$ 0.03 & -83.1$\pm$ 8.6 &  9.5 & 1 \\ 
7957-12703 & 258.190224 & 36.278856 &  0.13$\pm$ 0.06 &  0.22$\pm$ 0.11 &  0.05$\pm$ 0.07 & -0.88$\pm$ 0.19 & -5.9$\pm$ 2.8 &  4.2 & 2 \\ 
7960-3701 & 257.085763 & 31.746915 &  0.18$\pm$ 0.04 &  0.44$\pm$ 0.06 &  0.13$\pm$ 0.03 & -1.06$\pm$ 0.28 & -3.4$\pm$ 0.3 &  1.1 & 2 \\ 
7991-6104 & 258.827410 & 57.658770 &  0.28$\pm$ 0.01 &  0.71$\pm$ 0.05 & -0.02$\pm$ 0.04 & -1.35$\pm$ 0.08 & -11.6$\pm$ 2.2 &  5.3 & 1 \\ 
7992-6101 & 253.405559 & 63.031270 &  0.16$\pm$ 0.11 &  0.13$\pm$ 0.15 &  0.10$\pm$ 0.17 & -1.12$\pm$ 0.24 & -5.6$\pm$ 1.4 &  5.8 & 1 \\ 
7992-9102 & 254.542084 & 62.415648 &  0.18$\pm$ 0.02 &  0.98$\pm$ 0.03 & -0.13$\pm$ 0.05 & -1.46$\pm$ 0.11 & -29.7$\pm$ 8.1 &  7.2 & 1 \\ 
8077-6101 & 41.699909 & 0.421577 &  0.03$\pm$ 0.02 &  0.54$\pm$ 0.07 &  0.23$\pm$ 0.03 & -1.18$\pm$ 0.23 & -3.6$\pm$ 0.8 &  0.0 & 2 \\ 
8078-12701 & 40.880466 & 0.306822 &  0.16$\pm$ 0.02 &  0.28$\pm$ 0.06 &  0.07$\pm$ 0.05 & -1.12$\pm$ 0.27 & -4.3$\pm$ 1.1 &  0.0 & 2 \\ 
8081-6102 & 49.940137 & -0.077189 &  0.12$\pm$ 0.03 &  0.57$\pm$ 0.11 &  0.03$\pm$ 0.05 & -1.40$\pm$ 0.10 & -3.8$\pm$ 0.4 &  2.1 & 2 \\ 
8086-12705 & 57.243039 & -1.144831 &  0.31$\pm$ 0.09 &  0.47$\pm$ 0.05 &  0.41$\pm$ 0.14 & -0.92$\pm$ 0.22 & -6.9$\pm$ 3.5 &  5.8 & 1 \\ 
8131-6104 & 112.416704 & 41.072316 &  0.09$\pm$ 0.15 &  0.38$\pm$ 0.13 & -0.06$\pm$ 0.15 & -0.98$\pm$ 0.23 & -3.1$\pm$ 0.9 &  1.6 & 2 \\ 
8132-6101 & 111.733682 & 41.026691 & -0.02$\pm$ 0.01 &  0.42$\pm$ 0.06 & -0.43$\pm$ 0.06 & -1.51$\pm$ 0.19 & -23.4$\pm$ 6.0 &  3.6 & 1 \\ 
8134-9102 & 116.280207 & 46.072421 &  0.17$\pm$ 0.05 &  0.42$\pm$ 0.10 &  0.37$\pm$ 0.11 & -1.01$\pm$ 0.35 & -4.9$\pm$ 2.4 &  5.8 & 1 \\ 
8135-12701 & 113.472275 & 37.025905 &  0.29$\pm$ 0.04 &  0.42$\pm$ 0.06 &  0.23$\pm$ 0.05 & -1.17$\pm$ 0.27 & -4.9$\pm$ 1.6 &  4.8 & 2 \\ 
8137-3702 & 115.368720 & 44.408794 &  0.13$\pm$ 0.02 &  0.81$\pm$ 0.01 & -0.20$\pm$ 0.02 & -1.74$\pm$ 0.07 & -37.9$\pm$ 6.3 &  5.1 & 1 \\ 
8141-1901 & 117.472421 & 45.248483 & -0.17$\pm$ 0.02 &  0.80$\pm$ 0.03 & -0.05$\pm$ 0.03 & -1.70$\pm$ 0.05 & -19.5$\pm$ 1.8 &  3.8 & 2 \\ 
8141-6102 & 118.648986 & 44.151813 & -0.18$\pm$ 0.04 &  0.51$\pm$ 0.10 &  0.10$\pm$ 0.08 & -1.16$\pm$ 0.27 & -3.6$\pm$ 1.0 &  6.9 & 1 \\ 
8143-6101 & 121.014201 & 40.802613 &  0.26$\pm$ 0.02 &  0.79$\pm$ 0.02 & -0.19$\pm$ 0.05 & -1.40$\pm$ 0.09 & -26.8$\pm$ 6.2 &  5.7 & 1 \\ 
8146-12705 & 118.053214 & 28.772580 &  0.36$\pm$ 0.03 &  0.35$\pm$ 0.06 &  0.24$\pm$ 0.07 & -1.09$\pm$ 0.15 & -5.0$\pm$ 0.6 &  4.1 & 2 \\ 
8146-6104 & 118.307047 & 28.828298 & -0.05$\pm$ 0.12 &  0.34$\pm$ 0.07 & -0.05$\pm$ 0.14 & -0.72$\pm$ 0.06 & -3.2$\pm$ 0.7 &  2.1 & 2 \\ 
8147-6102 & 118.627843 & 25.815986 &  0.21$\pm$ 0.05 &  0.18$\pm$ 0.08 &  0.20$\pm$ 0.04 & -1.07$\pm$ 0.19 & -9.1$\pm$ 2.3 &  6.0 & 1 \\ 
8241-6102 & 126.059633 & 17.331951 & -0.05$\pm$ 0.01 &  0.95$\pm$ 0.02 & -0.14$\pm$ 0.03 & -1.17$\pm$ 0.03 & -66.4$\pm$12.1 & 11.3 & 1 \\ 
8241-9102 & 127.170800 & 17.581400 & -0.13$\pm$ 0.04 &  0.51$\pm$ 0.03 &  0.01$\pm$ 0.04 & -1.16$\pm$ 0.13 & -19.1$\pm$ 6.3 &  5.9 & 1 \\ 
8243-12701 & 128.687741 & 52.715686 &  0.10$\pm$ 0.13 &  0.38$\pm$ 0.21 & -0.27$\pm$ 0.10 & -0.95$\pm$ 0.14 & -3.6$\pm$ 0.7 &  1.2 & 2 \\ 
8243-9102 & 130.821739 & 52.757929 &  0.25$\pm$ 0.08 &  0.26$\pm$ 0.25 & -0.10$\pm$ 0.06 & -1.08$\pm$ 0.31 & -3.9$\pm$ 0.9 &  3.2 & 2 \\ 
8247-6103 & 136.719982 & 41.408252 & -0.01$\pm$ 0.02 &  0.30$\pm$ 0.04 &  0.29$\pm$ 0.06 & -0.62$\pm$ 0.07 & -3.0$\pm$ 0.4 &  4.7 & 2 \\ 
8249-3704 & 137.874763 & 45.468320 & -0.14$\pm$ 0.02 &  0.75$\pm$ 0.07 & -0.26$\pm$ 0.02 & -1.28$\pm$ 0.09 & -18.3$\pm$ 1.1 &  0.0 & 2 \\ 
8250-1902 & 140.218473 & 42.708802 & -0.13$\pm$ 0.10 &  0.31$\pm$ 0.11 & -0.05$\pm$ 0.07 & -1.03$\pm$ 0.19 & -3.1$\pm$ 0.3 &  3.0 & 2 \\ 
8255-6101 & 166.509879 & 43.173473 & -0.03$\pm$ 0.00 &  0.45$\pm$ 0.04 &  0.06$\pm$ 0.05 & -0.99$\pm$ 0.05 & -17.7$\pm$ 3.8 &  8.0 & 1 \\ 
8256-12704 & 166.129408 & 42.624554 & -0.03$\pm$ 0.02 &  0.51$\pm$ 0.12 &  0.08$\pm$ 0.03 & -1.07$\pm$ 0.11 & -17.5$\pm$ 5.1 &  4.7 & 2 \\ 
8257-6103 & 164.642096 & 45.812535 &  0.23$\pm$ 0.05 &  0.67$\pm$ 0.11 &  0.17$\pm$ 0.07 & -0.97$\pm$ 1.00 & -3.2$\pm$ 0.7 &  2.4 & 2 \\ 
8258-6102 & 167.103856 & 43.012962 &  0.13$\pm$ 0.05 &  0.27$\pm$ 0.08 &  0.26$\pm$ 0.07 & -1.08$\pm$ 0.28 & -4.5$\pm$ 2.0 &  4.7 & 2 \\ 
8261-12701 & 182.356280 & 46.549357 &  0.11$\pm$ 0.06 &  0.41$\pm$ 0.08 &  0.22$\pm$ 0.12 & -0.94$\pm$ 1.00 & -3.6$\pm$ 1.8 &  4.9 & 2 \\ 
8262-9101 & 184.543787 & 44.400460 &  0.33$\pm$ 0.08 &  0.66$\pm$ 0.08 &  0.29$\pm$ 0.07 & -1.32$\pm$ 0.26 & -3.4$\pm$ 0.7 &  3.5 & 2 \\ 
8274-12704 & 166.129408 & 42.624554 & -0.02$\pm$ 0.03 &  0.43$\pm$ 0.14 &  0.13$\pm$ 0.06 & -0.86$\pm$ 0.17 & -12.5$\pm$ 3.2 &  4.4 & 2 \\ 
8313-6101 & 240.658054 & 41.293427 &  0.31$\pm$ 0.03 &  0.95$\pm$ 0.11 & -0.14$\pm$ 0.06 & -0.90$\pm$ 0.14 & -5.4$\pm$ 0.6 &  3.2 & 2 \\ 
8315-6103 & 235.057231 & 39.904137 &  0.17$\pm$ 0.05 &  0.23$\pm$ 0.09 &  0.08$\pm$ 0.05 & -0.87$\pm$ 0.12 & -3.1$\pm$ 1.5 &  5.2 & 1 \\ 
8317-12704 & 193.703990 & 44.155566 &  0.21$\pm$ 0.18 &  0.55$\pm$ 0.05 &  0.25$\pm$ 0.23 & -1.05$\pm$ 0.30 & -10.1$\pm$ 5.0 &  5.4 & 1 \\ 
8318-1901 & 196.755055 & 46.309431 &  0.00$\pm$ 0.08 &  0.68$\pm$ 0.09 & -0.01$\pm$ 0.08 & -0.90$\pm$ 0.19 & -4.2$\pm$ 0.6 &  0.0 & 2 \\ 
8318-3703 & 198.491577 & 45.704463 & -0.10$\pm$ 0.03 &  0.72$\pm$ 0.10 & -0.22$\pm$ 0.04 & -1.46$\pm$ 0.10 & -6.1$\pm$ 0.4 &  0.0 & 2 \\ 
8318-3704 & 197.891834 & 44.933078 & -0.04$\pm$ 0.03 &  0.74$\pm$ 0.05 &  0.02$\pm$ 0.07 & -0.89$\pm$ 0.21 & -11.7$\pm$ 2.9 &  6.6 & 1 \\ 
8320-3704 & 206.612456 & 22.076742 & -0.09$\pm$ 0.02 &  0.97$\pm$ 0.03 & -0.11$\pm$ 0.02 & -1.47$\pm$ 0.09 & -26.1$\pm$ 3.5 &  0.0 & 2 \\ 
\hline
\end{tabular}
\end{table*}
\clearpage
\addtocounter{table}{-1}
\begin{table*}
\caption{List of AGN candidates ({\it continue})}
%\label{tab:agns}
\begin{tabular}{lllrrrrrrc}
\hline
MaNGAID & RA & DEC & [\ION{N}{ii}]/H$\alpha$ & [\ION{O}{iii}]/H$\beta$ & [\ION{S}{ii}]/H$\alpha$ & [\ION{O}{i}]/H$\alpha$ & EW(H$\alpha$) & S/N & AGN\\ 
        & (deg) & (deg)  & log10  & log10 & log10 & log10 & \AA & H${\alpha} B$ & type \\ 
\hline
8325-6101 & 210.054964 & 46.432207 &  0.01$\pm$ 0.09 &  0.18$\pm$ 0.10 &  0.29$\pm$ 0.08 & -0.65$\pm$ 0.16 & -3.9$\pm$ 2.3 &  3.7 & 2 \\ 
8329-6102 & 211.904865 & 44.482269 &  0.02$\pm$ 0.13 &  0.39$\pm$ 0.12 & -0.04$\pm$ 0.05 & -0.93$\pm$ 0.08 & -10.4$\pm$ 5.3 &  0.0 & 2 \\ 
8332-12702 & 207.928807 & 43.166603 &  0.06$\pm$ 0.07 &  0.27$\pm$ 0.11 & -0.05$\pm$ 0.11 & -1.10$\pm$ 0.19 & -3.3$\pm$ 0.3 &  0.0 & 2 \\ 
8341-12704 & 189.213253 & 45.651170 & -0.04$\pm$ 0.04 &  0.66$\pm$ 0.02 & -0.18$\pm$ 0.06 & -1.27$\pm$ 0.09 & -45.3$\pm$12.4 & 18.1 & 1 \\ 
8440-12704 & 136.142338 & 41.397827 &  0.04$\pm$ 0.02 &  0.15$\pm$ 0.09 &  0.07$\pm$ 0.04 & -0.95$\pm$ 0.19 & -3.0$\pm$ 0.2 &  0.0 & 2 \\ 
8450-6104 & 172.607538 & 22.216530 &  0.24$\pm$ 0.07 &  0.22$\pm$ 0.03 &  0.25$\pm$ 0.09 & -1.09$\pm$ 0.10 & -12.5$\pm$ 3.9 &  5.6 & 1 \\ 
8451-12701 & 166.129408 & 42.624554 & -0.01$\pm$ 0.04 &  0.39$\pm$ 0.13 &  0.09$\pm$ 0.07 & -0.70$\pm$ 0.11 & -12.0$\pm$ 1.7 &  4.8 & 2 \\ 
8452-12703 & 156.805685 & 48.244791 &  0.15$\pm$ 0.04 &  0.27$\pm$ 0.08 & -0.04$\pm$ 0.04 & -0.63$\pm$ 0.25 & -4.4$\pm$ 0.2 &  3.9 & 2 \\ 
8452-1901 & 155.885556 & 46.057755 & -0.19$\pm$ 0.02 &  0.87$\pm$ 0.06 & -0.20$\pm$ 0.03 & -1.13$\pm$ 0.10 & -8.8$\pm$ 0.2 &  0.0 & 2 \\ 
8454-6102 & 153.535479 & 44.175746 &  0.17$\pm$ 0.06 &  0.42$\pm$ 0.15 &  0.17$\pm$ 0.11 & -0.97$\pm$ 0.22 & -13.0$\pm$ 4.6 &  5.0 & 1 \\ 
8454-9102 & 154.594514 & 45.954645 &  0.15$\pm$ 0.04 &  0.25$\pm$ 0.07 & -0.02$\pm$ 0.06 & -1.22$\pm$ 0.08 & -3.5$\pm$ 0.4 &  0.0 & 2 \\ 
8455-12703 & 157.723268 & 41.221095 &  0.12$\pm$ 0.13 &  0.04$\pm$ 0.27 &  0.17$\pm$ 0.21 & -1.03$\pm$ 0.36 & -3.4$\pm$ 2.5 &  6.4 & 1 \\ 
8456-6101 & 151.220914 & 44.636123 &  0.09$\pm$ 0.02 &  0.66$\pm$ 0.08 & -0.05$\pm$ 0.03 & -1.12$\pm$ 0.05 & -3.9$\pm$ 0.3 &  0.0 & 2 \\ 
8482-12703 & 245.503111 & 49.520790 &  0.32$\pm$ 0.03 &  0.46$\pm$ 0.04 &  0.06$\pm$ 0.11 & -1.20$\pm$ 0.01 & -3.3$\pm$ 0.5 &  2.5 & 2 \\ 
8482-12704 & 243.581821 & 50.465611 &  0.06$\pm$ 0.06 &  0.48$\pm$ 0.04 &  0.17$\pm$ 0.07 & -1.14$\pm$ 0.17 & -13.5$\pm$ 4.4 &  5.4 & 1 \\ 
8482-3704 & 245.412402 & 49.448843 &  0.10$\pm$ 0.07 &  0.18$\pm$ 0.09 & -0.07$\pm$ 0.05 & -0.76$\pm$ 0.22 & -3.9$\pm$ 0.7 &  2.1 & 2 \\ 
8483-12703 & 245.248314 & 49.001777 &  0.08$\pm$ 0.02 &  0.54$\pm$ 0.08 &  0.16$\pm$ 0.03 & -1.14$\pm$ 0.07 & -11.7$\pm$ 2.7 &  4.0 & 2 \\ 
8484-6101 & 248.055742 & 44.403296 & -0.14$\pm$ 0.04 &  0.36$\pm$ 0.12 & -0.30$\pm$ 0.07 & -1.01$\pm$ 0.29 & -3.5$\pm$ 0.3 &  0.0 & 2 \\ 
8549-12701 & 240.470871 & 45.351940 & -0.07$\pm$ 0.02 &  0.63$\pm$ 0.01 &  0.03$\pm$ 0.06 & -1.52$\pm$ 0.15 & -33.3$\pm$ 6.9 &  8.9 & 1 \\ 
8549-12702 & 241.271447 & 45.442992 &  0.04$\pm$ 0.02 &  0.47$\pm$ 0.08 & -0.24$\pm$ 0.17 & -0.95$\pm$ 0.11 & -6.4$\pm$ 3.0 & 10.3 & 1 \\ 
8549-9101 & 242.276472 & 46.671205 &  0.19$\pm$ 0.08 &  0.43$\pm$ 0.13 &  0.30$\pm$ 0.09 & -0.96$\pm$ 0.18 & -3.2$\pm$ 1.0 &  3.2 & 2 \\ 
8550-12702 & 247.620046 & 39.626044 &  0.32$\pm$ 0.05 &  0.31$\pm$ 0.09 &  0.08$\pm$ 0.13 & -0.61$\pm$ 1.00 & -3.4$\pm$ 1.0 &  0.0 & 2 \\ 
8550-3704 & 248.426386 & 39.185120 &  0.28$\pm$ 0.05 &  0.67$\pm$ 0.02 &  0.21$\pm$ 0.07 & -1.00$\pm$ 0.13 & -14.1$\pm$ 5.1 &  5.3 & 1 \\ 
8550-6103 & 247.638691 & 39.830726 &  0.06$\pm$ 0.04 &  0.25$\pm$ 0.09 &  0.06$\pm$ 0.04 & -1.15$\pm$ 0.17 & -3.9$\pm$ 0.3 &  0.0 & 2 \\ 
8588-12704 & 249.557306 & 40.146821 &  0.05$\pm$ 0.01 &  0.83$\pm$ 0.04 & -0.10$\pm$ 0.03 & -1.62$\pm$ 0.23 & -13.4$\pm$ 1.4 &  5.8 & 1 \\ 
8588-3701 & 248.140561 & 39.131021 &  0.17$\pm$ 0.02 &  0.59$\pm$ 0.15 &  0.00$\pm$ 0.05 & -1.14$\pm$ 0.23 & -3.4$\pm$ 0.3 &  0.6 & 2 \\ 
8597-12703 & 225.388974 & 49.112429 &  0.03$\pm$ 0.01 &  0.27$\pm$ 0.09 & -0.13$\pm$ 0.05 & -1.03$\pm$ 0.22 & -3.4$\pm$ 0.3 &  0.0 & 2 \\ 
8602-12701 & 247.048171 & 39.821898 &  0.27$\pm$ 0.03 &  0.34$\pm$ 0.02 &  0.08$\pm$ 0.05 & -1.05$\pm$ 0.07 & -18.0$\pm$ 3.8 &  7.1 & 1 \\ 
8606-12701 & 255.029870 & 37.839502 &  0.37$\pm$ 0.09 &  0.60$\pm$ 0.03 &  0.27$\pm$ 0.14 & -0.88$\pm$ 0.14 & -6.7$\pm$ 2.0 &  6.2 & 1 \\ 
8612-12704 & 254.564575 & 39.391464 &  0.01$\pm$ 0.01 &  1.02$\pm$ 0.01 & -0.10$\pm$ 0.02 & -1.34$\pm$ 0.04 & -56.1$\pm$ 8.7 &  8.8 & 1 \\ 
8612-6102 & 252.927152 & 39.235833 &  0.03$\pm$ 0.04 &  0.15$\pm$ 0.16 &  0.21$\pm$ 0.07 & -1.04$\pm$ 0.20 & -3.3$\pm$ 0.6 &  3.7 & 2 \\ 
8623-12704 & 311.829452 & 0.320795 &  0.07$\pm$ 0.02 &  0.34$\pm$ 0.08 &  0.17$\pm$ 0.04 & -1.31$\pm$ 0.30 & -3.7$\pm$ 0.5 &  3.5 & 2 \\ 
8655-6103 & 355.825111 & 0.442475 & -0.23$\pm$ 0.02 &  0.68$\pm$ 0.05 & -0.42$\pm$ 0.09 & -1.35$\pm$ 0.11 & -7.8$\pm$ 1.3 &  0.0 & 2 \\ 
8712-12704 & 121.054830 & 55.397665 &  0.29$\pm$ 0.05 &  0.36$\pm$ 0.12 &  0.16$\pm$ 0.06 & -1.32$\pm$ 0.24 & -3.5$\pm$ 0.7 &  1.9 & 2 \\ 
8714-3704 & 118.184153 & 45.949276 &  0.34$\pm$ 0.10 &  0.65$\pm$ 0.06 &  0.04$\pm$ 0.15 & -0.70$\pm$ 0.20 & -6.6$\pm$ 4.0 &  5.3 & 1 \\ 
8715-3701 & 119.120001 & 50.287866 &  0.33$\pm$ 0.12 &  0.30$\pm$ 0.07 &  0.23$\pm$ 0.22 & -0.60$\pm$ 0.34 & -4.1$\pm$ 2.7 &  5.8 & 1 \\ 
8715-3702 & 119.920672 & 50.839973 &  0.01$\pm$ 0.01 &  1.07$\pm$ 0.01 & -0.28$\pm$ 0.03 & -1.30$\pm$ 0.02 & -176.0$\pm$13.0 &  9.7 & 1 \\ 
8717-1902 & 118.091110 & 34.326570 & -0.15$\pm$ 0.06 &  0.30$\pm$ 0.07 &  0.14$\pm$ 0.16 & -1.10$\pm$ 0.28 & -8.7$\pm$ 4.6 &  6.0 & 1 \\ 
8718-12701 & 119.182152 & 44.856709 & -0.06$\pm$ 0.03 &  0.72$\pm$ 0.05 & -0.14$\pm$ 0.02 & -1.23$\pm$ 0.06 & -21.3$\pm$ 2.5 &  3.0 & 2 \\ 
8718-12702 & 120.700706 & 45.034554 &  0.09$\pm$ 0.04 &  0.87$\pm$ 0.06 & -0.00$\pm$ 0.02 & -1.35$\pm$ 0.13 & -20.4$\pm$ 1.4 &  5.7 & 1 \\ 
8720-1901 & 121.147928 & 50.708556 & -0.33$\pm$ 0.03 &  0.98$\pm$ 0.05 & -0.33$\pm$ 0.10 & -1.37$\pm$ 0.12 & -7.9$\pm$ 0.8 &  0.0 & 2 \\ 
8725-9102 & 127.178094 & 45.742555 & -0.29$\pm$ 0.02 &  0.72$\pm$ 0.04 & -0.39$\pm$ 0.03 & -1.51$\pm$ 0.12 & -34.0$\pm$ 2.1 &  8.8 & 1 \\ 
8939-9101 & 125.227739 & 23.729200 &  0.11$\pm$ 0.03 &  0.41$\pm$ 0.18 &  0.19$\pm$ 0.10 & -1.00$\pm$ 0.16 & -5.3$\pm$ 2.2 &  4.9 & 2 \\ 
8943-9101 & 156.403128 & 37.222305 & -0.25$\pm$ 0.03 &  0.46$\pm$ 0.08 & -0.08$\pm$ 0.03 & -1.26$\pm$ 0.08 & -11.5$\pm$ 1.7 &  3.0 & 2 \\ 
8946-3703 & 170.588145 & 46.430504 &  0.07$\pm$ 0.05 &  0.22$\pm$ 0.06 &  0.15$\pm$ 0.06 & -1.23$\pm$ 0.17 & -4.0$\pm$ 1.0 &  5.8 & 1 \\ 
\hline
\end{tabular}
\end{table*}
\clearpage
\addtocounter{table}{-1}
\begin{table*}
\caption{List of AGN candidates ({\it continue})}
%\label{tab:agns}
\begin{tabular}{lllrrrrrrc}
\hline
MaNGAID & RA & DEC & [\ION{N}{ii}]/H$\alpha$ & [\ION{O}{iii}]/H$\beta$ & [\ION{S}{ii}]/H$\alpha$ & [\ION{O}{i}]/H$\alpha$ & EW(H$\alpha$) & S/N & AGN\\ 
        & (deg) & (deg)  & log10  & log10 & log10 & log10 & \AA & H${\alpha} B$ & type \\ 
\hline
8947-12703 & 172.886686 & 49.857504 &  0.08$\pm$ 0.06 &  0.56$\pm$ 0.13 & -0.11$\pm$ 0.07 & -1.17$\pm$ 0.21 & -4.1$\pm$ 0.2 &  0.3 & 2 \\ 
8947-12704 & 171.102654 & 51.234941 &  0.07$\pm$ 0.06 &  0.22$\pm$ 0.08 &  0.01$\pm$ 0.06 & -1.02$\pm$ 0.15 & -3.2$\pm$ 0.3 &  0.0 & 2 \\ 
8947-3701 & 168.947800 & 50.401634 & -0.44$\pm$ 0.01 &  0.89$\pm$ 0.01 & -0.29$\pm$ 0.01 & -1.63$\pm$ 0.08 & -45.5$\pm$ 4.5 &  0.0 & 2 \\ 
8978-9101 & 247.907996 & 41.493643 &  0.19$\pm$ 0.02 &  0.33$\pm$ 0.03 &  0.19$\pm$ 0.02 & -1.04$\pm$ 0.07 & -10.1$\pm$ 2.0 &  4.2 & 2 \\ 
8979-6102 & 241.823389 & 41.403604 &  0.06$\pm$ 0.02 &  0.58$\pm$ 0.05 &  0.10$\pm$ 0.03 & -1.33$\pm$ 0.20 & -8.8$\pm$ 2.0 &  5.0 & 2 \\ 
9002-1901 & 223.612368 & 30.908509 & -0.03$\pm$ 0.04 &  0.60$\pm$ 0.06 &  0.10$\pm$ 0.02 & -1.41$\pm$ 0.18 & -3.3$\pm$ 0.4 &  0.0 & 2 \\ 
9026-9101 & 249.318419 & 44.418230 &  0.20$\pm$ 0.01 &  0.81$\pm$ 0.05 & -0.06$\pm$ 0.04 & -1.54$\pm$ 0.08 & -11.4$\pm$ 1.6 &  4.0 & 2 \\ 
9029-12704 & 247.216953 & 42.812011 &  0.10$\pm$ 0.04 &  0.37$\pm$ 0.18 &  0.02$\pm$ 0.05 & -1.07$\pm$ 0.39 & -5.3$\pm$ 1.2 &  0.0 & 2 \\ 
9029-9101 & 247.476832 & 41.604523 &  0.04$\pm$ 0.02 &  0.32$\pm$ 0.13 &  0.16$\pm$ 0.05 & -0.61$\pm$ 0.01 & -3.3$\pm$ 1.2 &  0.0 & 2 \\ 
\hline
\end{tabular}
\end{table*}
%\clearpage

\subsection{AGN candidates catalog}
\label{app:agns}

Table \ref{tab:agns} show the list of 98 AGN cantidates selected along this study including the main properties used to selected them as AGNs: (1) the MaNGAID-flag, defined as the [plate]-[ifudsgn] (Sec. \ref{app:Pipe3D}, (2) the right accession of the target, (3) the declination of the target, (3) the [\ION{N}{ii}]/H$\alpha$ line ratio in the central regions in logarithm scales and its error, (4) the [\ION{O}{iii}]/H$\beta$ line ratio in the central regions in logarithm scales and its error, (5) the [\ION{S}{ii}]/H$\alpha$ line ratio in the central regions in logarithm scales and its error, (6) the [\ION{O}{i}]/H$\alpha$ line ratio in the central regions in logarithm scales and its error, (7) the EW(H$\alpha$) in the central regions and its error, (8) the signal-to-noise of the peak intensity of the fitted broad component to the H$\alpha$ emission line described in \ref{sec:type_I}, and finally, (9) the AGN classification (type-I or type-II).

%\url{https://internal.sdss.org/dr14/datamodel/files/MANGA_PIPE3D/MANGADRP_VER/PIPE3D_VER/PLATE/manga.Pipe3D.cube.html}
%\url{https://sas.sdss.org/resources/dr14/manga/spectro/pipe3d/v2_1_2/2.1.2/list/}

% All you wanted to know of MPL-5.

\bibliography{CALIFAI,references-VAR,Alenka,ref_raom}

\begin{thebibliography}
\expandafter\ifx\csname natexlab\endcsname\relax\def\natexlab#1{#1}\fi
\expandafter\ifx\csname href\endcsname\relax
  \def\href#1#2{}\fi
\expandafter\ifx\csname urllinklabel\endcsname\relax
  \def\urllinklabel{[LINK]}\fi
\expandafter\ifx\csname adsurllinklabel\endcsname\relax
  \def\adsurllinklabel{[ADS]}\fi

\bibitem[{{Abolfathi} {et~al.}(2017){Abolfathi}, {Aguado}, {Aguilar}, {Allende
  Prieto}, {Almeida}, {Tasnim Ananna}, {Anders}, {Anderson}, {Andrews},
  {Anguiano}, \& et~al.}]{SDSSDR14}
{Abolfathi}, B., {Aguado}, D.~S., {Aguilar}, G., {Allende Prieto}, C.,
  {Almeida}, A., {Tasnim Ananna}, T., {Anders}, F., {Anderson}, S.~F.,
  {Andrews}, B.~H., {Anguiano}, B., \& et~al. 2017, ArXiv e-prints


\bibitem[{{Baldry} {et~al.}(2004){Baldry}, {Glazebrook}, {Brinkmann},
  {Ivezi{\'c}}, {Lupton}, {Nichol}, \& {Szalay}}]{Baldry+2004}
{Baldry}, I.~K., {Glazebrook}, K., {Brinkmann}, J., {Ivezi{\'c}}, {\v Z}.,
  {Lupton}, R.~H., {Nichol}, R.~C., \& {Szalay}, A.~S. 2004, \apj, 600, 681


\bibitem[{{Baldwin} {et~al.}(1981){Baldwin}, {Phillips}, \&
  {Terlevich}}]{baldwin81}
{Baldwin}, J.~A., {Phillips}, M.~M., \& {Terlevich}, R. 1981, \pasp, 93, 5


\bibitem[{{Barrera-Ballesteros} {et~al.}(2017){Barrera-Ballesteros},
  {S{\'a}nchez}, {Heckman}, {Blanc}, \& {The MaNGA Team}}]{bb17}
{Barrera-Ballesteros}, J.~K., {S{\'a}nchez}, S.~F., {Heckman}, T., {Blanc},
  G.~A., \& {The MaNGA Team}. 2017, \apj, 844, 80


\bibitem[{{Belfiore} {et~al.}(2017){Belfiore}, {Maiolino}, {Maraston},
  {Emsellem}, {Bershady}, {Masters}, {Bizyaev}, {Boquien}, {Brownstein},
  {Bundy}, {Diamond-Stanic}, {Drory}, {Heckman}, {Law}, {Malanushenko},
  {Oravetz}, {Pan}, {Roman-Lopes}, {Thomas}, {Weijmans}, {Westfall}, \&
  {Yan}}]{Belfiore17a}
{Belfiore}, F., {Maiolino}, R., {Maraston}, C., {Emsellem}, E., {Bershady},
  M.~A., {Masters}, K.~L., {Bizyaev}, D., {Boquien}, M., {Brownstein}, J.~R.,
  {Bundy}, K., {Diamond-Stanic}, A.~M., {Drory}, N., {Heckman}, T.~M., {Law},
  D.~R., {Malanushenko}, O., {Oravetz}, A., {Pan}, K., {Roman-Lopes}, A.,
  {Thomas}, D., {Weijmans}, A.-M., {Westfall}, K.~B., \& {Yan}, R. 2017,
  \mnras, 466, 2570


\bibitem[{{Bell} {et~al.}(2004){Bell}, {Wolf}, {Meisenheimer}, {Rix}, {Borch},
  {Dye}, {Kleinheinrich}, {Wisotzki}, \& {McIntosh}}]{Bell+2004}
{Bell}, E.~F., {Wolf}, C., {Meisenheimer}, K., {Rix}, H.-W., {Borch}, A.,
  {Dye}, S., {Kleinheinrich}, M., {Wisotzki}, L., \& {McIntosh}, D.~H. 2004,
  \apj, 608, 752


\bibitem[{{Benn} {et~al.}(1998){Benn}, {Vigotti}, {Carballo},
  {Gonzalez-Serrano}, \& {S{\'a}nchez}}]{benn98}
{Benn}, C.~R., {Vigotti}, M., {Carballo}, R., {Gonzalez-Serrano}, J.~I., \&
  {S{\'a}nchez}, S.~F. 1998, \mnras, 295, 451


\bibitem[{{Binette} {et~al.}(2009){Binette}, {Flores-Fajardo}, {Raga},
  {Drissen}, \& {Morisset}}]{binn09}
{Binette}, L., {Flores-Fajardo}, N., {Raga}, A.~C., {Drissen}, L., \&
  {Morisset}, C. 2009, \apj, 695, 552


\bibitem[{{Binette} {et~al.}(1994){Binette}, {Magris}, {Stasi{\'n}ska}, \&
  {Bruzual}}]{binn94}
{Binette}, L., {Magris}, C.~G., {Stasi{\'n}ska}, G., \& {Bruzual}, A.~G. 1994,
  \aap, 292, 13


\bibitem[{{Bitsakis} {et~al.}(2016){Bitsakis}, {Dultzin}, {Ciesla},
  {D{\'{\i}}az-Santos}, {Appleton}, {Charmandaris}, {Krongold}, {Guillard},
  {Alatalo}, {Zezas}, {Gonz{\'a}lez}, \& {Lanz}}]{Bitsakis+2016}
{Bitsakis}, T., {Dultzin}, D., {Ciesla}, L., {D{\'{\i}}az-Santos}, T.,
  {Appleton}, P.~N., {Charmandaris}, V., {Krongold}, Y., {Guillard}, P.,
  {Alatalo}, K., {Zezas}, A., {Gonz{\'a}lez}, J., \& {Lanz}, L. 2016, \mnras,
  459, 957


\bibitem[{{Blanton} {et~al.}(2017){Blanton}, {Bershady}, {Abolfathi},
  {Albareti}, {Allende Prieto}, {Almeida}, {Alonso-Garc{\'{\i}}a}, {Anders},
  {Anderson}, {Andrews}, \& et~al.}]{blanton17}
{Blanton}, M.~R., {Bershady}, M.~A., {Abolfathi}, B., {Albareti}, F.~D.,
  {Allende Prieto}, C., {Almeida}, A., {Alonso-Garc{\'{\i}}a}, J., {Anders},
  F., {Anderson}, S.~F., {Andrews}, B., \& et~al. 2017, \aj, 154, 28


\bibitem[{{Blanton} {et~al.}(2003){Blanton}, {Hogg}, {Bahcall}, {Brinkmann},
  {Britton}, {Connolly}, {Csabai}, {Fukugita}, {Loveday}, {Meiksin}, {Munn},
  {Nichol}, {Okamura}, {Quinn}, {Schneider}, {Shimasaku}, {Strauss}, {Tegmark},
  {Vogeley}, \& {Weinberg}}]{blanton03}
{Blanton}, M.~R., {Hogg}, D.~W., {Bahcall}, N.~A., {Brinkmann}, J., {Britton},
  M., {Connolly}, A.~J., {Csabai}, I., {Fukugita}, M., {Loveday}, J.,
  {Meiksin}, A., {Munn}, J.~A., {Nichol}, R.~C., {Okamura}, S., {Quinn}, T.,
  {Schneider}, D.~P., {Shimasaku}, K., {Strauss}, M.~A., {Tegmark}, M.,
  {Vogeley}, M.~S., \& {Weinberg}, D.~H. 2003, \apj, 592, 819


\bibitem[{{Blanton} \& {Moustakas}(2009)}]{Blanton+2009}
{Blanton}, M.~R. \& {Moustakas}, J. 2009, \araa, 47, 159


\bibitem[{{Blanton} {et~al.}(2005){Blanton}, {Schlegel}, {Strauss},
  {Brinkmann}, {Finkbeiner}, {Fukugita}, {Gunn}, {Hogg}, {Ivezi{\'c}}, {Knapp},
  {Lupton}, {Munn}, {Schneider}, {Tegmark}, \& {Zehavi}}]{blanton05}
{Blanton}, M.~R., {Schlegel}, D.~J., {Strauss}, M.~A., {Brinkmann}, J.,
  {Finkbeiner}, D., {Fukugita}, M., {Gunn}, J.~E., {Hogg}, D.~W., {Ivezi{\'c}},
  {\v Z}., {Knapp}, G.~R., {Lupton}, R.~H., {Munn}, J.~A., {Schneider}, D.~P.,
  {Tegmark}, M., \& {Zehavi}, I. 2005, \aj, 129, 2562


\bibitem[{{Bohlin} {et~al.}(1978){Bohlin}, {Savage}, \& {Drake}}]{bohlin1978}
{Bohlin}, R.~C., {Savage}, B.~D., \& {Drake}, J.~F. 1978, \apj, 224, 132


\bibitem[{{B{\"o}hm} {et~al.}(2013){B{\"o}hm}, {Wisotzki}, {Bell}, {Jahnke},
  {Wolf}, {Bacon}, {Barden}, {Gray}, {Hoeppe}, {Jogee}, {McIntosh}, {Peng},
  {Robaina}, {Balogh}, {Barazza}, {Caldwell}, {Heymans}, {H{\"a}u{\ss}ler},
  {van Kampen}, {Lane}, {Meisenheimer}, {S{\'a}nchez}, {Taylor}, \&
  {Zheng}}]{bohm13}
{B{\"o}hm}, A., {Wisotzki}, L., {Bell}, E.~F., {Jahnke}, K., {Wolf}, C.,
  {Bacon}, D., {Barden}, M., {Gray}, M.~E., {Hoeppe}, G., {Jogee}, S.,
  {McIntosh}, D.~H., {Peng}, C.~Y., {Robaina}, A.~R., {Balogh}, M., {Barazza},
  F.~D., {Caldwell}, J.~A.~R., {Heymans}, C., {H{\"a}u{\ss}ler}, B., {van
  Kampen}, E., {Lane}, K., {Meisenheimer}, K., {S{\'a}nchez}, S.~F., {Taylor},
  A.~N., \& {Zheng}, X. 2013, \aap, 549, A46


\bibitem[{{Bolatto} {et~al.}(2013){Bolatto}, {Wolfire}, \& {Leroy}}]{bolatto13}
{Bolatto}, A.~D., {Wolfire}, M., \& {Leroy}, A.~K. 2013, \araa, 51, 207


\bibitem[{{Bolatto} {et~al.}(2017){Bolatto}, {Wong}, {Utomo}, {Blitz}, {Vogel},
  {S{\'a}nchez}, {Barrera-Ballesteros}, {Cao}, {Colombo}, {Dannerbauer},
  {Garc{\'{\i}}a-Benito}, {Herrera-Camus}, {Husemann}, {Kalinova}, {Leroy},
  {Leung}, {Levy}, {Mast}, {Ostriker}, {Rosolowsky}, {Sandstrom}, {Teuben},
  {van de Ven}, \& {Walter}}]{bolatto17}
{Bolatto}, A.~D., {Wong}, T., {Utomo}, D., {Blitz}, L., {Vogel}, S.~N.,
  {S{\'a}nchez}, S.~F., {Barrera-Ballesteros}, J., {Cao}, Y., {Colombo}, D.,
  {Dannerbauer}, H., {Garc{\'{\i}}a-Benito}, R., {Herrera-Camus}, R.,
  {Husemann}, B., {Kalinova}, V., {Leroy}, A.~K., {Leung}, G., {Levy}, R.~C.,
  {Mast}, D., {Ostriker}, E., {Rosolowsky}, E., {Sandstrom}, K.~M., {Teuben},
  P., {van de Ven}, G., \& {Walter}, F. 2017, \apj, 846, 159


\bibitem[{{Boquien} {et~al.}(2013){Boquien}, {Boselli}, {Buat}, {Baes},
  {Bendo}, {Boissier}, {Ciesla}, {Cooray}, {Cortese}, {Eales}, {Koda},
  {Lebouteiller}, {De Looze}, {Smith}, {Spinoglio}, \& {Wilson}}]{boquien13}
{Boquien}, M., {Boselli}, A., {Buat}, V., {Baes}, M., {Bendo}, G., {Boissier},
  S., {Ciesla}, L., {Cooray}, A., {Cortese}, L., {Eales}, S., {Koda}, J.,
  {Lebouteiller}, V., {De Looze}, I., {Smith}, M.~W.~L., {Spinoglio}, L., \&
  {Wilson}, C.~D. 2013, \aap, 554, A14


\bibitem[{{Bower} {et~al.}(2006){Bower}, {Benson}, {Malbon}, {Helly}, {Frenk},
  {Baugh}, {Cole}, \& {Lacey}}]{Bower+2006}
{Bower}, R.~G., {Benson}, A.~J., {Malbon}, R., {Helly}, J.~C., {Frenk}, C.~S.,
  {Baugh}, C.~M., {Cole}, S., \& {Lacey}, C.~G. 2006, \mnras, 370, 645


\bibitem[{{Brinchmann} {et~al.}(2013){Brinchmann}, {Charlot}, {Kauffmann},
  {Heckman}, {White}, \& {Tremonti}}]{brin14}
{Brinchmann}, J., {Charlot}, S., {Kauffmann}, G., {Heckman}, T., {White},
  S.~D.~M., \& {Tremonti}, C. 2013, \mnras, 432, 2112


\bibitem[{{Brinchmann} {et~al.}(2004){Brinchmann}, {Charlot}, {White},
  {Tremonti}, {Kauffmann}, {Heckman}, \& {Brinkmann}}]{Brinchmann04}
{Brinchmann}, J., {Charlot}, S., {White}, S.~D.~M., {Tremonti}, C.,
  {Kauffmann}, G., {Heckman}, T., \& {Brinkmann}, J. 2004, \mnras, 351, 1151


\bibitem[{{Bundy} {et~al.}(2015){Bundy}, {Bershady}, {Law}, {Yan}, {Drory},
  {MacDonald}, {Wake}, {Cherinka}, {S{\'a}nchez-Gallego}, {Weijmans}, {Thomas},
  {Tremonti}, {Masters}, {Coccato}, {Diamond-Stanic}, {Arag{\'o}n-Salamanca},
  {Avila-Reese}, {Badenes}, {Falc{\'o}n-Barroso}, {Belfiore}, {Bizyaev},
  {Blanc}, {Bland-Hawthorn}, {Blanton}, {Brownstein}, {Byler}, {Cappellari},
  {Conroy}, {Dutton}, {Emsellem}, {Etherington}, {Frinchaboy}, {Fu}, {Gunn},
  {Harding}, {Johnston}, {Kauffmann}, {Kinemuchi}, {Klaene}, {Knapen},
  {Leauthaud}, {Li}, {Lin}, {Maiolino}, {Malanushenko}, {Malanushenko}, {Mao},
  {Maraston}, {McDermid}, {Merrifield}, {Nichol}, {Oravetz}, {Pan}, {Parejko},
  {Sanchez}, {Schlegel}, {Simmons}, {Steele}, {Steinmetz}, {Thanjavur},
  {Thompson}, {Tinker}, {van den Bosch}, {Westfall}, {Wilkinson}, {Wright},
  {Xiao}, \& {Zhang}}]{manga}
{Bundy}, K., {Bershady}, M.~A., {Law}, D.~R., {Yan}, R., {Drory}, N.,
  {MacDonald}, N., {Wake}, D.~A., {Cherinka}, B., {S{\'a}nchez-Gallego}, J.~R.,
  {Weijmans}, A.-M., {Thomas}, D., {Tremonti}, C., {Masters}, K., {Coccato},
  L., {Diamond-Stanic}, A.~M., {Arag{\'o}n-Salamanca}, A., {Avila-Reese}, V.,
  {Badenes}, C., {Falc{\'o}n-Barroso}, J., {Belfiore}, F., {Bizyaev}, D.,
  {Blanc}, G.~A., {Bland-Hawthorn}, J., {Blanton}, M.~R., {Brownstein}, J.~R.,
  {Byler}, N., {Cappellari}, M., {Conroy}, C., {Dutton}, A.~A., {Emsellem}, E.,
  {Etherington}, J., {Frinchaboy}, P.~M., {Fu}, H., {Gunn}, J.~E., {Harding},
  P., {Johnston}, E.~J., {Kauffmann}, G., {Kinemuchi}, K., {Klaene}, M.~A.,
  {Knapen}, J.~H., {Leauthaud}, A., {Li}, C., {Lin}, L., {Maiolino}, R.,
  {Malanushenko}, V., {Malanushenko}, E., {Mao}, S., {Maraston}, C.,
  {McDermid}, R.~M., {Merrifield}, M.~R., {Nichol}, R.~C., {Oravetz}, D.,
  {Pan}, K., {Parejko}, J.~K., {Sanchez}, S.~F., {Schlegel}, D., {Simmons}, A.,
  {Steele}, O., {Steinmetz}, M., {Thanjavur}, K., {Thompson}, B.~A., {Tinker},
  J.~L., {van den Bosch}, R.~C.~E., {Westfall}, K.~B., {Wilkinson}, D.,
  {Wright}, S., {Xiao}, T., \& {Zhang}, K. 2015, \apj, 798, 7


\bibitem[{{Bundy} {et~al.}(2006){Bundy}, {Ellis}, {Conselice}, {Taylor},
  {Cooper}, {Willmer}, {Weiner}, {Coil}, {Noeske}, \& {Eisenhardt}}]{bundy06}
{Bundy}, K., {Ellis}, R.~S., {Conselice}, C.~J., {Taylor}, J.~E., {Cooper},
  M.~C., {Willmer}, C.~N.~A., {Weiner}, B.~J., {Coil}, A.~L., {Noeske}, K.~G.,
  \& {Eisenhardt}, P.~R.~M. 2006, \apj, 651, 120


\bibitem[{{Bundy} {et~al.}(2009){Bundy}, {Fukugita}, {Ellis}, {Targett},
  {Belli}, \& {Kodama}}]{bundy09}
{Bundy}, K., {Fukugita}, M., {Ellis}, R.~S., {Targett}, T.~A., {Belli}, S., \&
  {Kodama}, T. 2009, \apj, 697, 1369


\bibitem[{{Bundy} {et~al.}(2010){Bundy}, {Scarlata}, {Carollo}, {Ellis},
  {Drory}, {Hopkins}, {Salvato}, {Leauthaud}, {Koekemoer}, {Murray}, {Ilbert},
  {Oesch}, {Ma}, {Capak}, {Pozzetti}, \& {Scoville}}]{bundy10}
{Bundy}, K., {Scarlata}, C., {Carollo}, C.~M., {Ellis}, R.~S., {Drory}, N.,
  {Hopkins}, P., {Salvato}, M., {Leauthaud}, A., {Koekemoer}, A.~M., {Murray},
  N., {Ilbert}, O., {Oesch}, P., {Ma}, C.-P., {Capak}, P., {Pozzetti}, L., \&
  {Scoville}, N. 2010, \apj, 719, 1969


\bibitem[{{Butcher} \& {Oemler}(1984)}]{butcher84}
{Butcher}, H. \& {Oemler}, Jr., A. 1984, \apj, 285, 426


\bibitem[{{Butcher} {et~al.}(1980){Butcher}, {van Breugel}, \&
  {Miley}}]{Butcher1980}
{Butcher}, H.~R., {van Breugel}, W., \& {Miley}, G.~K. 1980, \apj, 235, 749


\bibitem[{{Buttiglione} {et~al.}(2010){Buttiglione}, {Capetti}, {Celotti},
  {Axon}, {Chiaberge}, {Macchetto}, \& {Sparks}}]{Buttiglione10}
{Buttiglione}, S., {Capetti}, A., {Celotti}, A., {Axon}, D.~J., {Chiaberge},
  M., {Macchetto}, F.~D., \& {Sparks}, W.~B. 2010, \aap, 509, A6


\bibitem[{{Calette} {et~al.}(2017){Calette}, {Avila-Reese},
  {Rodr{\'{\i}}guez-Puebla}, {Hern{\'a}ndez-Toledo}, \&
  {Papastergis}}]{Calette+2017}
{Calette}, R., {Avila-Reese}, V., {Rodr{\'{\i}}guez-Puebla}, A.,
  {Hern{\'a}ndez-Toledo}, H.~M., \& {Papastergis}, E. 2017, \rmxaa, submitted


\bibitem[{{Calvi} {et~al.}(2012){Calvi}, {Poggianti}, {Fasano}, \&
  {Vulcani}}]{calvi12}
{Calvi}, R., {Poggianti}, B.~M., {Fasano}, G., \& {Vulcani}, B. 2012, \mnras,
  419, L14


\bibitem[{{Cano-D{\'{\i}}az} {et~al.}(2016){Cano-D{\'{\i}}az}, {S{\'a}nchez},
  {Zibetti}, {Ascasibar}, {Bland-Hawthorn}, {Ziegler}, {Gonz{\'a}lez Delgado},
  {Walcher}, {Garc{\'{\i}}a-Benito}, {Mast}, {Mendoza-P{\'e}rez},
  {Falc{\'o}n-Barroso}, {Galbany}, {Husemann}, {Kehrig}, {Marino},
  {S{\'a}nchez-Bl{\'a}zquez}, {L{\'o}pez-Cob{\'a}}, {L{\'o}pez-S{\'a}nchez}, \&
  {Vilchez}}]{mariana16}
{Cano-D{\'{\i}}az}, M., {S{\'a}nchez}, S.~F., {Zibetti}, S., {Ascasibar}, Y.,
  {Bland-Hawthorn}, J., {Ziegler}, B., {Gonz{\'a}lez Delgado}, R.~M.,
  {Walcher}, C.~J., {Garc{\'{\i}}a-Benito}, R., {Mast}, D.,
  {Mendoza-P{\'e}rez}, M.~A., {Falc{\'o}n-Barroso}, J., {Galbany}, L.,
  {Husemann}, B., {Kehrig}, C., {Marino}, R.~A., {S{\'a}nchez-Bl{\'a}zquez},
  P., {L{\'o}pez-Cob{\'a}}, C., {L{\'o}pez-S{\'a}nchez}, {\'A}.~R., \&
  {Vilchez}, J.~M. 2016, \apjl, 821, L26


\bibitem[{{Cappellari} {et~al.}(2013){Cappellari}, {McDermid}, {Alatalo},
  {Blitz}, {Bois}, {Bournaud}, {Bureau}, {Crocker}, {Davies}, {Davis}, {de
  Zeeuw}, {Duc}, {Emsellem}, {Khochfar}, {Krajnovi{\'c}}, {Kuntschner},
  {Morganti}, {Naab}, {Oosterloo}, {Sarzi}, {Scott}, {Serra}, {Weijmans}, \&
  {Young}}]{cappellari13}
{Cappellari}, M., {McDermid}, R.~M., {Alatalo}, K., {Blitz}, L., {Bois}, M.,
  {Bournaud}, F., {Bureau}, M., {Crocker}, A.~F., {Davies}, R.~L., {Davis},
  T.~A., {de Zeeuw}, P.~T., {Duc}, P.-A., {Emsellem}, E., {Khochfar}, S.,
  {Krajnovi{\'c}}, D., {Kuntschner}, H., {Morganti}, R., {Naab}, T.,
  {Oosterloo}, T., {Sarzi}, M., {Scott}, N., {Serra}, P., {Weijmans}, A.-M., \&
  {Young}, L.~M. 2013, \mnras, 432, 1862


\bibitem[{{Cardelli} {et~al.}(1989){Cardelli}, {Clayton}, \&
  {Mathis}}]{cardelli89}
{Cardelli}, J.~A., {Clayton}, G.~C., \& {Mathis}, J.~S. 1989, \apj, 345, 245


\bibitem[{{Catal{\'a}n-Torrecilla} {et~al.}(2015){Catal{\'a}n-Torrecilla}, {Gil
  de Paz}, {Castillo-Morales}, {Iglesias-P{\'a}ramo}, {S{\'a}nchez},
  {Kennicutt}, {P{\'e}rez-Gonz{\'a}lez}, {Marino}, {Walcher}, {Husemann},
  {Garc{\'{\i}}a-Benito}, {Mast}, {Gonz{\'a}lez Delgado}, {Mu{\~n}oz-Mateos},
  {Bland-Hawthorn}, {Bomans}, {Del Olmo}, {Galbany}, {Gomes}, {Kehrig},
  {L{\'o}pez-S{\'a}nchez}, {Mendoza}, {Monreal-Ibero}, {P{\'e}rez-Torres},
  {S{\'a}nchez-Bl{\'a}zquez}, {Vilchez}, \& {Califa Collaboration}}]{catalan15}
{Catal{\'a}n-Torrecilla}, C., {Gil de Paz}, A., {Castillo-Morales}, A.,
  {Iglesias-P{\'a}ramo}, J., {S{\'a}nchez}, S.~F., {Kennicutt}, R.~C.,
  {P{\'e}rez-Gonz{\'a}lez}, P.~G., {Marino}, R.~A., {Walcher}, C.~J.,
  {Husemann}, B., {Garc{\'{\i}}a-Benito}, R., {Mast}, D., {Gonz{\'a}lez
  Delgado}, R.~M., {Mu{\~n}oz-Mateos}, J.~C., {Bland-Hawthorn}, J., {Bomans},
  D.~J., {Del Olmo}, A., {Galbany}, L., {Gomes}, J.~M., {Kehrig}, C.,
  {L{\'o}pez-S{\'a}nchez}, {\'A}.~R., {Mendoza}, M.~A., {Monreal-Ibero}, A.,
  {P{\'e}rez-Torres}, M., {S{\'a}nchez-Bl{\'a}zquez}, P., {Vilchez}, J.~M., \&
  {Califa Collaboration}. 2015, \aap, 584, A87


\bibitem[{{Catal{\'a}n-Torrecilla} {et~al.}(2017){Catal{\'a}n-Torrecilla}, {Gil
  de Paz}, {Castillo-Morales}, {M{\'e}ndez-Abreu}, {Falc{\'o}n-Barroso},
  {Bekeraite}, {Costantin}, {de Lorenzo-C{\'a}ceres}, {Florido},
  {Garc{\'{\i}}a-Benito}, {Husemann}, {Iglesias-P{\'a}ramo}, {Kennicutt},
  {Mast}, {Pascual}, {Ruiz-Lara}, {S{\'a}nchez-Menguiano}, {S{\'a}nchez},
  {Walcher}, {Bland-Hawthorn}, {Duarte Puertas}, {Marino}, {Masegosa},
  {S{\'a}nchez-Bl{\'a}zquez}, \& {CALIFA Collaboration}}]{catalan17}
{Catal{\'a}n-Torrecilla}, C., {Gil de Paz}, A., {Castillo-Morales}, A.,
  {M{\'e}ndez-Abreu}, J., {Falc{\'o}n-Barroso}, J., {Bekeraite}, S.,
  {Costantin}, L., {de Lorenzo-C{\'a}ceres}, A., {Florido}, E.,
  {Garc{\'{\i}}a-Benito}, R., {Husemann}, B., {Iglesias-P{\'a}ramo}, J.,
  {Kennicutt}, R.~C., {Mast}, D., {Pascual}, S., {Ruiz-Lara}, T.,
  {S{\'a}nchez-Menguiano}, L., {S{\'a}nchez}, S.~F., {Walcher}, C.~J.,
  {Bland-Hawthorn}, J., {Duarte Puertas}, S., {Marino}, R.~A., {Masegosa}, J.,
  {S{\'a}nchez-Bl{\'a}zquez}, P., \& {CALIFA Collaboration}. 2017, \apj, 848,
  87


\bibitem[{{Cid Fernandes} {et~al.}(2013){Cid Fernandes}, {P{\'e}rez},
  {Garc{\'{\i}}a Benito}, {Gonz{\'a}lez Delgado}, {de Amorim}, {S{\'a}nchez},
  {Husemann}, {Falc{\'o}n Barroso}, {S{\'a}nchez-Bl{\'a}zquez}, {Walcher}, \&
  {Mast}}]{cid-fernandes13}
{Cid Fernandes}, R., {P{\'e}rez}, E., {Garc{\'{\i}}a Benito}, R., {Gonz{\'a}lez
  Delgado}, R.~M., {de Amorim}, A.~L., {S{\'a}nchez}, S.~F., {Husemann}, B.,
  {Falc{\'o}n Barroso}, J., {S{\'a}nchez-Bl{\'a}zquez}, P., {Walcher}, C.~J.,
  \& {Mast}, D. 2013, \aap, 557, A86


\bibitem[{{Cid Fernandes} {et~al.}(2011){Cid Fernandes}, {Stasi{\'n}ska},
  {Mateus}, \& {Vale Asari}}]{cid11}
{Cid Fernandes}, R., {Stasi{\'n}ska}, G., {Mateus}, A., \& {Vale Asari}, N.
  2011, \mnras, 413, 1687


\bibitem[{{Cid Fernandes} {et~al.}(2010){Cid Fernandes}, {Stasi{\'n}ska},
  {Schlickmann}, {Mateus}, {Vale Asari}, {Schoenell}, \&
  {Sodr{\'e}}}]{cid-fernandes10}
{Cid Fernandes}, R., {Stasi{\'n}ska}, G., {Schlickmann}, M.~S., {Mateus}, A.,
  {Vale Asari}, N., {Schoenell}, W., \& {Sodr{\'e}}, L. 2010, \mnras, 403, 1036


\bibitem[{{Cisternas} {et~al.}(2015){Cisternas}, {Sheth}, {Salvato}, {Knapen},
  {Civano}, \& {Santini}}]{Cisternas15}
{Cisternas}, M., {Sheth}, K., {Salvato}, M., {Knapen}, J.~H., {Civano}, F., \&
  {Santini}, P. 2015, \apj, 802, 137


\bibitem[{{Colombo} {et~al.}(2017){Colombo}, {Kalinova}, {Utomo}, {Rosolowsky},
  {Bolatto}, {Levy}, {Wong}, {Sanchez}, {Leroy}, {Ostriker}, {Blitz}, {Vogel},
  {Mast}, {Garcia-Benito}, {Husemann}, {Dannerbauer}, {Ellmeier}, \&
  {Cao}}]{colombo17}
{Colombo}, D., {Kalinova}, V., {Utomo}, D., {Rosolowsky}, E., {Bolatto}, A.~D.,
  {Levy}, R.~C., {Wong}, T., {Sanchez}, S.~F., {Leroy}, A.~K., {Ostriker}, E.,
  {Blitz}, L., {Vogel}, S., {Mast}, D., {Garcia-Benito}, R., {Husemann}, B.,
  {Dannerbauer}, H., {Ellmeier}, L., \& {Cao}, Y. 2017, ArXiv e-prints


\bibitem[{{Croton} {et~al.}(2006){Croton}, {Springel}, {White}, {De Lucia},
  {Frenk}, {Gao}, {Jenkins}, {Kauffmann}, {Navarro}, \&
  {Yoshida}}]{Croton+2006}
{Croton}, D.~J., {Springel}, V., {White}, S.~D.~M., {De Lucia}, G., {Frenk},
  C.~S., {Gao}, L., {Jenkins}, A., {Kauffmann}, G., {Navarro}, J.~F., \&
  {Yoshida}, N. 2006, \mnras, 365, 11


\bibitem[{{De Lucia} \& {Blaizot}(2007)}]{DeLucia+2007}
{De Lucia}, G. \& {Blaizot}, J. 2007, \mnras, 375, 2


\bibitem[{{de Vaucouleurs}(1959)}]{deVauc59}
{de Vaucouleurs}, G. 1959, Handbuch der Physik, 53, 311


\bibitem[{{Dekel} {et~al.}(2009){Dekel}, {Birnboim}, {Engel}, {Freundlich},
  {Goerdt}, {Mumcuoglu}, {Neistein}, {Pichon}, {Teyssier}, \&
  {Zinger}}]{Dekel+2009}
{Dekel}, A., {Birnboim}, Y., {Engel}, G., {Freundlich}, J., {Goerdt}, T.,
  {Mumcuoglu}, M., {Neistein}, E., {Pichon}, C., {Teyssier}, R., \& {Zinger},
  E. 2009, \nat, 457, 451


\bibitem[{{Drory} {et~al.}(2015){Drory}, {MacDonald}, {Bershady}, {Bundy},
  {Gunn}, {Law}, {Smith}, {Stoll}, {Tremonti}, {Wake}, {Yan}, {Weijmans},
  {Byler}, {Cherinka}, {Cope}, {Eigenbrot}, {Harding}, {Holder}, {Huehnerhoff},
  {Jaehnig}, {Jansen}, {Klaene}, {Paat}, {Percival}, \&
  {Sayres}}]{2015AJ....149...77D}
{Drory}, N., {MacDonald}, N., {Bershady}, M.~A., {Bundy}, K., {Gunn}, J.,
  {Law}, D.~R., {Smith}, M., {Stoll}, R., {Tremonti}, C.~A., {Wake}, D.~A.,
  {Yan}, R., {Weijmans}, A.~M., {Byler}, N., {Cherinka}, B., {Cope}, F.,
  {Eigenbrot}, A., {Harding}, P., {Holder}, D., {Huehnerhoff}, J., {Jaehnig},
  K., {Jansen}, T.~C., {Klaene}, M., {Paat}, A.~M., {Percival}, J., \&
  {Sayres}, C. 2015, \aj, 149, 77


\bibitem[{{Dubois} {et~al.}(2016){Dubois}, {Peirani}, {Pichon}, {Devriendt},
  {Gavazzi}, {Welker}, \& {Volonteri}}]{Dubois+2016}
{Dubois}, Y., {Peirani}, S., {Pichon}, C., {Devriendt}, J., {Gavazzi}, R.,
  {Welker}, C., \& {Volonteri}, M. 2016, \mnras, 463, 3948


\bibitem[{{Ellison} {et~al.}(2018){Ellison}, {S{\'a}nchez}, {Ibarra-Medel},
  {Antonio}, {Mendel}, \& {Barrera-Ballesteros}}]{elli17}
{Ellison}, S.~L., {S{\'a}nchez}, S.~F., {Ibarra-Medel}, H., {Antonio}, B.,
  {Mendel}, J.~T., \& {Barrera-Ballesteros}, J. 2018, \mnras, 474, 2039


\bibitem[{{Faber} {et~al.}(2007){Faber}, {Willmer}, {Wolf}, {Koo}, {Weiner},
  {Newman}, {Im}, {Coil}, {Conroy}, {Cooper}, {Davis}, {Finkbeiner}, {Gerke},
  {Gebhardt}, {Groth}, {Guhathakurta}, {Harker}, {Kaiser}, {Kassin},
  {Kleinheinrich}, {Konidaris}, {Kron}, {Lin}, {Luppino}, {Madgwick},
  {Meisenheimer}, {Noeske}, {Phillips}, {Sarajedini}, {Schiavon}, {Simard},
  {Szalay}, {Vogt}, \& {Yan}}]{Faber+2007}
{Faber}, S.~M., {Willmer}, C.~N.~A., {Wolf}, C., {Koo}, D.~C., {Weiner}, B.~J.,
  {Newman}, J.~A., {Im}, M., {Coil}, A.~L., {Conroy}, C., {Cooper}, M.~C.,
  {Davis}, M., {Finkbeiner}, D.~P., {Gerke}, B.~F., {Gebhardt}, K., {Groth},
  E.~J., {Guhathakurta}, P., {Harker}, J., {Kaiser}, N., {Kassin}, S.,
  {Kleinheinrich}, M., {Konidaris}, N.~P., {Kron}, R.~G., {Lin}, L., {Luppino},
  G., {Madgwick}, D.~S., {Meisenheimer}, K., {Noeske}, K.~G., {Phillips},
  A.~C., {Sarajedini}, V.~L., {Schiavon}, R.~P., {Simard}, L., {Szalay}, A.~S.,
  {Vogt}, N.~P., \& {Yan}, R. 2007, \apj, 665, 265


\bibitem[{{Fabian}(2012)}]{fabian12}
{Fabian}, A.~C. 2012, \araa, 50, 455


\bibitem[{{Freeman}(1970)}]{free70}
{Freeman}, K.~C. 1970, \apj, 160, 811


\bibitem[{{Galbany} {et~al.}(2017){Galbany}, {Mora}, {Gonz{\'a}lez-Gait{\'a}n},
  {Bolatto}, {Dannerbauer}, {L{\'o}pez-S{\'a}nchez}, {Maeda}, {P{\'e}rez},
  {P{\'e}rez-Torres}, {S{\'a}nchez}, {Wong}, {Badenes}, {Blitz}, {Marino},
  {Utomo}, \& {Van de Ven}}]{galbany17}
{Galbany}, L., {Mora}, L., {Gonz{\'a}lez-Gait{\'a}n}, S., {Bolatto}, A.,
  {Dannerbauer}, H., {L{\'o}pez-S{\'a}nchez}, {\'A}.~R., {Maeda}, K.,
  {P{\'e}rez}, S., {P{\'e}rez-Torres}, M.~A., {S{\'a}nchez}, S.~F., {Wong}, T.,
  {Badenes}, C., {Blitz}, L., {Marino}, R.~A., {Utomo}, D., \& {Van de Ven}, G.
  2017, \mnras, 468, 628


\bibitem[{{Gao} \& {Solomon}(2004)}]{gao04}
{Gao}, Y. \& {Solomon}, P.~M. 2004, \apj, 606, 271


\bibitem[{{Garc{\'{\i}}a-Benito} {et~al.}(2017){Garc{\'{\i}}a-Benito},
  {Gonz{\'a}lez Delgado}, {P{\'e}rez}, {Cid Fernandes}, {Cortijo-Ferrero},
  {L{\'o}pez Fern{\'a}ndez}, {de Amorim}, {Lacerda}, {Vale Asari}, \&
  {S{\'a}nchez}}]{rgb17}
{Garc{\'{\i}}a-Benito}, R., {Gonz{\'a}lez Delgado}, R.~M., {P{\'e}rez}, E.,
  {Cid Fernandes}, R., {Cortijo-Ferrero}, C., {L{\'o}pez Fern{\'a}ndez}, R.,
  {de Amorim}, A.~L., {Lacerda}, E.~A.~D., {Vale Asari}, N., \& {S{\'a}nchez},
  S.~F. 2017, \aap, 608, A27


\bibitem[{{Garc{\'{\i}}a-Lorenzo}
  {et~al.}(2005{\natexlab{a}}){Garc{\'{\i}}a-Lorenzo}, {S{\'a}nchez},
  {Mediavilla}, {Gonz{\'a}lez-Serrano}, \& {Christensen}}]{bego05}
{Garc{\'{\i}}a-Lorenzo}, B., {S{\'a}nchez}, S.~F., {Mediavilla}, E.,
  {Gonz{\'a}lez-Serrano}, J.~I., \& {Christensen}, L. 2005{\natexlab{a}}, \apj,
  621, 146


\bibitem[{{Garc{\'{\i}}a-Lorenzo}
  {et~al.}(2005{\natexlab{b}}){Garc{\'{\i}}a-Lorenzo}, {S{\'a}nchez},
  {Mediavilla}, {Gonz{\'a}lez-Serrano}, \& {Christensen}}]{Begona2005}
---. 2005{\natexlab{b}}, \apj, 621, 146


\bibitem[{{Gebhardt} {et~al.}(2000){Gebhardt}, {Bender}, {Bower}, {Dressler},
  {Faber}, {Filippenko}, {Green}, {Grillmair}, {Ho}, {Kormendy}, {Lauer},
  {Magorrian}, {Pinkney}, {Richstone}, \& {Tremaine}}]{gebhardt2000}
{Gebhardt}, K., {Bender}, R., {Bower}, G., {Dressler}, A., {Faber}, S.~M.,
  {Filippenko}, A.~V., {Green}, R., {Grillmair}, C., {Ho}, L.~C., {Kormendy},
  J., {Lauer}, T.~R., {Magorrian}, J., {Pinkney}, J., {Richstone}, D., \&
  {Tremaine}, S. 2000, \apjl, 539, L13


\bibitem[{{Georgantopoulos} \& {Akylas}(2010)}]{Georgantopoulos2010}
{Georgantopoulos}, I. \& {Akylas}, A. 2010, \aap, 509, A38


\bibitem[{{Gomes} {et~al.}(2016{\natexlab{a}}){Gomes}, {Papaderos}, {Kehrig},
  {V{\'{\i}}lchez}, {Lehnert}, {S{\'a}nchez}, {Ziegler}, {Breda}, {Dos Reis},
  {Iglesias-P{\'a}ramo}, {Bland-Hawthorn}, {Galbany}, {Bomans},
  {Rosales-Ortega}, {Cid Fernandes}, {Walcher}, {Falc{\'o}n-Barroso},
  {Garc{\'{\i}}a-Benito}, {M{\'a}rquez}, {Del Olmo}, {Masegosa}, {Moll{\'a}},
  {Marino}, {Gonz{\'a}lez Delgado}, {L{\'o}pez-S{\'a}nchez}, \& {CALIFA
  Collaboration}}]{Gomes16a}
{Gomes}, J.~M., {Papaderos}, P., {Kehrig}, C., {V{\'{\i}}lchez}, J.~M.,
  {Lehnert}, M.~D., {S{\'a}nchez}, S.~F., {Ziegler}, B., {Breda}, I., {Dos
  Reis}, S.~N., {Iglesias-P{\'a}ramo}, J., {Bland-Hawthorn}, J., {Galbany}, L.,
  {Bomans}, D.~J., {Rosales-Ortega}, F.~F., {Cid Fernandes}, R., {Walcher},
  C.~J., {Falc{\'o}n-Barroso}, J., {Garc{\'{\i}}a-Benito}, R., {M{\'a}rquez},
  I., {Del Olmo}, A., {Masegosa}, J., {Moll{\'a}}, M., {Marino}, R.~A.,
  {Gonz{\'a}lez Delgado}, R.~M., {L{\'o}pez-S{\'a}nchez}, {\'A}.~R., \& {CALIFA
  Collaboration}. 2016{\natexlab{a}}, \aap, 588, A68


\bibitem[{{Gomes} {et~al.}(2016{\natexlab{b}}){Gomes}, {Papaderos},
  {V{\'{\i}}lchez}, {Kehrig}, {Iglesias-P{\'a}ramo}, {Breda}, {Lehnert},
  {S{\'a}nchez}, {Ziegler}, {Dos Reis}, {Bland-Hawthorn}, {Galbany}, {Bomans},
  {Rosales-Ortega}, {Walcher}, {Garc{\'{\i}}a-Benito}, {M{\'a}rquez}, {Del
  Olmo}, {Moll{\'a}}, {Marino}, {Catal{\'a}n-Torrecilla}, {Gonz{\'a}lez
  Delgado}, {L{\'o}pez-S{\'a}nchez}, \& {Califa Collaboration}}]{Gomes16b}
{Gomes}, J.~M., {Papaderos}, P., {V{\'{\i}}lchez}, J.~M., {Kehrig}, C.,
  {Iglesias-P{\'a}ramo}, J., {Breda}, I., {Lehnert}, M.~D., {S{\'a}nchez},
  S.~F., {Ziegler}, B., {Dos Reis}, S.~N., {Bland-Hawthorn}, J., {Galbany}, L.,
  {Bomans}, D.~J., {Rosales-Ortega}, F.~F., {Walcher}, C.~J.,
  {Garc{\'{\i}}a-Benito}, R., {M{\'a}rquez}, I., {Del Olmo}, A., {Moll{\'a}},
  M., {Marino}, R.~A., {Catal{\'a}n-Torrecilla}, C., {Gonz{\'a}lez Delgado},
  R.~M., {L{\'o}pez-S{\'a}nchez}, {\'A}.~R., \& {Califa Collaboration}.
  2016{\natexlab{b}}, \aap, 585, A92


\bibitem[{{Gon{\c c}alves} {et~al.}(2012){Gon{\c c}alves}, {Martin},
  {Men{\'e}ndez-Delmestre}, {Wyder}, \& {Koekemoer}}]{Goncalves+2012}
{Gon{\c c}alves}, T.~S., {Martin}, D.~C., {Men{\'e}ndez-Delmestre}, K.,
  {Wyder}, T.~K., \& {Koekemoer}, A. 2012, \apj, 759, 67


\bibitem[{{Gonz{\'a}lez Delgado} {et~al.}(2014{\natexlab{a}}){Gonz{\'a}lez
  Delgado}, {Cid Fernandes}, {Garc{\'{\i}}a-Benito}, {P{\'e}rez}, {de Amorim},
  {Cortijo-Ferrero}, {Lacerda}, {L{\'o}pez Fern{\'a}ndez}, {S{\'a}nchez}, {Vale
  Asari}, {Alves}, {Bland-Hawthorn}, {Galbany}, {Gallazzi}, {Husemann},
  {Bekeraite}, {Jungwiert}, {L{\'o}pez-S{\'a}nchez}, {de Lorenzo-C{\'a}ceres},
  {Marino}, {Mast}, {Moll{\'a}}, {del Olmo}, {S{\'a}nchez-Bl{\'a}zquez}, {van
  de Ven}, {V{\'{\i}}lchez}, {Walcher}, {Wisotzki}, {Ziegler}, \&
  {collaboration920}}]{rosa14b}
{Gonz{\'a}lez Delgado}, R.~M., {Cid Fernandes}, R., {Garc{\'{\i}}a-Benito}, R.,
  {P{\'e}rez}, E., {de Amorim}, A.~L., {Cortijo-Ferrero}, C., {Lacerda},
  E.~A.~D., {L{\'o}pez Fern{\'a}ndez}, R., {S{\'a}nchez}, S.~F., {Vale Asari},
  N., {Alves}, J., {Bland-Hawthorn}, J., {Galbany}, L., {Gallazzi}, A.,
  {Husemann}, B., {Bekeraite}, S., {Jungwiert}, B., {L{\'o}pez-S{\'a}nchez},
  A.~R., {de Lorenzo-C{\'a}ceres}, A., {Marino}, R.~A., {Mast}, D.,
  {Moll{\'a}}, M., {del Olmo}, A., {S{\'a}nchez-Bl{\'a}zquez}, P., {van de
  Ven}, G., {V{\'{\i}}lchez}, J.~M., {Walcher}, C.~J., {Wisotzki}, L.,
  {Ziegler}, B., \& {collaboration920}, C. 2014{\natexlab{a}}, \apjl, 791, L16


\bibitem[{{Gonz{\'a}lez Delgado} {et~al.}(2016){Gonz{\'a}lez Delgado}, {Cid
  Fernandes}, {P{\'e}rez}, {Garc{\'{\i}}a-Benito}, {L{\'o}pez Fern{\'a}ndez},
  {Lacerda}, {Cortijo-Ferrero}, {de Amorim}, {Vale Asari}, {S{\'a}nchez},
  {Walcher}, {Wisotzki}, {Mast}, {Alves}, {Ascasibar}, {Bland-Hawthorn},
  {Galbany}, {Kennicutt}, {M{\'a}rquez}, {Masegosa}, {Moll{\'a}},
  {S{\'a}nchez-Bl{\'a}zquez}, \& {V{\'{\i}}lchez}}]{rosa16a}
{Gonz{\'a}lez Delgado}, R.~M., {Cid Fernandes}, R., {P{\'e}rez}, E.,
  {Garc{\'{\i}}a-Benito}, R., {L{\'o}pez Fern{\'a}ndez}, R., {Lacerda},
  E.~A.~D., {Cortijo-Ferrero}, C., {de Amorim}, A.~L., {Vale Asari}, N.,
  {S{\'a}nchez}, S.~F., {Walcher}, C.~J., {Wisotzki}, L., {Mast}, D., {Alves},
  J., {Ascasibar}, Y., {Bland-Hawthorn}, J., {Galbany}, L., {Kennicutt}, R.~C.,
  {M{\'a}rquez}, I., {Masegosa}, J., {Moll{\'a}}, M.,
  {S{\'a}nchez-Bl{\'a}zquez}, P., \& {V{\'{\i}}lchez}, J.~M. 2016, \aap, 590,
  A44


\bibitem[{{Gonz{\'a}lez Delgado} {et~al.}(2015){Gonz{\'a}lez Delgado},
  {Garc{\'{\i}}a-Benito}, {P{\'e}rez}, {Cid Fernandes}, {de Amorim},
  {Cortijo-Ferrero}, {Lacerda}, {L{\'o}pez Fern{\'a}ndez}, {Vale-Asari},
  {S{\'a}nchez}, {Moll{\'a}}, {Ruiz-Lara}, {S{\'a}nchez-Bl{\'a}zquez},
  {Walcher}, {Alves}, {Aguerri}, {Bekerait{\'e}}, {Bland-Hawthorn}, {Galbany},
  {Gallazzi}, {Husemann}, {Iglesias-P{\'a}ramo}, {Kalinova},
  {L{\'o}pez-S{\'a}nchez}, {Marino}, {M{\'a}rquez}, {Masegosa}, {Mast},
  {M{\'e}ndez-Abreu}, {Mendoza}, {del Olmo}, {P{\'e}rez}, {Quirrenbach}, \&
  {Zibetti}}]{rosa15a}
{Gonz{\'a}lez Delgado}, R.~M., {Garc{\'{\i}}a-Benito}, R., {P{\'e}rez}, E.,
  {Cid Fernandes}, R., {de Amorim}, A.~L., {Cortijo-Ferrero}, C., {Lacerda},
  E.~A.~D., {L{\'o}pez Fern{\'a}ndez}, R., {Vale-Asari}, N., {S{\'a}nchez},
  S.~F., {Moll{\'a}}, M., {Ruiz-Lara}, T., {S{\'a}nchez-Bl{\'a}zquez}, P.,
  {Walcher}, C.~J., {Alves}, J., {Aguerri}, J.~A.~L., {Bekerait{\'e}}, S.,
  {Bland-Hawthorn}, J., {Galbany}, L., {Gallazzi}, A., {Husemann}, B.,
  {Iglesias-P{\'a}ramo}, J., {Kalinova}, V., {L{\'o}pez-S{\'a}nchez}, A.~R.,
  {Marino}, R.~A., {M{\'a}rquez}, I., {Masegosa}, J., {Mast}, D.,
  {M{\'e}ndez-Abreu}, J., {Mendoza}, A., {del Olmo}, A., {P{\'e}rez}, I.,
  {Quirrenbach}, A., \& {Zibetti}, S. 2015, \aap, 581, A103


\bibitem[{{Gonz{\'a}lez Delgado} {et~al.}(2014{\natexlab{b}}){Gonz{\'a}lez
  Delgado}, {P{\'e}rez}, {Cid Fernandes}, {Garc{\'{\i}}a-Benito}, {de Amorim},
  {S{\'a}nchez}, {Husemann}, {Cortijo-Ferrero}, {L{\'o}pez Fern{\'a}ndez},
  {S{\'a}nchez-Bl{\'a}zquez}, {Bekeraite}, {Walcher}, {Falc{\'o}n-Barroso},
  {Gallazzi}, {van de Ven}, {Alves}, {Bland-Hawthorn}, {Kennicutt}, {Kupko},
  {Lyubenova}, {Mast}, {Moll{\'a}}, {Marino}, {Quirrenbach}, {V{\'{\i}}lchez},
  \& {Wisotzki}}]{rosa14}
{Gonz{\'a}lez Delgado}, R.~M., {P{\'e}rez}, E., {Cid Fernandes}, R.,
  {Garc{\'{\i}}a-Benito}, R., {de Amorim}, A.~L., {S{\'a}nchez}, S.~F.,
  {Husemann}, B., {Cortijo-Ferrero}, C., {L{\'o}pez Fern{\'a}ndez}, R.,
  {S{\'a}nchez-Bl{\'a}zquez}, P., {Bekeraite}, S., {Walcher}, C.~J.,
  {Falc{\'o}n-Barroso}, J., {Gallazzi}, A., {van de Ven}, G., {Alves}, J.,
  {Bland-Hawthorn}, J., {Kennicutt}, R.~C., {Kupko}, D., {Lyubenova}, M.,
  {Mast}, D., {Moll{\'a}}, M., {Marino}, R.~A., {Quirrenbach}, A.,
  {V{\'{\i}}lchez}, J.~M., \& {Wisotzki}, L. 2014{\natexlab{b}}, \aap, 562, A47


\bibitem[{{Gonz{\'a}lez Delgado} {et~al.}(2017){Gonz{\'a}lez Delgado},
  {P{\'e}rez}, {Cid Fernandes}, {Garc{\'{\i}}a-Benito}, {L{\'o}pez
  Fern{\'a}ndez}, {Vale Asari}, {Cortijo-Ferrero}, {de Amorim}, {Lacerda},
  {S{\'a}nchez}, {Lehnert}, \& {Walcher}}]{rosa17}
{Gonz{\'a}lez Delgado}, R.~M., {P{\'e}rez}, E., {Cid Fernandes}, R.,
  {Garc{\'{\i}}a-Benito}, R., {L{\'o}pez Fern{\'a}ndez}, R., {Vale Asari}, N.,
  {Cortijo-Ferrero}, C., {de Amorim}, A.~L., {Lacerda}, E.~A.~D.,
  {S{\'a}nchez}, S.~F., {Lehnert}, M.~D., \& {Walcher}, C.~J. 2017, \aap, 607,
  A128


\bibitem[{{Graham}(2016)}]{Graham2016}
{Graham}, A.~W. 2016, Galactic Bulges, 418, 263


\bibitem[{{Gunn} {et~al.}(2006){Gunn}, {Siegmund}, {Mannery}, {Owen}, {Hull},
  {Leger}, {Carey}, {Knapp}, {York}, {Boroski}, {Kent}, {Lupton}, {Rockosi},
  {Evans}, {Waddell}, {Anderson}, {Annis}, {Barentine}, {Bartoszek}, {Bastian},
  {Bracker}, {Brewington}, {Briegel}, {Brinkmann}, {Brown}, {Carr},
  {Czarapata}, {Drennan}, {Dombeck}, {Federwitz}, {Gillespie}, {Gonzales},
  {Hansen}, {Harvanek}, {Hayes}, {Jordan}, {Kinney}, {Klaene}, {Kleinman},
  {Kron}, {Kresinski}, {Lee}, {Limmongkol}, {Lindenmeyer}, {Long}, {Loomis},
  {McGehee}, {Mantsch}, {Neilsen}, {Neswold}, {Newman}, {Nitta}, {Peoples},
  {Pier}, {Prieto}, {Prosapio}, {Rivetta}, {Schneider}, {Snedden}, \&
  {Wang}}]{2006AJ....131.2332G}
{Gunn}, J.~E., {Siegmund}, W.~A., {Mannery}, E.~J., {Owen}, R.~E., {Hull},
  C.~L., {Leger}, R.~F., {Carey}, L.~N., {Knapp}, G.~R., {York}, D.~G.,
  {Boroski}, W.~N., {Kent}, S.~M., {Lupton}, R.~H., {Rockosi}, C.~M., {Evans},
  M.~L., {Waddell}, P., {Anderson}, J.~E., {Annis}, J., {Barentine}, J.~C.,
  {Bartoszek}, L.~M., {Bastian}, S., {Bracker}, S.~B., {Brewington}, H.~J.,
  {Briegel}, C.~I., {Brinkmann}, J., {Brown}, Y.~J., {Carr}, M.~A.,
  {Czarapata}, P.~C., {Drennan}, C.~C., {Dombeck}, T., {Federwitz}, G.~R.,
  {Gillespie}, B.~A., {Gonzales}, C., {Hansen}, S.~U., {Harvanek}, M., {Hayes},
  J., {Jordan}, W., {Kinney}, E., {Klaene}, M., {Kleinman}, S.~J., {Kron},
  R.~G., {Kresinski}, J., {Lee}, G., {Limmongkol}, S., {Lindenmeyer}, C.~W.,
  {Long}, D.~C., {Loomis}, C.~L., {McGehee}, P.~M., {Mantsch}, P.~M.,
  {Neilsen}, Jr., E.~H., {Neswold}, R.~M., {Newman}, P.~R., {Nitta}, A.,
  {Peoples}, Jr., J., {Pier}, J.~R., {Prieto}, P.~S., {Prosapio}, A.,
  {Rivetta}, C., {Schneider}, D.~P., {Snedden}, S., \& {Wang}, S.-i. 2006, \aj,
  131, 2332


\bibitem[{{H{\"a}ring} \& {Rix}(2004)}]{haring04}
{H{\"a}ring}, N. \& {Rix}, H.-W. 2004, \apjl, 604, L89


\bibitem[{{Heiderman} {et~al.}(2010){Heiderman}, {Evans}, {Allen}, {Huard}, \&
  {Heyer}}]{Heiderman2010}
{Heiderman}, A., {Evans}, II, N.~J., {Allen}, L.~E., {Huard}, T., \& {Heyer},
  M. 2010, \apj, 723, 1019


\bibitem[{{Hopkins} {et~al.}(2010){Hopkins}, {Bundy}, {Croton}, {Hernquist},
  {Keres}, {Khochfar}, {Stewart}, {Wetzel}, \& {Younger}}]{Hopkins+2010}
{Hopkins}, P.~F., {Bundy}, K., {Croton}, D., {Hernquist}, L., {Keres}, D.,
  {Khochfar}, S., {Stewart}, K., {Wetzel}, A., \& {Younger}, J.~D. 2010, \apj,
  715, 202


\bibitem[{{Hopkins} {et~al.}(2009){Hopkins}, {Cox}, {Younger}, \&
  {Hernquist}}]{Hopkins+2009}
{Hopkins}, P.~F., {Cox}, T.~J., {Younger}, J.~D., \& {Hernquist}, L. 2009,
  \apj, 691, 1168


\bibitem[{{Hopkins} \&
  {Hernquist}(2009{\natexlab{a}})}]{Hopkins_Hernquist2009a}
{Hopkins}, P.~F. \& {Hernquist}, L. 2009{\natexlab{a}}, \apj, 694, 599


\bibitem[{{Hopkins} \& {Hernquist}(2009{\natexlab{b}})}]{hopkins09}
---. 2009{\natexlab{b}}, \apj, 694, 599


\bibitem[{{Hopkins} {et~al.}(2006){Hopkins}, {Hernquist}, {Cox}, {Di Matteo},
  {Robertson}, \& {Springel}}]{Hopkins2006ApJS..163....1H}
{Hopkins}, P.~F., {Hernquist}, L., {Cox}, T.~J., {Di Matteo}, T., {Robertson},
  B., \& {Springel}, V. 2006, \apjs, 163, 1


\bibitem[{{Hunt} \& {Malkan}(1999)}]{hunt99}
{Hunt}, L.~K. \& {Malkan}, M.~A. 1999, \apj, 516, 660


\bibitem[{{Husemann} {et~al.}(2017){Husemann}, {Davis}, {Jahnke},
  {Dannerbauer}, {Urrutia}, \& {Hodge}}]{husemann17}
{Husemann}, B., {Davis}, T.~A., {Jahnke}, K., {Dannerbauer}, H., {Urrutia}, T.,
  \& {Hodge}, J. 2017, \mnras, 470, 1570


\bibitem[{{Husemann} {et~al.}(2010){Husemann}, {S{\'a}nchez}, {Wisotzki},
  {Jahnke}, {Kupko}, {Nugroho}, \& {Schramm}}]{husemann10}
{Husemann}, B., {S{\'a}nchez}, S.~F., {Wisotzki}, L., {Jahnke}, K., {Kupko},
  D., {Nugroho}, D., \& {Schramm}, M. 2010, \aap, 519, A115+


\bibitem[{{Ibarra-Medel} {et~al.}(2016){Ibarra-Medel}, {S{\'a}nchez},
  {Avila-Reese}, {Hern{\'a}ndez-Toledo}, {Gonz{\'a}lez}, {Drory}, {Bundy},
  {Bizyaev}, {Cano-D{\'{\i}}az}, {Malanushenko}, {Pan}, {Roman-Lopes}, \&
  {Thomas}}]{ibarra16}
{Ibarra-Medel}, H.~J., {S{\'a}nchez}, S.~F., {Avila-Reese}, V.,
  {Hern{\'a}ndez-Toledo}, H.~M., {Gonz{\'a}lez}, J.~J., {Drory}, N., {Bundy},
  K., {Bizyaev}, D., {Cano-D{\'{\i}}az}, M., {Malanushenko}, E., {Pan}, K.,
  {Roman-Lopes}, A., \& {Thomas}, D. 2016, \mnras, 463, 2799


\bibitem[{{Jahnke} {et~al.}(2004{\natexlab{a}}){Jahnke}, {Kuhlbrodt}, \&
  {Wisotzki}}]{jahnke04}
{Jahnke}, K., {Kuhlbrodt}, B., \& {Wisotzki}, L. 2004{\natexlab{a}}, \mnras,
  352, 399


\bibitem[{{Jahnke} {et~al.}(2004{\natexlab{b}}){Jahnke}, {S{\'a}nchez},
  {Wisotzki}, {Barden}, {Beckwith}, {Bell}, {Borch}, {Caldwell},
  {H{\"a}ussler}, {Heymans}, {Jogee}, {McIntosh}, {Meisenheimer}, {Peng},
  {Rix}, {Somerville}, \& {Wolf}}]{jahnke04b}
{Jahnke}, K., {S{\'a}nchez}, S.~F., {Wisotzki}, L., {Barden}, M., {Beckwith},
  S.~V.~W., {Bell}, E.~F., {Borch}, A., {Caldwell}, J.~A.~R., {H{\"a}ussler},
  B., {Heymans}, C., {Jogee}, S., {McIntosh}, D.~H., {Meisenheimer}, K.,
  {Peng}, C.~Y., {Rix}, H.-W., {Somerville}, R.~S., \& {Wolf}, C.
  2004{\natexlab{b}}, \apj, 614, 568


\bibitem[{{Jogee} {et~al.}(2004){Jogee}, {Barazza}, {Rix}, {Shlosman},
  {Barden}, {Wolf}, {Davies}, {Heyer}, {Beckwith}, {Bell}, {Borch}, {Caldwell},
  {Conselice}, {Dahlen}, {H{\"a}ussler}, {Heymans}, {Jahnke}, {Knapen},
  {Laine}, {Lubell}, {Mobasher}, {McIntosh}, {Meisenheimer}, {Peng},
  {Ravindranath}, {Sanchez}, {Somerville}, \& {Wisotzki}}]{jogee04}
{Jogee}, S., {Barazza}, F.~D., {Rix}, H.-W., {Shlosman}, I., {Barden}, M.,
  {Wolf}, C., {Davies}, J., {Heyer}, I., {Beckwith}, S.~V.~W., {Bell}, E.~F.,
  {Borch}, A., {Caldwell}, J.~A.~R., {Conselice}, C.~J., {Dahlen}, T.,
  {H{\"a}ussler}, B., {Heymans}, C., {Jahnke}, K., {Knapen}, J.~H., {Laine},
  S., {Lubell}, G.~M., {Mobasher}, B., {McIntosh}, D.~H., {Meisenheimer}, K.,
  {Peng}, C.~Y., {Ravindranath}, S., {Sanchez}, S.~F., {Somerville}, R.~S., \&
  {Wisotzki}, L. 2004, \apjl, 615, L105


\bibitem[{{Kannappan} {et~al.}(2009){Kannappan}, {Guie}, \&
  {Baker}}]{Kannappan+2009}
{Kannappan}, S.~J., {Guie}, J.~M., \& {Baker}, A.~J. 2009, \aj, 138, 579


\bibitem[{{Katsianis} {et~al.}(2015){Katsianis}, {Tescari}, \&
  {Wyithe}}]{Katsianis15}
{Katsianis}, A., {Tescari}, E., \& {Wyithe}, J.~S.~B. 2015, \mnras, 448, 3001


\bibitem[{{Kauffmann} \& {Haehnelt}(2000)}]{Kauffmann+2000}
{Kauffmann}, G. \& {Haehnelt}, M. 2000, \mnras, 311, 576


\bibitem[{{Kauffmann} {et~al.}(2003{\natexlab{a}}){Kauffmann}, {Heckman},
  {Tremonti}, {Brinchmann}, {Charlot}, {White}, {Ridgway}, {Brinkmann},
  {Fukugita}, {Hall}, {Ivezi{\'c}}, {Richards}, \& {Schneider}}]{kauffmann03}
{Kauffmann}, G., {Heckman}, T.~M., {Tremonti}, C., {Brinchmann}, J., {Charlot},
  S., {White}, S.~D.~M., {Ridgway}, S.~E., {Brinkmann}, J., {Fukugita}, M.,
  {Hall}, P.~B., {Ivezi{\'c}}, {\v Z}., {Richards}, G.~T., \& {Schneider},
  D.~P. 2003{\natexlab{a}}, \mnras, 346, 1055


\bibitem[{{Kauffmann} {et~al.}(2003{\natexlab{b}}){Kauffmann}, {Heckman},
  {Tremonti}, {Brinchmann}, {Charlot}, {White}, {Ridgway}, {Brinkmann},
  {Fukugita}, {Hall}, {Ivezi{\'c}}, {Richards}, \&
  {Schneider}}]{Kauffmann+2003}
---. 2003{\natexlab{b}}, \mnras, 346, 1055


\bibitem[{{Keel}(1983)}]{keel83}
{Keel}, W.~C. 1983, \apj, 268, 632


\bibitem[{{Kennicutt} \& {Evans}(2012)}]{Kennicutt+2012}
{Kennicutt}, R.~C. \& {Evans}, N.~J. 2012, \araa, 50, 531


\bibitem[{{Kennicutt}(1998)}]{kennicutt98}
{Kennicutt}, Jr., R.~C. 1998, \apj, 498, 541


\bibitem[{{Kennicutt} {et~al.}(2007){Kennicutt}, {Calzetti}, {Walter}, {Helou},
  {Hollenbach}, {Armus}, {Bendo}, {Dale}, {Draine}, {Engelbracht}, {Gordon},
  {Prescott}, {Regan}, {Thornley}, {Bot}, {Brinks}, {de Blok}, {de Mello},
  {Meyer}, {Moustakas}, {Murphy}, {Sheth}, \& {Smith}}]{Kennicutt07}
{Kennicutt}, Jr., R.~C., {Calzetti}, D., {Walter}, F., {Helou}, G.,
  {Hollenbach}, D.~J., {Armus}, L., {Bendo}, G., {Dale}, D.~A., {Draine},
  B.~T., {Engelbracht}, C.~W., {Gordon}, K.~D., {Prescott}, M.~K.~M., {Regan},
  M.~W., {Thornley}, M.~D., {Bot}, C., {Brinks}, E., {de Blok}, E., {de Mello},
  D., {Meyer}, M., {Moustakas}, J., {Murphy}, E.~J., {Sheth}, K., \& {Smith},
  J.~D.~T. 2007, \apj, 671, 333


\bibitem[{{Kewley} {et~al.}(2001){Kewley}, {Dopita}, {Sutherland}, {Heisler},
  \& {Trevena}}]{kewley01}
{Kewley}, L.~J., {Dopita}, M.~A., {Sutherland}, R.~S., {Heisler}, C.~A., \&
  {Trevena}, J. 2001, \apj, 556, 121


\bibitem[{{Kewley} {et~al.}(2006){Kewley}, {Groves}, {Kauffmann}, \&
  {Heckman}}]{kewley06}
{Kewley}, L.~J., {Groves}, B., {Kauffmann}, G., \& {Heckman}, T. 2006, \mnras,
  372, 961


\bibitem[{{Komugi} {et~al.}(2012){Komugi}, {Tateuchi}, {Motohara}, {Takagi},
  {Iono}, {Kaneko}, {Ueda}, {Saitoh}, {Kato}, {Konishi}, {Koshida}, {Morokuma},
  {Takahashi}, {Tanab{\'e}}, \& {Yoshii}}]{komu12}
{Komugi}, S., {Tateuchi}, K., {Motohara}, K., {Takagi}, T., {Iono}, D.,
  {Kaneko}, H., {Ueda}, J., {Saitoh}, T.~R., {Kato}, N., {Konishi}, M.,
  {Koshida}, S., {Morokuma}, T., {Takahashi}, H., {Tanab{\'e}}, T., \&
  {Yoshii}, Y. 2012, \apj, 757, 138


\bibitem[{{Kormendy} \& {Ho}(2013)}]{Kormendy+2013}
{Kormendy}, J. \& {Ho}, L.~C. 2013, \araa, 51, 511


\bibitem[{{Kraft} {et~al.}(2012){Kraft}, {Birkinshaw}, {Nulsen}, {Worrall},
  {Croston}, {Forman}, {Hardcastle}, {Jones}, \& {Murray}}]{kraft12}
{Kraft}, R.~P., {Birkinshaw}, M., {Nulsen}, P.~E.~J., {Worrall}, D.~M.,
  {Croston}, J.~H., {Forman}, W.~R., {Hardcastle}, M.~J., {Jones}, C., \&
  {Murray}, S.~S. 2012, \apj, 749, 19


\bibitem[{{Krongold} {et~al.}(2002){Krongold}, {Dultzin-Hacyan}, \&
  {Marziani}}]{krongold02}
{Krongold}, Y., {Dultzin-Hacyan}, D., \& {Marziani}, P. 2002, \apj, 572, 169


\bibitem[{{Krumholz} {et~al.}(2012){Krumholz}, {Dekel}, \&
  {McKee}}]{Krumholz+2012}
{Krumholz}, M.~R., {Dekel}, A., \& {McKee}, C.~F. 2012, \apj, 745, 69


\bibitem[{{Lacerda} {et~al.}(2018){Lacerda}, {Cid Fernandes}, {Couto},
  {Stasi{\'n}ska}, {Garc{\'{\i}}a-Benito}, {Vale Asari}, {P{\'e}rez},
  {Gonz{\'a}lez Delgado}, {S{\'a}nchez}, \& {de Amorim}}]{lacerda17}
{Lacerda}, E.~A.~D., {Cid Fernandes}, R., {Couto}, G.~S., {Stasi{\'n}ska}, G.,
  {Garc{\'{\i}}a-Benito}, R., {Vale Asari}, N., {P{\'e}rez}, E., {Gonz{\'a}lez
  Delgado}, R.~M., {S{\'a}nchez}, S.~F., \& {de Amorim}, A.~L. 2018, \mnras,
  474, 3727


\bibitem[{{Lacerna} {et~al.}(2016){Lacerna}, {Hern{\'a}ndez-Toledo},
  {Avila-Reese}, {Abonza-Sane}, \& {del Olmo}}]{Lacerna+2016}
{Lacerna}, I., {Hern{\'a}ndez-Toledo}, H.~M., {Avila-Reese}, V., {Abonza-Sane},
  J., \& {del Olmo}, A. 2016, \aap, 588, A79


\bibitem[{{Law} {et~al.}(2016){Law}, {Cherinka}, {Yan}, {Andrews}, {Bershady},
  {Bizyaev}, {Blanc}, {Blanton}, {Bolton}, {Brownstein}, {Bundy}, {Chen},
  {Drory}, {D'Souza}, {Fu}, {Jones}, {Kauffmann}, {MacDonald}, {Masters},
  {Newman}, {Parejko}, {S{\'a}nchez-Gallego}, {S{\'a}nchez}, {Schlegel},
  {Thomas}, {Wake}, {Weijmans}, {Westfall}, \& {Zhang}}]{2016AJ....152...83L}
{Law}, D.~R., {Cherinka}, B., {Yan}, R., {Andrews}, B.~H., {Bershady}, M.~A.,
  {Bizyaev}, D., {Blanc}, G.~A., {Blanton}, M.~R., {Bolton}, A.~S.,
  {Brownstein}, J.~R., {Bundy}, K., {Chen}, Y., {Drory}, N., {D'Souza}, R.,
  {Fu}, H., {Jones}, A., {Kauffmann}, G., {MacDonald}, N., {Masters}, K.~L.,
  {Newman}, J.~A., {Parejko}, J.~K., {S{\'a}nchez-Gallego}, J.~R.,
  {S{\'a}nchez}, S.~F., {Schlegel}, D.~J., {Thomas}, D., {Wake}, D.~A.,
  {Weijmans}, A.-M., {Westfall}, K.~B., \& {Zhang}, K. 2016, \aj, 152, 83


\bibitem[{{Law} {et~al.}(2015){Law}, {Yan}, {Bershady}, {Bundy}, {Cherinka},
  {Drory}, {MacDonald}, {S{\'a}nchez-Gallego}, {Wake}, {Weijmans}, {Blanton},
  {Klaene}, {Moran}, {Sanchez}, \& {Zhang}}]{law15}
{Law}, D.~R., {Yan}, R., {Bershady}, M.~A., {Bundy}, K., {Cherinka}, B.,
  {Drory}, N., {MacDonald}, N., {S{\'a}nchez-Gallego}, J.~R., {Wake}, D.~A.,
  {Weijmans}, A.-M., {Blanton}, M.~R., {Klaene}, M.~A., {Moran}, S.~M.,
  {Sanchez}, S.~F., \& {Zhang}, K. 2015, \aj, 150, 19


\bibitem[{{Leroy} {et~al.}(2013){Leroy}, {Walter}, {Sandstrom}, {Schruba},
  {Munoz-Mateos}, {Bigiel}, {Bolatto}, {Brinks}, {de Blok}, {Meidt}, {Rix},
  {Rosolowsky}, {Schinnerer}, {Schuster}, \& {Usero}}]{leroy13}
{Leroy}, A.~K., {Walter}, F., {Sandstrom}, K., {Schruba}, A., {Munoz-Mateos},
  J.-C., {Bigiel}, F., {Bolatto}, A., {Brinks}, E., {de Blok}, W.~J.~G.,
  {Meidt}, S., {Rix}, H.-W., {Rosolowsky}, E., {Schinnerer}, E., {Schuster},
  K.-F., \& {Usero}, A. 2013, \aj, 146, 19


\bibitem[{{Lian} {et~al.}(2016){Lian}, {Yan}, {Zhang}, \& {Kong}}]{Lian+2016}
{Lian}, J., {Yan}, R., {Zhang}, K., \& {Kong}, X. 2016, \apj, 832, 29


\bibitem[{{Lilly} {et~al.}(2013{\natexlab{a}}){Lilly}, {Carollo}, {Pipino},
  {Renzini}, \& {Peng}}]{2013ApJ...772..119L}
{Lilly}, S.~J., {Carollo}, C.~M., {Pipino}, A., {Renzini}, A., \& {Peng}, Y.
  2013{\natexlab{a}}, \apj, 772, 119


\bibitem[{{Lilly} {et~al.}(2013{\natexlab{b}}){Lilly}, {Carollo}, {Pipino},
  {Renzini}, \& {Peng}}]{Lilly+2013}
---. 2013{\natexlab{b}}, \apj, 772, 119


\bibitem[{{Lin} {et~al.}(2017){Lin}, {Belfiore}, {Pan}, {Bothwell}, {Hsieh},
  {Huang}, {Xiao}, {S{\'a}nchez}, {Hsieh}, {Masters}, {Ramya}, {Lin}, {Hsu},
  {Li}, {Maiolino}, {Bundy}, {Bizyaev}, {Drory}, {Ibarra-Medel}, {Lacerna},
  {Haines}, {Smethurst}, {Stark}, \& {Thomas}}]{lin17}
{Lin}, L., {Belfiore}, F., {Pan}, H.-A., {Bothwell}, M.~S., {Hsieh}, P.-Y.,
  {Huang}, S., {Xiao}, T., {S{\'a}nchez}, S.~F., {Hsieh}, B.-C., {Masters}, K.,
  {Ramya}, S., {Lin}, J.-H., {Hsu}, C.-H., {Li}, C., {Maiolino}, R., {Bundy},
  K., {Bizyaev}, D., {Drory}, N., {Ibarra-Medel}, H., {Lacerna}, I., {Haines},
  T., {Smethurst}, R., {Stark}, D.~V., \& {Thomas}, D. 2017, \apj, 851, 18


\bibitem[{{Lintott} {et~al.}(2011){Lintott}, {Schawinski}, {Bamford}, {Slosar},
  {Land}, {Thomas}, {Edmondson}, {Masters}, {Nichol}, {Raddick}, {Szalay},
  {Andreescu}, {Murray}, \& {Vandenberg}}]{galzoo}
{Lintott}, C., {Schawinski}, K., {Bamford}, S., {Slosar}, A., {Land}, K.,
  {Thomas}, D., {Edmondson}, E., {Masters}, K., {Nichol}, R.~C., {Raddick},
  M.~J., {Szalay}, A., {Andreescu}, D., {Murray}, P., \& {Vandenberg}, J. 2011,
  \mnras, 410, 166


\bibitem[{{Lipari} {et~al.}(1994){Lipari}, {Colina}, \& {Macchetto}}]{lipari94}
{Lipari}, S., {Colina}, L., \& {Macchetto}, F. 1994, \apj, 427, 174


\bibitem[{{L{\'o}pez-Cob{\'a}} {et~al.}(2017){L{\'o}pez-Cob{\'a}},
  {S{\'a}nchez}, {Moiseev}, {Oparin}, {Bitsakis}, {Cruz-Gonz{\'a}lez},
  {Morisset}, {Galbany}, {Bland-Hawthorn}, {Roth}, {Dettmar}, {Bomans},
  {Gonz{\'a}lez Delgado}, {Cano-D{\'{\i}}az}, {Marino}, {Kehrig}, {Monreal
  Ibero}, \& {Abril-Melgarejo}}]{carlos16}
{L{\'o}pez-Cob{\'a}}, C., {S{\'a}nchez}, S.~F., {Moiseev}, A.~V., {Oparin},
  D.~V., {Bitsakis}, T., {Cruz-Gonz{\'a}lez}, I., {Morisset}, C., {Galbany},
  L., {Bland-Hawthorn}, J., {Roth}, M.~M., {Dettmar}, R.-J., {Bomans}, D.~J.,
  {Gonz{\'a}lez Delgado}, R.~M., {Cano-D{\'{\i}}az}, M., {Marino}, R.~A.,
  {Kehrig}, C., {Monreal Ibero}, A., \& {Abril-Melgarejo}, V. 2017, \mnras,
  467, 4951


\bibitem[{{Magorrian} {et~al.}(1998){Magorrian}, {Tremaine}, {Richstone},
  {Bender}, {Bower}, {Dressler}, {Faber}, {Gebhardt}, {Green}, {Grillmair},
  {Kormendy}, \& {Lauer}}]{magorrian98}
{Magorrian}, J., {Tremaine}, S., {Richstone}, D., {Bender}, R., {Bower}, G.,
  {Dressler}, A., {Faber}, S.~M., {Gebhardt}, K., {Green}, R., {Grillmair}, C.,
  {Kormendy}, J., \& {Lauer}, T. 1998, \aj, 115, 2285


\bibitem[{{Marino} {et~al.}(2013){Marino}, {Rosales-Ortega}, {S{\'a}nchez},
  {Gil de Paz}, {V{\'{\i}}lchez}, {Miralles-Caballero}, {Kehrig},
  {P{\'e}rez-Montero}, {Stanishev}, {Iglesias-P{\'a}ramo}, {D{\'{\i}}az},
  {Castillo-Morales}, {Kennicutt}, {L{\'o}pez-S{\'a}nchez}, {Galbany},
  {Garc{\'{\i}}a-Benito}, {Mast}, {Mendez-Abreu}, {Monreal-Ibero}, {Husemann},
  {Walcher}, {Garc{\'{\i}}a-Lorenzo}, {Masegosa}, {Del Olmo Orozco},
  {Mour{\~a}o}, {Ziegler}, {Moll{\'a}}, {Papaderos},
  {S{\'a}nchez-Bl{\'a}zquez}, {Gonz{\'a}lez Delgado}, {Falc{\'o}n-Barroso},
  {Roth}, {van de Ven}, \& {Califa Team}}]{marino13}
{Marino}, R.~A., {Rosales-Ortega}, F.~F., {S{\'a}nchez}, S.~F., {Gil de Paz},
  A., {V{\'{\i}}lchez}, J., {Miralles-Caballero}, D., {Kehrig}, C.,
  {P{\'e}rez-Montero}, E., {Stanishev}, V., {Iglesias-P{\'a}ramo}, J.,
  {D{\'{\i}}az}, A.~I., {Castillo-Morales}, A., {Kennicutt}, R.,
  {L{\'o}pez-S{\'a}nchez}, A.~R., {Galbany}, L., {Garc{\'{\i}}a-Benito}, R.,
  {Mast}, D., {Mendez-Abreu}, J., {Monreal-Ibero}, A., {Husemann}, B.,
  {Walcher}, C.~J., {Garc{\'{\i}}a-Lorenzo}, B., {Masegosa}, J., {Del Olmo
  Orozco}, A., {Mour{\~a}o}, A.~M., {Ziegler}, B., {Moll{\'a}}, M.,
  {Papaderos}, P., {S{\'a}nchez-Bl{\'a}zquez}, P., {Gonz{\'a}lez Delgado},
  R.~M., {Falc{\'o}n-Barroso}, J., {Roth}, M.~M., {van de Ven}, G., \& {Califa
  Team}. 2013, \aap, 559, A114


\bibitem[{{Martig} {et~al.}(2009){Martig}, {Bournaud}, {Teyssier}, \&
  {Dekel}}]{martig09}
{Martig}, M., {Bournaud}, F., {Teyssier}, R., \& {Dekel}, A. 2009, \apj, 707,
  250


\bibitem[{{Martin} {et~al.}(2007){Martin}, {Wyder}, {Schiminovich}, {Barlow},
  {Forster}, {Friedman}, {Morrissey}, {Neff}, {Seibert}, {Small}, {Welsh},
  {Bianchi}, {Donas}, {Heckman}, {Lee}, {Madore}, {Milliard}, {Rich}, {Szalay},
  \& {Yi}}]{Martin+2007}
{Martin}, D.~C., {Wyder}, T.~K., {Schiminovich}, D., {Barlow}, T.~A.,
  {Forster}, K., {Friedman}, P.~G., {Morrissey}, P., {Neff}, S.~G., {Seibert},
  M., {Small}, T., {Welsh}, B.~Y., {Bianchi}, L., {Donas}, J., {Heckman},
  T.~M., {Lee}, Y.-W., {Madore}, B.~F., {Milliard}, B., {Rich}, R.~M.,
  {Szalay}, A.~S., \& {Yi}, S.~K. 2007, \apjs, 173, 342


\bibitem[{{McIntosh} {et~al.}(2014){McIntosh}, {Wagner}, {Cooper}, {Bell},
  {Kere{\v s}}, {Bosch}, {Gallazzi}, {Haines}, {Mann}, {Pasquali}, \&
  {Christian}}]{McIntosh+2014}
{McIntosh}, D.~H., {Wagner}, C., {Cooper}, A., {Bell}, E.~F., {Kere{\v s}}, D.,
  {Bosch}, F.~C.~v.~d., {Gallazzi}, A., {Haines}, T., {Mann}, J., {Pasquali},
  A., \& {Christian}, A.~M. 2014, \mnras, 442, 533


\bibitem[{{Men{\'e}ndez-Delmestre} {et~al.}(2007){Men{\'e}ndez-Delmestre},
  {Sheth}, {Schinnerer}, {Jarrett}, \& {Scoville}}]{Menendez+2007}
{Men{\'e}ndez-Delmestre}, K., {Sheth}, K., {Schinnerer}, E., {Jarrett}, T.~H.,
  \& {Scoville}, N.~Z. 2007, \apj, 657, 790


\bibitem[{{Mingo} {et~al.}(2012){Mingo}, {Hardcastle}, {Croston}, {Evans},
  {Kharb}, {Kraft}, \& {Lenc}}]{mingo12}
{Mingo}, B., {Hardcastle}, M.~J., {Croston}, J.~H., {Evans}, D.~A., {Kharb},
  P., {Kraft}, R.~P., \& {Lenc}, E. 2012, \apj, 758, 95


\bibitem[{{Morisset} {et~al.}(2016){Morisset}, {Delgado-Inglada},
  {S{\'a}nchez}, {Galbany}, {Garc{\'{\i}}a-Benito}, {Husemann}, {Marino},
  {Mast}, \& {Roth}}]{mori16}
{Morisset}, C., {Delgado-Inglada}, G., {S{\'a}nchez}, S.~F., {Galbany}, L.,
  {Garc{\'{\i}}a-Benito}, R., {Husemann}, B., {Marino}, R.~A., {Mast}, D., \&
  {Roth}, M.~M. 2016, \aap, 594, A37


\bibitem[{{Nair} \& {Abraham}(2010)}]{nair2010}
{Nair}, P.~B. \& {Abraham}, R.~G. 2010, \apjs, 186, 427


\bibitem[{{Narayanan} {et~al.}(2012){Narayanan}, {Krumholz}, {Ostriker}, \&
  {Hernquist}}]{nara12}
{Narayanan}, D., {Krumholz}, M.~R., {Ostriker}, E.~C., \& {Hernquist}, L. 2012,
  \mnras, 421, 3127


\bibitem[{{Noeske} {et~al.}(2007){Noeske}, {Weiner}, {Faber}, {Papovich},
  {Koo}, {Somerville}, {Bundy}, {Conselice}, {Newman}, {Schiminovich}, {Le
  Floc'h}, {Coil}, {Rieke}, {Lotz}, {Primack}, {Barmby}, {Cooper}, {Davis},
  {Ellis}, {Fazio}, {Guhathakurta}, {Huang}, {Kassin}, {Martin}, {Phillips},
  {Rich}, {Small}, {Willmer}, \& {Wilson}}]{Noeske+2007}
{Noeske}, K.~G., {Weiner}, B.~J., {Faber}, S.~M., {Papovich}, C., {Koo}, D.~C.,
  {Somerville}, R.~S., {Bundy}, K., {Conselice}, C.~J., {Newman}, J.~A.,
  {Schiminovich}, D., {Le Floc'h}, E., {Coil}, A.~L., {Rieke}, G.~H., {Lotz},
  J.~M., {Primack}, J.~R., {Barmby}, P., {Cooper}, M.~C., {Davis}, M., {Ellis},
  R.~S., {Fazio}, G.~G., {Guhathakurta}, P., {Huang}, J., {Kassin}, S.~A.,
  {Martin}, D.~C., {Phillips}, A.~C., {Rich}, R.~M., {Small}, T.~A., {Willmer},
  C.~N.~A., \& {Wilson}, G. 2007, \apjl, 660, L43


\bibitem[{{Ortega-Minakata}(2015)}]{ortega-minakata15}
{Ortega-Minakata}, R.~A. 2015, PhD thesis, Universidad de Guanajuato


\bibitem[{{Osterbrock}(1989)}]{osterbrock89}
{Osterbrock}, D.~E. 1989, {Astrophysics of gaseous nebulae and active galactic
  nuclei} (University Science Books)


\bibitem[{{Papaderos} {et~al.}(2013){Papaderos}, {Gomes}, {V{\'{\i}}lchez},
  {Kehrig}, {Lehnert}, {Ziegler}, {S{\'a}nchez}, {Husemann}, {Monreal-Ibero},
  {Garc{\'{\i}}a-Benito}, {Bland-Hawthorn}, {Cortijo-Ferrero}, {de
  Lorenzo-C{\'a}ceres}, {del Olmo}, {Falc{\'o}n-Barroso}, {Galbany},
  {Iglesias-P{\'a}ramo}, {L{\'o}pez-S{\'a}nchez}, {Marquez}, {Moll{\'a}},
  {Mast}, {van de Ven}, \& {Wisotzki}}]{papa13}
{Papaderos}, P., {Gomes}, J.~M., {V{\'{\i}}lchez}, J.~M., {Kehrig}, C.,
  {Lehnert}, M.~D., {Ziegler}, B., {S{\'a}nchez}, S.~F., {Husemann}, B.,
  {Monreal-Ibero}, A., {Garc{\'{\i}}a-Benito}, R., {Bland-Hawthorn}, J.,
  {Cortijo-Ferrero}, C., {de Lorenzo-C{\'a}ceres}, A., {del Olmo}, A.,
  {Falc{\'o}n-Barroso}, J., {Galbany}, L., {Iglesias-P{\'a}ramo}, J.,
  {L{\'o}pez-S{\'a}nchez}, {\'A}.~R., {Marquez}, I., {Moll{\'a}}, M., {Mast},
  D., {van de Ven}, G., \& {Wisotzki}, L. 2013, \aap, 555, L1


\bibitem[{{Parma} {et~al.}(2007){Parma}, {Murgia}, {de Ruiter}, {Fanti},
  {Mack}, \& {Govoni}}]{parma07}
{Parma}, P., {Murgia}, M., {de Ruiter}, H.~R., {Fanti}, R., {Mack}, K.-H., \&
  {Govoni}, F. 2007, \aap, 470, 875


\bibitem[{{Peimbert} \& {Peimbert}(2006)}]{2006RMxAC..26R.163P}
{Peimbert}, M. \& {Peimbert}, A. 2006, in Revista Mexicana de Astronomia y
  Astrofisica Conference Series, Vol.~26, Revista Mexicana de Astronomia y
  Astrofisica Conference Series, 163


\bibitem[{{P{\'e}rez} {et~al.}(2013){P{\'e}rez}, {Cid Fernandes}, {Gonz{\'a}lez
  Delgado}, {Garc{\'{\i}}a-Benito}, {S{\'a}nchez}, {Husemann}, {Mast},
  {Rod{\'o}n}, {Kupko}, {Backsmann}, {de Amorim}, {van de Ven}, {Walcher},
  {Wisotzki}, {Cortijo-Ferrero}, \& {collaboration6}}]{perez13}
{P{\'e}rez}, E., {Cid Fernandes}, R., {Gonz{\'a}lez Delgado}, R.~M.,
  {Garc{\'{\i}}a-Benito}, R., {S{\'a}nchez}, S.~F., {Husemann}, B., {Mast}, D.,
  {Rod{\'o}n}, J.~R., {Kupko}, D., {Backsmann}, N., {de Amorim}, A.~L., {van de
  Ven}, G., {Walcher}, J., {Wisotzki}, L., {Cortijo-Ferrero}, C., \&
  {collaboration6}, C. 2013, \apjl, 764, L1


\bibitem[{{P{\'e}rez-Gonz{\'a}lez} {et~al.}(2008){P{\'e}rez-Gonz{\'a}lez},
  {Rieke}, {Villar}, {Barro}, {Blaylock}, {Egami}, {Gallego}, {Gil de Paz},
  {Pascual}, {Zamorano}, \& {Donley}}]{per-gon08}
{P{\'e}rez-Gonz{\'a}lez}, P.~G., {Rieke}, G.~H., {Villar}, V., {Barro}, G.,
  {Blaylock}, M., {Egami}, E., {Gallego}, J., {Gil de Paz}, A., {Pascual}, S.,
  {Zamorano}, J., \& {Donley}, J.~L. 2008, \apj, 675, 234


\bibitem[{{Peterson} {et~al.}(2004){Peterson}, {Ferrarese}, {Gilbert}, {Kaspi},
  {Malkan}, {Maoz}, {Merritt}, {Netzer}, {Onken}, {Pogge}, {Vestergaard}, \&
  {Wandel}}]{peterson04}
{Peterson}, B.~M., {Ferrarese}, L., {Gilbert}, K.~M., {Kaspi}, S., {Malkan},
  M.~A., {Maoz}, D., {Merritt}, D., {Netzer}, H., {Onken}, C.~A., {Pogge},
  R.~W., {Vestergaard}, M., \& {Wandel}, A. 2004, \apj, 613, 682


\bibitem[{{Pilyugin} {et~al.}(2010){Pilyugin}, {V{\'{\i}}lchez}, \&
  {Thuan}}]{pilyugin10}
{Pilyugin}, L.~S., {V{\'{\i}}lchez}, J.~M., \& {Thuan}, T.~X. 2010, \apj, 720,
  1738


\bibitem[{{Renzini} \& {Peng}(2015)}]{Renzini15}
{Renzini}, A. \& {Peng}, Y.-j. 2015, \apjl, 801, L29


\bibitem[{{Rich} {et~al.}(2010){Rich}, {Dopita}, {Kewley}, \& {Rupke}}]{rich10}
{Rich}, J.~A., {Dopita}, M.~A., {Kewley}, L.~J., \& {Rupke}, D.~S.~N. 2010,
  \apj, 721, 505


\bibitem[{{Rodr{\'{\i}}guez-Puebla} {et~al.}(2017){Rodr{\'{\i}}guez-Puebla},
  {Primack}, {Avila-Reese}, \& {Faber}}]{Rodriguez-Puebla+2017}
{Rodr{\'{\i}}guez-Puebla}, A., {Primack}, J.~R., {Avila-Reese}, V., \& {Faber},
  S.~M. 2017, \mnras, 470, 651


\bibitem[{{Rosas-Guevara} {et~al.}(2016){Rosas-Guevara}, {Bower}, {Schaye},
  {McAlpine}, {Dalla Vecchia}, {Frenk}, {Schaller}, \&
  {Theuns}}]{Rosas-Guevara+2016}
{Rosas-Guevara}, Y., {Bower}, R.~G., {Schaye}, J., {McAlpine}, S., {Dalla
  Vecchia}, C., {Frenk}, C.~S., {Schaller}, M., \& {Theuns}, T. 2016, \mnras,
  462, 190


\bibitem[{{Saintonge} {et~al.}(2011){Saintonge}, {Kauffmann}, {Kramer},
  {Tacconi}, {Buchbender}, {Catinella}, {Fabello}, {Graci{\'a}-Carpio}, {Wang},
  {Cortese}, {Fu}, {Genzel}, {Giovanelli}, {Guo}, {Haynes}, {Heckman},
  {Krumholz}, {Lemonias}, {Li}, {Moran}, {Rodriguez-Fernandez}, {Schiminovich},
  {Schuster}, \& {Sievers}}]{Saintonge+2011}
{Saintonge}, A., {Kauffmann}, G., {Kramer}, C., {Tacconi}, L.~J., {Buchbender},
  C., {Catinella}, B., {Fabello}, S., {Graci{\'a}-Carpio}, J., {Wang}, J.,
  {Cortese}, L., {Fu}, J., {Genzel}, R., {Giovanelli}, R., {Guo}, Q., {Haynes},
  M.~P., {Heckman}, T.~M., {Krumholz}, M.~R., {Lemonias}, J., {Li}, C.,
  {Moran}, S., {Rodriguez-Fernandez}, N., {Schiminovich}, D., {Schuster}, K.,
  \& {Sievers}, A. 2011, \mnras, 415, 32


\bibitem[{{Salim} {et~al.}(2007){Salim}, {Rich}, {Charlot}, {Brinchmann},
  {Johnson}, {Schiminovich}, {Seibert}, {Mallery}, {Heckman}, {Forster},
  {Friedman}, {Martin}, {Morrissey}, {Neff}, {Small}, {Wyder}, {Bianchi},
  {Donas}, {Lee}, {Madore}, {Milliard}, {Szalay}, {Welsh}, \& {Yi}}]{Salim07}
{Salim}, S., {Rich}, R.~M., {Charlot}, S., {Brinchmann}, J., {Johnson}, B.~D.,
  {Schiminovich}, D., {Seibert}, M., {Mallery}, R., {Heckman}, T.~M.,
  {Forster}, K., {Friedman}, P.~G., {Martin}, D.~C., {Morrissey}, P., {Neff},
  S.~G., {Small}, T., {Wyder}, T.~K., {Bianchi}, L., {Donas}, J., {Lee}, Y.-W.,
  {Madore}, B.~F., {Milliard}, B., {Szalay}, A.~S., {Welsh}, B.~Y., \& {Yi},
  S.~K. 2007, \apjs, 173, 267


\bibitem[{{S{\'a}nchez} {et~al.}(2017){S{\'a}nchez}, {Barrera-Ballesteros},
  {S{\'a}nchez-Menguiano}, {Walcher}, {Marino}, {Galbany}, {Bland-Hawthorn},
  {Cano-D{\'{\i}}az}, {Garc{\'{\i}}a-Benito}, {L{\'o}pez-Cob{\'a}}, {Zibetti},
  {Vilchez}, {Igl{\'e}sias-P{\'a}ramo}, {Kehrig}, {L{\'o}pez S{\'a}nchez},
  {Duarte Puertas}, \& {Ziegler}}]{sanchez2017a}
{S{\'a}nchez}, S.~F., {Barrera-Ballesteros}, J.~K., {S{\'a}nchez-Menguiano},
  L., {Walcher}, C.~J., {Marino}, R.~A., {Galbany}, L., {Bland-Hawthorn}, J.,
  {Cano-D{\'{\i}}az}, M., {Garc{\'{\i}}a-Benito}, R., {L{\'o}pez-Cob{\'a}}, C.,
  {Zibetti}, S., {Vilchez}, J.~M., {Igl{\'e}sias-P{\'a}ramo}, J., {Kehrig}, C.,
  {L{\'o}pez S{\'a}nchez}, A.~R., {Duarte Puertas}, S., \& {Ziegler}, B. 2017,
  \mnras, 469, 2121


\bibitem[{{S{\'a}nchez} {et~al.}(2007b){S{\'a}nchez}, {Cardiel}, {Verheijen},
  {Pedraz}, \& {Covone}}]{sanchez07b}
{S{\'a}nchez}, S.~F., {Cardiel}, N., {Verheijen}, M.~A.~W., {Pedraz}, S., \&
  {Covone}, G. 2007b, \mnras, 376, 125


\bibitem[{{S{\'a}nchez} {et~al.}(2004{\natexlab{a}}){S{\'a}nchez},
  {Garcia-Lorenzo}, {Mediavilla}, {Gonz{\'a}lez-Serrano}, \&
  {Christensen}}]{Sanchez2004}
{S{\'a}nchez}, S.~F., {Garcia-Lorenzo}, B., {Mediavilla}, E.,
  {Gonz{\'a}lez-Serrano}, J.~I., \& {Christensen}, L. 2004{\natexlab{a}}, \apj,
  615, 156


\bibitem[{{S{\'a}nchez} \& {Gonz{\'a}lez-Serrano}(2002)}]{sanchez02}
{S{\'a}nchez}, S.~F. \& {Gonz{\'a}lez-Serrano}, J.~I. 2002, \aap, 396, 773


\bibitem[{{S{\'a}nchez} \& {Gonz{\'a}lez-Serrano}(2003)}]{sanchez03}
---. 2003, \aap, 406, 435


\bibitem[{{S{\'a}nchez} {et~al.}(2004{\natexlab{b}}){S{\'a}nchez}, {Jahnke},
  {Wisotzki}, {McIntosh}, {Bell}, {Barden}, {Beckwith}, {Borch}, {Caldwell},
  {H{\"a}ussler}, {Jogee}, {Meisenheimer}, {Peng}, {Rix}, {Somerville}, \&
  {Wolf}}]{sanchez04}
{S{\'a}nchez}, S.~F., {Jahnke}, K., {Wisotzki}, L., {McIntosh}, D.~H., {Bell},
  E.~F., {Barden}, M., {Beckwith}, S.~V.~W., {Borch}, A., {Caldwell}, J.~A.~R.,
  {H{\"a}ussler}, B., {Jogee}, S., {Meisenheimer}, K., {Peng}, C.~Y., {Rix},
  H., {Somerville}, R.~S., \& {Wolf}, C. 2004{\natexlab{b}}, \apj, 614, 586


\bibitem[{{S{\'a}nchez} {et~al.}(2016{\natexlab{a}}){S{\'a}nchez}, {P{\'e}rez},
  {S{\'a}nchez-Bl{\'a}zquez}, {Garc{\'{\i}}a-Benito}, {Ibarra-Mede},
  {Gonz{\'a}lez}, {Rosales-Ortega}, {S{\'a}nchez-Menguiano}, {Ascasibar},
  {Bitsakis}, {Law}, {Cano-D{\'{\i}}az}, {L{\'o}pez-Cob{\'a}}, {Marino}, {Gil
  de Paz}, {L{\'o}pez-S{\'a}nchez}, {Barrera-Ballesteros}, {Galbany}, {Mast},
  {Abril-Melgarejo}, \& {Roman-Lopes}}]{Pipe3D_II}
{S{\'a}nchez}, S.~F., {P{\'e}rez}, E., {S{\'a}nchez-Bl{\'a}zquez}, P.,
  {Garc{\'{\i}}a-Benito}, R., {Ibarra-Mede}, H.~J., {Gonz{\'a}lez}, J.~J.,
  {Rosales-Ortega}, F.~F., {S{\'a}nchez-Menguiano}, L., {Ascasibar}, Y.,
  {Bitsakis}, T., {Law}, D., {Cano-D{\'{\i}}az}, M., {L{\'o}pez-Cob{\'a}}, C.,
  {Marino}, R.~A., {Gil de Paz}, A., {L{\'o}pez-S{\'a}nchez}, A.~R.,
  {Barrera-Ballesteros}, J., {Galbany}, L., {Mast}, D., {Abril-Melgarejo}, V.,
  \& {Roman-Lopes}, A. 2016{\natexlab{a}}, \rmxaa, 52, 171


\bibitem[{{S{\'a}nchez} {et~al.}(2016{\natexlab{b}}){S{\'a}nchez}, {P{\'e}rez},
  {S{\'a}nchez-Bl{\'a}zquez}, {Gonz{\'a}lez}, {Ros{\'a}lez-Ortega},
  {Cano-D{\'{\i}} az}, {L{\'o}pez-Cob{\'a}}, {Marino}, {Gil de Paz},
  {Moll{\'a}}, {L{\'o}pez-S{\'a}nchez}, {Ascasibar}, \&
  {Barrera-Ballesteros}}]{Pipe3D_I}
{S{\'a}nchez}, S.~F., {P{\'e}rez}, E., {S{\'a}nchez-Bl{\'a}zquez}, P.,
  {Gonz{\'a}lez}, J.~J., {Ros{\'a}lez-Ortega}, F.~F., {Cano-D{\'{\i}} az}, M.,
  {L{\'o}pez-Cob{\'a}}, C., {Marino}, R.~A., {Gil de Paz}, A., {Moll{\'a}}, M.,
  {L{\'o}pez-S{\'a}nchez}, A.~R., {Ascasibar}, Y., \& {Barrera-Ballesteros}, J.
  2016{\natexlab{b}}, \rmxaa, 52, 21


\bibitem[{{S{\'a}nchez} {et~al.}(2014){S{\'a}nchez}, {Rosales-Ortega},
  {Iglesias-P{\'a}ramo}, {Moll{\'a}}, {Barrera-Ballesteros}, {Marino},
  {P{\'e}rez}, {S{\'a}nchez-Blazquez}, {Gonz{\'a}lez Delgado}, {Cid Fernandes},
  {de Lorenzo-C{\'a}ceres}, {Mendez-Abreu}, {Galbany}, {Falcon-Barroso},
  {Miralles-Caballero}, {Husemann}, {Garc{\'{\i}}a-Benito}, {Mast}, {Walcher},
  {Gil de Paz}, {Garc{\'{\i}}a-Lorenzo}, {Jungwiert}, {V{\'{\i}}lchez},
  {J{\'{\i}}lkov{\'a}}, {Lyubenova}, {Cortijo-Ferrero}, {D{\'{\i}}az},
  {Wisotzki}, {M{\'a}rquez}, {Bland-Hawthorn}, {Ellis}, {van de Ven}, {Jahnke},
  {Papaderos}, {Gomes}, {Mendoza}, \& {L{\'o}pez-S{\'a}nchez}}]{sanchez14}
{S{\'a}nchez}, S.~F., {Rosales-Ortega}, F.~F., {Iglesias-P{\'a}ramo}, J.,
  {Moll{\'a}}, M., {Barrera-Ballesteros}, J., {Marino}, R.~A., {P{\'e}rez}, E.,
  {S{\'a}nchez-Blazquez}, P., {Gonz{\'a}lez Delgado}, R., {Cid Fernandes}, R.,
  {de Lorenzo-C{\'a}ceres}, A., {Mendez-Abreu}, J., {Galbany}, L.,
  {Falcon-Barroso}, J., {Miralles-Caballero}, D., {Husemann}, B.,
  {Garc{\'{\i}}a-Benito}, R., {Mast}, D., {Walcher}, C.~J., {Gil de Paz}, A.,
  {Garc{\'{\i}}a-Lorenzo}, B., {Jungwiert}, B., {V{\'{\i}}lchez}, J.~M.,
  {J{\'{\i}}lkov{\'a}}, L., {Lyubenova}, M., {Cortijo-Ferrero}, C.,
  {D{\'{\i}}az}, A.~I., {Wisotzki}, L., {M{\'a}rquez}, I., {Bland-Hawthorn},
  J., {Ellis}, S., {van de Ven}, G., {Jahnke}, K., {Papaderos}, P., {Gomes},
  J.~M., {Mendoza}, M.~A., \& {L{\'o}pez-S{\'a}nchez}, {\'A}.~R. 2014, \aap,
  563, A49


\bibitem[{{S{\'a}nchez} {et~al.}(2013){S{\'a}nchez}, {Rosales-Ortega},
  {Jungwiert}, {Iglesias-P{\'a}ramo}, {V{\'{\i}}lchez}, {Marino}, {Walcher},
  {Husemann}, {Mast}, {Monreal-Ibero}, {Cid Fernandes}, {P{\'e}rez},
  {Gonz{\'a}lez Delgado}, {Garc{\'{\i}}a-Benito}, {Galbany}, {van de Ven},
  {Jahnke}, {Flores}, {Bland-Hawthorn}, {L{\'o}pez-S{\'a}nchez}, {Stanishev},
  {Miralles-Caballero}, {D{\'{\i}}az}, {S{\'a}nchez-Blazquez}, {Moll{\'a}},
  {Gallazzi}, {Papaderos}, {Gomes}, {Gruel}, {P{\'e}rez}, {Ruiz-Lara},
  {Florido}, {de Lorenzo-C{\'a}ceres}, {Mendez-Abreu}, {Kehrig}, {Roth},
  {Ziegler}, {Alves}, {Wisotzki}, {Kupko}, {Quirrenbach}, {Bomans}, \& {Califa
  Collaboration}}]{sanchez13}
{S{\'a}nchez}, S.~F., {Rosales-Ortega}, F.~F., {Jungwiert}, B.,
  {Iglesias-P{\'a}ramo}, J., {V{\'{\i}}lchez}, J.~M., {Marino}, R.~A.,
  {Walcher}, C.~J., {Husemann}, B., {Mast}, D., {Monreal-Ibero}, A., {Cid
  Fernandes}, R., {P{\'e}rez}, E., {Gonz{\'a}lez Delgado}, R.,
  {Garc{\'{\i}}a-Benito}, R., {Galbany}, L., {van de Ven}, G., {Jahnke}, K.,
  {Flores}, H., {Bland-Hawthorn}, J., {L{\'o}pez-S{\'a}nchez}, A.~R.,
  {Stanishev}, V., {Miralles-Caballero}, D., {D{\'{\i}}az}, A.~I.,
  {S{\'a}nchez-Blazquez}, P., {Moll{\'a}}, M., {Gallazzi}, A., {Papaderos}, P.,
  {Gomes}, J.~M., {Gruel}, N., {P{\'e}rez}, I., {Ruiz-Lara}, T., {Florido}, E.,
  {de Lorenzo-C{\'a}ceres}, A., {Mendez-Abreu}, J., {Kehrig}, C., {Roth},
  M.~M., {Ziegler}, B., {Alves}, J., {Wisotzki}, L., {Kupko}, D.,
  {Quirrenbach}, A., {Bomans}, D., \& {Califa Collaboration}. 2013, \aap, 554,
  A58


\bibitem[{{S{\'a}nchez} {et~al.}(2012){S{\'a}nchez}, {Rosales-Ortega},
  {Marino}, {Iglesias-P{\'a}ramo}, {V{\'{\i}}lchez}, {Kennicutt},
  {D{\'{\i}}az}, {Mast}, {Monreal-Ibero}, {Garc{\'{\i}}a-Benito},
  {Bland-Hawthorn}, {P{\'e}rez}, {Gonz{\'a}lez Delgado}, {Husemann},
  {L{\'o}pez-S{\'a}nchez}, {Cid Fernandes}, {Kehrig}, {Walcher}, {Gil de Paz},
  \& {Ellis}}]{sanchez12b}
{S{\'a}nchez}, S.~F., {Rosales-Ortega}, F.~F., {Marino}, R.~A.,
  {Iglesias-P{\'a}ramo}, J., {V{\'{\i}}lchez}, J.~M., {Kennicutt}, R.~C.,
  {D{\'{\i}}az}, A.~I., {Mast}, D., {Monreal-Ibero}, A.,
  {Garc{\'{\i}}a-Benito}, R., {Bland-Hawthorn}, J., {P{\'e}rez}, E.,
  {Gonz{\'a}lez Delgado}, R., {Husemann}, B., {L{\'o}pez-S{\'a}nchez},
  {\'A}.~R., {Cid Fernandes}, R., {Kehrig}, C., {Walcher}, C.~J., {Gil de Paz},
  A., \& {Ellis}, S. 2012, \aap, 546, A2


\bibitem[{{S{\'a}nchez-Menguiano} {et~al.}(2016){S{\'a}nchez-Menguiano},
  {S{\'a}nchez}, {P{\'e}rez}, {Garc{\'{\i}}a-Benito}, {Husemann}, {Mast},
  {Mendoza}, {Ruiz-Lara}, {Ascasibar}, {Bland-Hawthorn}, {Cavichia},
  {D{\'{\i}}az}, {Florido}, {Galbany}, {G{\'o}nzalez Delgado}, {Kehrig},
  {Marino}, {M{\'a}rquez}, {Masegosa}, {M{\'e}ndez-Abreu}, {Moll{\'a}}, {Del
  Olmo}, {P{\'e}rez}, {S{\'a}nchez-Bl{\'a}zquez}, {Stanishev}, {Walcher},
  {L{\'o}pez-S{\'a}nchez}, \& {Califa Collaboration}}]{laura16}
{S{\'a}nchez-Menguiano}, L., {S{\'a}nchez}, S.~F., {P{\'e}rez}, I.,
  {Garc{\'{\i}}a-Benito}, R., {Husemann}, B., {Mast}, D., {Mendoza}, A.,
  {Ruiz-Lara}, T., {Ascasibar}, Y., {Bland-Hawthorn}, J., {Cavichia}, O.,
  {D{\'{\i}}az}, A.~I., {Florido}, E., {Galbany}, L., {G{\'o}nzalez Delgado},
  R.~M., {Kehrig}, C., {Marino}, R.~A., {M{\'a}rquez}, I., {Masegosa}, J.,
  {M{\'e}ndez-Abreu}, J., {Moll{\'a}}, M., {Del Olmo}, A., {P{\'e}rez}, E.,
  {S{\'a}nchez-Bl{\'a}zquez}, P., {Stanishev}, V., {Walcher}, C.~J.,
  {L{\'o}pez-S{\'a}nchez}, {\'A}.~R., \& {Califa Collaboration}. 2016, \aap,
  587, A70


\bibitem[{{Sanders} \& {Mirabel}(1996)}]{sanders96}
{Sanders}, D.~B. \& {Mirabel}, I.~F. 1996, \araa, 34, 749


\bibitem[{{Sanders} {et~al.}(1988){Sanders}, {Soifer}, {Elias}, {Madore},
  {Matthews}, {Neugebauer}, \& {Scoville}}]{Sanders+1988}
{Sanders}, D.~B., {Soifer}, B.~T., {Elias}, J.~H., {Madore}, B.~F., {Matthews},
  K., {Neugebauer}, G., \& {Scoville}, N.~Z. 1988, \apj, 325, 74


\bibitem[{{Sarzi} {et~al.}(2010){Sarzi}, {Shields}, {Schawinski}, {Jeong},
  {Shapiro}, {Bacon}, {Bureau}, {Cappellari}, {Davies}, {de Zeeuw}, {Emsellem},
  {Falc{\'o}n-Barroso}, {Krajnovi{\'c}}, {Kuntschner}, {McDermid}, {Peletier},
  {van den Bosch}, {van de Ven}, \& {Yi}}]{sarzi10}
{Sarzi}, M., {Shields}, J.~C., {Schawinski}, K., {Jeong}, H., {Shapiro}, K.,
  {Bacon}, R., {Bureau}, M., {Cappellari}, M., {Davies}, R.~L., {de Zeeuw},
  P.~T., {Emsellem}, E., {Falc{\'o}n-Barroso}, J., {Krajnovi{\'c}}, D.,
  {Kuntschner}, H., {McDermid}, R.~M., {Peletier}, R.~F., {van den Bosch},
  R.~C.~E., {van de Ven}, G., \& {Yi}, S.~K. 2010, \mnras, 402, 2187


\bibitem[{{Schawinski} {et~al.}(2009){Schawinski}, {Lintott}, {Thomas},
  {Sarzi}, {Andreescu}, {Bamford}, {Kaviraj}, {Khochfar}, {Land}, {Murray},
  {Nichol}, {Raddick}, {Slosar}, {Szalay}, {Vandenberg}, \&
  {Yi}}]{Schawinski+2009}
{Schawinski}, K., {Lintott}, C., {Thomas}, D., {Sarzi}, M., {Andreescu}, D.,
  {Bamford}, S.~P., {Kaviraj}, S., {Khochfar}, S., {Land}, K., {Murray}, P.,
  {Nichol}, R.~C., {Raddick}, M.~J., {Slosar}, A., {Szalay}, A., {Vandenberg},
  J., \& {Yi}, S.~K. 2009, \mnras, 396, 818


\bibitem[{{Schawinski} {et~al.}(2014){Schawinski}, {Urry}, {Simmons},
  {Fortson}, {Kaviraj}, {Keel}, {Lintott}, {Masters}, {Nichol}, {Sarzi},
  {Skibba}, {Treister}, {Willett}, {Wong}, \& {Yi}}]{Schawinski+2014}
{Schawinski}, K., {Urry}, C.~M., {Simmons}, B.~D., {Fortson}, L., {Kaviraj},
  S., {Keel}, W.~C., {Lintott}, C.~J., {Masters}, K.~L., {Nichol}, R.~C.,
  {Sarzi}, M., {Skibba}, R., {Treister}, E., {Willett}, K.~W., {Wong}, O.~I.,
  \& {Yi}, S.~K. 2014, \mnras, 440, 889


\bibitem[{{Schawinski} {et~al.}(2010){Schawinski}, {Urry}, {Virani}, {Coppi},
  {Bamford}, {Treister}, {Lintott}, {Sarzi}, {Keel}, {Kaviraj}, {Cardamone},
  {Masters}, {Ross}, {Andreescu}, {Murray}, {Nichol}, {Raddick}, {Slosar},
  {Szalay}, {Thomas}, \& {Vandenberg}}]{Schawinski+2010}
{Schawinski}, K., {Urry}, C.~M., {Virani}, S., {Coppi}, P., {Bamford}, S.~P.,
  {Treister}, E., {Lintott}, C.~J., {Sarzi}, M., {Keel}, W.~C., {Kaviraj}, S.,
  {Cardamone}, C.~N., {Masters}, K.~L., {Ross}, N.~P., {Andreescu}, D.,
  {Murray}, P., {Nichol}, R.~C., {Raddick}, M.~J., {Slosar}, A., {Szalay},
  A.~S., {Thomas}, D., \& {Vandenberg}, J. 2010, \apj, 711, 284


\bibitem[{{Schiminovich} {et~al.}(2007){Schiminovich}, {Wyder}, {Martin},
  {Johnson}, {Salim}, {Seibert}, {Treyer}, {Budav{\'a}ri}, {Hoopes},
  {Zamojski}, {Barlow}, {Forster}, {Friedman}, {Morrissey}, {Neff}, {Small},
  {Bianchi}, {Donas}, {Heckman}, {Lee}, {Madore}, {Milliard}, {Rich}, {Szalay},
  {Welsh}, \& {Yi}}]{Schiminovich+2007}
{Schiminovich}, D., {Wyder}, T.~K., {Martin}, D.~C., {Johnson}, B.~D., {Salim},
  S., {Seibert}, M., {Treyer}, M.~A., {Budav{\'a}ri}, T., {Hoopes}, C.,
  {Zamojski}, M., {Barlow}, T.~A., {Forster}, K.~G., {Friedman}, P.~G.,
  {Morrissey}, P., {Neff}, S.~G., {Small}, T.~A., {Bianchi}, L., {Donas}, J.,
  {Heckman}, T.~M., {Lee}, Y.-W., {Madore}, B.~F., {Milliard}, B., {Rich},
  R.~M., {Szalay}, A.~S., {Welsh}, B.~Y., \& {Yi}, S. 2007, \apjs, 173, 315


\bibitem[{{Schmidt}(1959)}]{schmidt59}
{Schmidt}, M. 1959, \apj, 129, 243


\bibitem[{{Sersic}(1968)}]{sersic68}
{Sersic}, J.~L. 1968, {Atlas de galaxias australes}


\bibitem[{{Sheth} {et~al.}(2008){Sheth}, {Elmegreen}, {Elmegreen}, {Capak},
  {Abraham}, {Athanassoula}, {Ellis}, {Mobasher}, {Salvato}, {Schinnerer},
  {Scoville}, {Spalsbury}, {Strubbe}, {Carollo}, {Rich}, \&
  {West}}]{Sheth+2008}
{Sheth}, K., {Elmegreen}, D.~M., {Elmegreen}, B.~G., {Capak}, P., {Abraham},
  R.~G., {Athanassoula}, E., {Ellis}, R.~S., {Mobasher}, B., {Salvato}, M.,
  {Schinnerer}, E., {Scoville}, N.~Z., {Spalsbury}, L., {Strubbe}, L.,
  {Carollo}, M., {Rich}, M., \& {West}, A.~A. 2008, \apj, 675, 1141


\bibitem[{{Shulevski} {et~al.}(2015){Shulevski}, {Morganti}, {Barthel},
  {Harwood}, {Brunetti}, {van Weeren}, {R{\"o}ttgering}, {White}, {Horellou},
  {Kunert-Bajraszewska}, {Jamrozy}, {Chyzy}, {Mahony}, {Miley}, {Brienza},
  {B{\^i}rzan}, {Rafferty}, {Br{\"u}ggen}, {Wise}, {Conway}, {de Gasperin}, \&
  {Vilchez}}]{shu15}
{Shulevski}, A., {Morganti}, R., {Barthel}, P.~D., {Harwood}, J.~J.,
  {Brunetti}, G., {van Weeren}, R.~J., {R{\"o}ttgering}, H.~J.~A., {White},
  G.~J., {Horellou}, C., {Kunert-Bajraszewska}, M., {Jamrozy}, M., {Chyzy},
  K.~T., {Mahony}, E., {Miley}, G., {Brienza}, M., {B{\^i}rzan}, L.,
  {Rafferty}, D.~A., {Br{\"u}ggen}, M., {Wise}, M.~W., {Conway}, J., {de
  Gasperin}, F., \& {Vilchez}, N. 2015, \aap, 583, A89


\bibitem[{{Sijacki} {et~al.}(2015){Sijacki}, {Vogelsberger}, {Genel},
  {Springel}, {Torrey}, {Snyder}, {Nelson}, \& {Hernquist}}]{Sijacki+2015}
{Sijacki}, D., {Vogelsberger}, M., {Genel}, S., {Springel}, V., {Torrey}, P.,
  {Snyder}, G.~F., {Nelson}, D., \& {Hernquist}, L. 2015, \mnras, 452, 575


\bibitem[{{Silk}(2005)}]{Silk2005}
{Silk}, J. 2005, \mnras, 364, 1337


\bibitem[{{Silk} \& {Rees}(1998)}]{Silk+1998}
{Silk}, J. \& {Rees}, M.~J. 1998, \aap, 331, L1


\bibitem[{{Singh} {et~al.}(2013){Singh}, {van de Ven}, {Jahnke}, {Lyubenova},
  {Falc{\'o}n-Barroso}, {Alves}, {Cid Fernandes}, {Galbany},
  {Garc{\'{\i}}a-Benito}, {Husemann}, {Kennicutt}, {Marino}, {M{\'a}rquez},
  {Masegosa}, {Mast}, {Pasquali}, {S{\'a}nchez}, {Walcher}, {Wild}, {Wisotzki},
  \& {Ziegler}}]{sign13}
{Singh}, R., {van de Ven}, G., {Jahnke}, K., {Lyubenova}, M.,
  {Falc{\'o}n-Barroso}, J., {Alves}, J., {Cid Fernandes}, R., {Galbany}, L.,
  {Garc{\'{\i}}a-Benito}, R., {Husemann}, B., {Kennicutt}, R.~C., {Marino},
  R.~A., {M{\'a}rquez}, I., {Masegosa}, J., {Mast}, D., {Pasquali}, A.,
  {S{\'a}nchez}, S.~F., {Walcher}, J., {Wild}, V., {Wisotzki}, L., \&
  {Ziegler}, B. 2013, \aap, 558, A43


\bibitem[{{Smee} {et~al.}(2013){Smee}, {Gunn}, {Uomoto}, {Roe}, {Schlegel},
  {Rockosi}, {Carr}, {Leger}, {Dawson}, {Olmstead}, {Brinkmann}, {Owen},
  {Barkhouser}, {Honscheid}, {Harding}, {Long}, {Lupton}, {Loomis}, {Anderson},
  {Annis}, {Bernardi}, {Bhardwaj}, {Bizyaev}, {Bolton}, {Brewington}, {Briggs},
  {Burles}, {Burns}, {Castander}, {Connolly}, {Davenport}, {Ebelke}, {Epps},
  {Feldman}, {Friedman}, {Frieman}, {Heckman}, {Hull}, {Knapp}, {Lawrence},
  {Loveday}, {Mannery}, {Malanushenko}, {Malanushenko}, {Merrelli}, {Muna},
  {Newman}, {Nichol}, {Oravetz}, {Pan}, {Pope}, {Ricketts}, {Shelden},
  {Sandford}, {Siegmund}, {Simmons}, {Smith}, {Snedden}, {Schneider},
  {SubbaRao}, {Tremonti}, {Waddell}, \& {York}}]{2013AJ....146...32S}
{Smee}, S.~A., {Gunn}, J.~E., {Uomoto}, A., {Roe}, N., {Schlegel}, D.,
  {Rockosi}, C.~M., {Carr}, M.~A., {Leger}, F., {Dawson}, K.~S., {Olmstead},
  M.~D., {Brinkmann}, J., {Owen}, R., {Barkhouser}, R.~H., {Honscheid}, K.,
  {Harding}, P., {Long}, D., {Lupton}, R.~H., {Loomis}, C., {Anderson}, L.,
  {Annis}, J., {Bernardi}, M., {Bhardwaj}, V., {Bizyaev}, D., {Bolton}, A.~S.,
  {Brewington}, H., {Briggs}, J.~W., {Burles}, S., {Burns}, J.~G., {Castander},
  F.~J., {Connolly}, A., {Davenport}, J.~R.~A., {Ebelke}, G., {Epps}, H.,
  {Feldman}, P.~D., {Friedman}, S.~D., {Frieman}, J., {Heckman}, T., {Hull},
  C.~L., {Knapp}, G.~R., {Lawrence}, D.~M., {Loveday}, J., {Mannery}, E.~J.,
  {Malanushenko}, E., {Malanushenko}, V., {Merrelli}, A.~J., {Muna}, D.,
  {Newman}, P.~R., {Nichol}, R.~C., {Oravetz}, D., {Pan}, K., {Pope}, A.~C.,
  {Ricketts}, P.~G., {Shelden}, A., {Sandford}, D., {Siegmund}, W., {Simmons},
  A., {Smith}, D.~S., {Snedden}, S., {Schneider}, D.~P., {SubbaRao}, M.,
  {Tremonti}, C., {Waddell}, P., \& {York}, D.~G. 2013, \aj, 146, 32


\bibitem[{{Smethurst} {et~al.}(2015){Smethurst}, {Lintott}, {Simmons},
  {Schawinski}, {Marshall}, {Bamford}, {Fortson}, {Kaviraj}, {Masters},
  {Melvin}, {Nichol}, {Skibba}, \& {Willett}}]{Smethurst+2015}
{Smethurst}, R.~J., {Lintott}, C.~J., {Simmons}, B.~D., {Schawinski}, K.,
  {Marshall}, P.~J., {Bamford}, S., {Fortson}, L., {Kaviraj}, S., {Masters},
  K.~L., {Melvin}, T., {Nichol}, R.~C., {Skibba}, R.~A., \& {Willett}, K.~W.
  2015, \mnras, 450, 435


\bibitem[{{Somerville} {et~al.}(2008){Somerville}, {Hopkins}, {Cox},
  {Robertson}, \& {Hernquist}}]{Somerville+2008}
{Somerville}, R.~S., {Hopkins}, P.~F., {Cox}, T.~J., {Robertson}, B.~E., \&
  {Hernquist}, L. 2008, \mnras, 391, 481


\bibitem[{{Sparre} {et~al.}(2015){Sparre}, {Hayward}, {Springel},
  {Vogelsberger}, {Genel}, {Torrey}, {Nelson}, {Sijacki}, \&
  {Hernquist}}]{Sparre15}
{Sparre}, M., {Hayward}, C.~C., {Springel}, V., {Vogelsberger}, M., {Genel},
  S., {Torrey}, P., {Nelson}, D., {Sijacki}, D., \& {Hernquist}, L. 2015,
  \mnras, 447, 3548


\bibitem[{{Speagle} {et~al.}(2014){Speagle}, {Steinhardt}, {Capak}, \&
  {Silverman}}]{Speagle14}
{Speagle}, J.~S., {Steinhardt}, C.~L., {Capak}, P.~L., \& {Silverman}, J.~D.
  2014, \apjs, 214, 15


\bibitem[{{Stasi{\'n}ska} {et~al.}(2008){Stasi{\'n}ska}, {Vale Asari}, {Cid
  Fernandes}, {Gomes}, {Schlickmann}, {Mateus}, {Schoenell}, {Sodr{\'e}}, \&
  {Seagal Collaboration}}]{sta08}
{Stasi{\'n}ska}, G., {Vale Asari}, N., {Cid Fernandes}, R., {Gomes}, J.~M.,
  {Schlickmann}, M., {Mateus}, A., {Schoenell}, W., {Sodr{\'e}}, Jr., L., \&
  {Seagal Collaboration}. 2008, \mnras, 391, L29


\bibitem[{{Strateva} {et~al.}(2001){Strateva}, {Ivezi{\'c}}, {Knapp},
  {Narayanan}, {Strauss}, {Gunn}, {Lupton}, {Schlegel}, {Bahcall}, {Brinkmann},
  {Brunner}, {Budav{\'a}ri}, {Csabai}, {Castander}, {Doi}, {Fukugita}, {Gy{\H
  o}ry}, {Hamabe}, {Hennessy}, {Ichikawa}, {Kunszt}, {Lamb}, {McKay},
  {Okamura}, {Racusin}, {Sekiguchi}, {Schneider}, {Shimasaku}, \&
  {York}}]{Strateva+2001}
{Strateva}, I., {Ivezi{\'c}}, {\v Z}., {Knapp}, G.~R., {Narayanan}, V.~K.,
  {Strauss}, M.~A., {Gunn}, J.~E., {Lupton}, R.~H., {Schlegel}, D., {Bahcall},
  N.~A., {Brinkmann}, J., {Brunner}, R.~J., {Budav{\'a}ri}, T., {Csabai}, I.,
  {Castander}, F.~J., {Doi}, M., {Fukugita}, M., {Gy{\H o}ry}, Z., {Hamabe},
  M., {Hennessy}, G., {Ichikawa}, T., {Kunszt}, P.~Z., {Lamb}, D.~Q., {McKay},
  T.~A., {Okamura}, S., {Racusin}, J., {Sekiguchi}, M., {Schneider}, D.~P.,
  {Shimasaku}, K., \& {York}, D. 2001, \aj, 122, 1861


\bibitem[{{Tadhunter} {et~al.}(2012){Tadhunter}, {Ramos Almeida}, {Morganti},
  {Holt}, {Rose}, {Dicken}, \& {Inskip}}]{Tadhunter12}
{Tadhunter}, C.~N., {Ramos Almeida}, C., {Morganti}, R., {Holt}, J., {Rose},
  M., {Dicken}, D., \& {Inskip}, K. 2012, \mnras, 427, 1603


\bibitem[{{Thomas} {et~al.}(2010){Thomas}, {Maraston}, {Schawinski}, {Sarzi},
  \& {Silk}}]{Thomas+2010}
{Thomas}, D., {Maraston}, C., {Schawinski}, K., {Sarzi}, M., \& {Silk}, J.
  2010, \mnras, 404, 1775


\bibitem[{{Torres-Papaqui} {et~al.}(2012){Torres-Papaqui}, {Coziol},
  {Andernach}, {Ortega-Minakata}, {Neri-Larios}, \&
  {Plauchu-Frayn}}]{torres-papaqui12}
{Torres-Papaqui}, J.~P., {Coziol}, R., {Andernach}, H., {Ortega-Minakata},
  R.~A., {Neri-Larios}, D.~M., \& {Plauchu-Frayn}, I. 2012, \rmxaa, 48, 275


\bibitem[{{Torres-Papaqui} {et~al.}(2013){Torres-Papaqui}, {Coziol},
  {Plauchu-Frayn}, {Andernach}, \& {Ortega-Minakata}}]{torres-papaqui13}
{Torres-Papaqui}, J.~P., {Coziol}, R., {Plauchu-Frayn}, I., {Andernach}, H., \&
  {Ortega-Minakata}, R.~A. 2013, \rmxaa, 49, 311


\bibitem[{{Trump} {et~al.}(2015){Trump}, {Sun}, {Zeimann}, {Luck}, {Bridge},
  {Grier}, {Hagen}, {Juneau}, {Montero-Dorta}, {Rosario}, {Brandt},
  {Ciardullo}, \& {Schneider}}]{Trump+2015}
{Trump}, J.~R., {Sun}, M., {Zeimann}, G.~R., {Luck}, C., {Bridge}, J.~S.,
  {Grier}, C.~J., {Hagen}, A., {Juneau}, S., {Montero-Dorta}, A., {Rosario},
  D.~J., {Brandt}, W.~N., {Ciardullo}, R., \& {Schneider}, D.~P. 2015, \apj,
  811, 26


\bibitem[{{Urry} \& {Padovani}(1995)}]{urry95}
{Urry}, C.~M. \& {Padovani}, P. 1995, \pasp, 107, 803


\bibitem[{{Utomo} {et~al.}(2017){Utomo}, {Bolatto}, {Wong}, {Ostriker},
  {Blitz}, {Sanchez}, {Colombo}, {Leroy}, {Cao}, {Dannerbauer},
  {Garcia-Benito}, {Husemann}, {Kalinova}, {Levy}, {Mast}, {Rosolowsky}, \&
  {Vogel}}]{utomo17}
{Utomo}, D., {Bolatto}, A.~D., {Wong}, T., {Ostriker}, E.~C., {Blitz}, L.,
  {Sanchez}, S.~F., {Colombo}, D., {Leroy}, A.~K., {Cao}, Y., {Dannerbauer},
  H., {Garcia-Benito}, R., {Husemann}, B., {Kalinova}, V., {Levy}, R.~C.,
  {Mast}, D., {Rosolowsky}, E., \& {Vogel}, S.~N. 2017, \apj, 849, 26


\bibitem[{{Veilleux} {et~al.}(1995){Veilleux}, {Kim}, {Sanders}, {Mazzarella},
  \& {Soifer}}]{veil95}
{Veilleux}, S., {Kim}, D.-C., {Sanders}, D.~B., {Mazzarella}, J.~M., \&
  {Soifer}, B.~T. 1995, \apjs, 98, 171


\bibitem[{{Villarroel} \& {Korn}(2014)}]{villarroel14}
{Villarroel}, B. \& {Korn}, A.~J. 2014, Nature Physics, 10, 417


\bibitem[{{Vulcani} {et~al.}(2015){Vulcani}, {Poggianti}, {Fritz}, {Fasano},
  {Moretti}, {Calvi}, \& {Paccagnella}}]{Vulcani+2015}
{Vulcani}, B., {Poggianti}, B.~M., {Fritz}, J., {Fasano}, G., {Moretti}, A.,
  {Calvi}, R., \& {Paccagnella}, A. 2015, \apj, 798, 52


\bibitem[{{Wild} {et~al.}(2014){Wild}, {Rosales-Ortega}, {Falc{\'o}n-Barroso},
  {Garc{\'{\i}}a-Benito}, {Gallazzi}, {Gonz{\'a}lez Delgado}, {Bekerait{\'e}},
  {Pasquali}, {Johansson}, {Garc{\'{\i}}a Lorenzo}, {van de Ven}, {Pawlik},
  {Per{\'e}z}, {Monreal-Ibero}, {Lyubenova}, {Cid Fernandes},
  {M{\'e}ndez-Abreu}, {Barrera-Ballesteros}, {Kehrig}, {Iglesias-P{\'a}ramo},
  {Bomans}, {M{\'a}rquez}, {Johnson}, {Kennicutt}, {Husemann}, {Mast},
  {S{\'a}nchez}, {Walcher}, {Alves}, {Aguerri}, {Alonso Herrero},
  {Bland-Hawthorn}, {Catal{\'a}n-Torrecilla}, {Florido}, {Gomes}, {Jahnke},
  {L{\'o}pez-S{\'a}nchez}, {de Lorenzo-C{\'a}ceres}, {Marino},
  {M{\'a}rmol-Queralt{\'o}}, {Olden}, {del Olmo}, {Papaderos}, {Quirrenbach},
  {V{\'{\i}}lchez}, \& {Ziegler}}]{wild14}
{Wild}, V., {Rosales-Ortega}, F., {Falc{\'o}n-Barroso}, J.,
  {Garc{\'{\i}}a-Benito}, R., {Gallazzi}, A., {Gonz{\'a}lez Delgado}, R.~M.,
  {Bekerait{\'e}}, S., {Pasquali}, A., {Johansson}, P.~H., {Garc{\'{\i}}a
  Lorenzo}, B., {van de Ven}, G., {Pawlik}, M., {Per{\'e}z}, E.,
  {Monreal-Ibero}, A., {Lyubenova}, M., {Cid Fernandes}, R.,
  {M{\'e}ndez-Abreu}, J., {Barrera-Ballesteros}, J., {Kehrig}, C.,
  {Iglesias-P{\'a}ramo}, J., {Bomans}, D.~J., {M{\'a}rquez}, I., {Johnson},
  B.~D., {Kennicutt}, R.~C., {Husemann}, B., {Mast}, D., {S{\'a}nchez}, S.~F.,
  {Walcher}, C.~J., {Alves}, J., {Aguerri}, A.~L., {Alonso Herrero}, A.,
  {Bland-Hawthorn}, J., {Catal{\'a}n-Torrecilla}, C., {Florido}, E., {Gomes},
  J.~M., {Jahnke}, K., {L{\'o}pez-S{\'a}nchez}, {\'A}.~R., {de
  Lorenzo-C{\'a}ceres}, A., {Marino}, R.~A., {M{\'a}rmol-Queralt{\'o}}, E.,
  {Olden}, P., {del Olmo}, A., {Papaderos}, P., {Quirrenbach}, A.,
  {V{\'{\i}}lchez}, J.~M., \& {Ziegler}, B. 2014, \aap, 567, A132


\bibitem[{{Willott} {et~al.}(2001){Willott}, {Rawlings}, \&
  {Blundell}}]{willott01}
{Willott}, C.~J., {Rawlings}, S., \& {Blundell}, K.~M. 2001, \mnras, 324, 1


\bibitem[{{Wolf} {et~al.}(2005){Wolf}, {Bell}, {McIntosh}, {Rix}, {Barden},
  {Beckwith}, {Borch}, {Caldwell}, {H{\"a}ussler}, {Heymans}, {Jahnke},
  {Jogee}, {Meisenheimer}, {Peng}, {S{\'a}nchez}, {Somerville}, \&
  {Wisotzki}}]{wolf05}
{Wolf}, C., {Bell}, E.~F., {McIntosh}, D.~H., {Rix}, H., {Barden}, M.,
  {Beckwith}, S.~V.~W., {Borch}, A., {Caldwell}, J.~A.~R., {H{\"a}ussler}, B.,
  {Heymans}, C., {Jahnke}, K., {Jogee}, S., {Meisenheimer}, K., {Peng}, C.~Y.,
  {S{\'a}nchez}, S.~F., {Somerville}, R.~S., \& {Wisotzki}, L. 2005, \apj, 630,
  771


\bibitem[{{Xue} {et~al.}(2010){Xue}, {Brandt}, {Luo}, {Rafferty}, {Alexander},
  {Bauer}, {Lehmer}, {Schneider}, \& {Silverman}}]{Xue+2010}
{Xue}, Y.~Q., {Brandt}, W.~N., {Luo}, B., {Rafferty}, D.~A., {Alexander},
  D.~M., {Bauer}, F.~E., {Lehmer}, B.~D., {Schneider}, D.~P., \& {Silverman},
  J.~D. 2010, \apj, 720, 368


\bibitem[{{Yan} {et~al.}(2016{\natexlab{a}}){Yan}, {Bundy}, {Law}, {Bershady},
  {Andrews}, {Cherinka}, {Diamond-Stanic}, {Drory}, {MacDonald},
  {S{\'a}nchez-Gallego}, {Thomas}, {Wake}, {Weijmans}, {Westfall}, {Zhang},
  {Arag{\'o}n-Salamanca}, {Belfiore}, {Bizyaev}, {Blanc}, {Blanton},
  {Brownstein}, {Cappellari}, {D'Souza}, {Emsellem}, {Fu}, {Gaulme}, {Graham},
  {Goddard}, {Gunn}, {Harding}, {Jones}, {Kinemuchi}, {Li}, {Li}, {Maiolino},
  {Mao}, {Maraston}, {Masters}, {Merrifield}, {Oravetz}, {Pan}, {Parejko},
  {Sanchez}, {Schlegel}, {Simmons}, {Thanjavur}, {Tinker}, {Tremonti}, {van den
  Bosch}, \& {Zheng}}]{renbin16b}
{Yan}, R., {Bundy}, K., {Law}, D.~R., {Bershady}, M.~A., {Andrews}, B.,
  {Cherinka}, B., {Diamond-Stanic}, A.~M., {Drory}, N., {MacDonald}, N.,
  {S{\'a}nchez-Gallego}, J.~R., {Thomas}, D., {Wake}, D.~A., {Weijmans}, A.-M.,
  {Westfall}, K.~B., {Zhang}, K., {Arag{\'o}n-Salamanca}, A., {Belfiore}, F.,
  {Bizyaev}, D., {Blanc}, G.~A., {Blanton}, M.~R., {Brownstein}, J.,
  {Cappellari}, M., {D'Souza}, R., {Emsellem}, E., {Fu}, H., {Gaulme}, P.,
  {Graham}, M.~T., {Goddard}, D., {Gunn}, J.~E., {Harding}, P., {Jones}, A.,
  {Kinemuchi}, K., {Li}, C., {Li}, H., {Maiolino}, R., {Mao}, S., {Maraston},
  C., {Masters}, K., {Merrifield}, M.~R., {Oravetz}, D., {Pan}, K., {Parejko},
  J.~K., {Sanchez}, S.~F., {Schlegel}, D., {Simmons}, A., {Thanjavur}, K.,
  {Tinker}, J., {Tremonti}, C., {van den Bosch}, R., \& {Zheng}, Z.
  2016{\natexlab{a}}, \aj, 152, 197


\bibitem[{{Yan} {et~al.}(2016{\natexlab{b}}){Yan}, {Tremonti}, {Bershady},
  {Law}, {Schlegel}, {Bundy}, {Drory}, {MacDonald}, {Bizyaev}, {Blanc},
  {Blanton}, {Cherinka}, {Eigenbrot}, {Gunn}, {Harding}, {Hogg},
  {S{\'a}nchez-Gallego}, {S{\'a}nchez}, {Wake}, {Weijmans}, {Xiao}, \&
  {Zhang}}]{2016AJ....151....8Y}
{Yan}, R., {Tremonti}, C., {Bershady}, M.~A., {Law}, D.~R., {Schlegel}, D.~J.,
  {Bundy}, K., {Drory}, N., {MacDonald}, N., {Bizyaev}, D., {Blanc}, G.~A.,
  {Blanton}, M.~R., {Cherinka}, B., {Eigenbrot}, A., {Gunn}, J.~E., {Harding},
  P., {Hogg}, D.~W., {S{\'a}nchez-Gallego}, J.~R., {S{\'a}nchez}, S.~F.,
  {Wake}, D.~A., {Weijmans}, A.-M., {Xiao}, T., \& {Zhang}, K.
  2016{\natexlab{b}}, \aj, 151, 8


\bibitem[{{Yates} \& {Kauffmann}(2014)}]{yates14}
{Yates}, R.~M. \& {Kauffmann}, G. 2014, \mnras, 439, 3817


\bibitem[{{Young} {et~al.}(2011){Young}, {Bureau}, {Davis}, {Combes},
  {McDermid}, {Alatalo}, {Blitz}, {Bois}, {Bournaud}, {Cappellari}, {Davies},
  {de Zeeuw}, {Emsellem}, {Khochfar}, {Krajnovi{\'c}}, {Kuntschner},
  {Lablanche}, {Morganti}, {Naab}, {Oosterloo}, {Sarzi}, {Scott}, {Serra}, \&
  {Weijmans}}]{young11}
{Young}, L.~M., {Bureau}, M., {Davis}, T.~A., {Combes}, F., {McDermid}, R.~M.,
  {Alatalo}, K., {Blitz}, L., {Bois}, M., {Bournaud}, F., {Cappellari}, M.,
  {Davies}, R.~L., {de Zeeuw}, P.~T., {Emsellem}, E., {Khochfar}, S.,
  {Krajnovi{\'c}}, D., {Kuntschner}, H., {Lablanche}, P.-Y., {Morganti}, R.,
  {Naab}, T., {Oosterloo}, T., {Sarzi}, M., {Scott}, N., {Serra}, P., \&
  {Weijmans}, A.-M. 2011, \mnras, 414, 940


\bibitem[{{Zhang} {et~al.}(2016){Zhang}, {Shi}, {Rieke}, {Xia}, {Wang}, {Sun},
  \& {Wan}}]{zhang16}
{Zhang}, Z., {Shi}, Y., {Rieke}, G.~H., {Xia}, X., {Wang}, Y., {Sun}, B., \&
  {Wan}, L. 2016, \apjl, 819, L27


\end{thebibliography}

\end{document}